\documentclass[aps,prx,twocolumn,amsmath,amssymb,showpacs,superscriptaddress]{revtex4-2}
\usepackage{graphicx}
\usepackage{amssymb}
\usepackage{bm}
\usepackage{braket}
\usepackage{amsmath}
\usepackage{tabularx}
\usepackage{ltablex}
\usepackage{multirow}
\usepackage[dvipsnames]{xcolor}
\usepackage[normalem]{ulem}
\newcommand{\kk}{{\bf k}}
\newcommand{\nbands}{{\cal J}}

\usepackage[colorlinks=true,
                linkcolor=blue,
	        citecolor=blue,
	        urlcolor=blue]{hyperref}

\newcommand\mbf{\mathbf}
\newcommand{\sissa}{Scuola Internazionale Superiore di Studi Avanzati (SISSA), I-34136 Trieste, Italy}
\newcommand{\trieste}{Dipartimento di Fisica, Universit\`a di Trieste, I-34151 Trieste, Italy}
\newcommand{\psiaff}{Laboratory for Materials Simulations, and National Centre for Computational Design and Discovery of Novel Materials (MARVEL), Paul Scherrer Institut, 5232 Villigen, Switzerland}
\usepackage{verbatim}

\newcommand\code[1]{\texttt{#1}}

\AtBeginDocument{\usepackage{booktabs}}
\begin{document}
\title{Spin-dependent interactions in orbital-density-dependent functionals: \\ non-collinear Koopmans spectral functionals}
\date{\today}
\author{Antimo Marrazzo}
\email{amarrazz@sissa.it}
\affiliation{\sissa}
\affiliation{\trieste}

\author{Nicola Colonna}
\email{nicola.colonna@psi.ch}
\affiliation{\psiaff}

\begin{abstract}
The presence of spin-orbit coupling or non-collinear magnetic spin states can have dramatic effects on the ground-state and spectral properties of materials, in particular on the band structure. Here, we develop non-collinear Koopmans-compliant functionals based on Wannier functions and density-functional perturbation theory, targeting accurate spectral properties in the quasiparticle approximation. Our non-collinear Koopmans-compliant theory involves functionals of four-component orbitals densities, that can be obtained from the charge and spin-vector densities of Wannier functions. 
We validate our approach on four emblematic non-magnetic and magnetic semiconductors where the effect of spin-orbit coupling goes from small to very large: the III-IV semiconductor GaAs, the transition-metal dichalcogenide WSe$_2$, the cubic perovskite CsPbBr$_3$, and the ferromagnetic semiconductor CrI$_3$. 
The predicted band gaps are comparable in accuracy to state-of-the-art many-body perturbation theory and include spin-dependent interactions and screening effects that are missing in standard diagrammatic approaches based on the random phase approximation. While the inclusion of orbital- and spin-dependent interactions in many-body perturbation theory requires self-screening or vertex corrections, they emerge naturally in the Koopmans-functionals framework.
\end{abstract}

\maketitle
\section{Introduction}
The spin degree of freedom is crucial to describe electrons in materials and molecules. At a fundamental level, spin properties and spin-dependent interactions can play important roles and sometimes be dominant effects. Indeed, many technological applications related to the storage and transmission of information rely on pure spin phenomena or to interactions between the spin and the orbital degrees of freedom.
Remarkably, most electronic structure calculations often take into account spin only implicitly, neglecting spin-dependent couplings.
A prime example is density functional theory (DFT), where the total energy is a functional of the ground-state charge density~\cite{KS_pr_1965,martin_book_2020} and Kohn-Sham (KS) orbitals are scalar quantities. A simple extension of DFT for magnetic systems, named spinDFT, is based on a functional of two scalar densities corresponding to the spin-up and spin-down component of the ground-state electronic distribution, where a fixed quantization axis is aligned to the magnetization direction. SpinDFT allows studying ferromagnetic or antiferromagnetic orders, which are called collinear as there is a common quantization axis for the whole system.

While the majority of magnets can effectively be described by either ferromagnetic or antiferromagnetic collinear magnetic order, magnetic system can also exhibit non-collinear (NC) spin states, where the spin density\textemdash which in general is a vector\textemdash can change direction over space and there is no common spin quantization axis for the entire crystal~\cite{kubler_book_2017}. Beyond relatively simple co-planar spin states, more exotic NC ground states are spin spirals~\cite{kubler_book_2017}, skyrmions~\cite{nagaosa_naturenanotech_2013,fer_natrevmat_2017} and spin glasses~\cite{kamber_science_2020,verlhac_natphys_2022}.
A major source of non-collinearity is spin-orbit coupling (SOC), which is responsible for important effects such as the Dzyaloshinskii-Moriya interaction and the magnetic anisotropy~\cite{szilva_rmp_2023}. Crucially, SOC can lead to strong spin couplings even in the absence of magnetism, that is for time-reversal (TR) invariant systems.

Being a relativistic effect SOC is always present, but it is stronger in presence of heavy chemical elements: while often negligible for carbon, SOC has already tangible consequences in GaAs semiconductors and becomes a dominant factor for fifth-period elements. The impact of SOC on the electronic structure can be dramatic, including splitting of bands and renormalization of band gaps. For example, the band gap of organohalide perovskites is about 1 eV smaller than it would be in absence of SOC~\cite{Mosconi_PCCP_2016,galli_jctc_2016}. Finally, SOC can also affect the structural properties such as bond lengths, phonon frequencies~\cite{dalcorso_JPCM_2013} and even structural stability~\cite{marrazzo_PRL_2018,monserrat_JPM_2019}.

A NC extension~\cite{hedin_JPCSSP_1972,kubler_jpf_1988} of spinDFT can be obtained by considering a total energy functional that depends on the ground-state charge density as well as the ground-state spin-density vector, where the latter can change direction over space. In this framework, KS orbitals are two-component spinors and the KS Hamiltonian can be written in terms of the Pauli matrices in spin-space. Today, NC DFT calculations with SOC have become routine, although sensibly more computational intensive than their collinear counterparts. The higher computational cost of NC calculations is even more relevant in the context of many-body perturbation theory (MBPT), which is the gold standard for band structure calculations of solids. While one-shot $G_0W_0$ calculations with SOC have been performed in the literature (see e.g.,~\cite{sakuma_prb_2011,aguilera_prb_2013,aguilera_prb_2013bis,molina_prb_2013,chulkov_prb_2013,galli_jctc_2016,Mosconi_PCCP_2016,marrazzo_PRL_2018,marrazzo_nanolett_2019,marsili_PRB_2021,aguilera_prb_2022}), it is still very challenging to perform NC calculations that include SOC with more accurate methods such as self-consistent $GW$ (scGW)~\cite{zgid_prb_2022,zgid_fd_2024} or quasi-particle self-consistent $GW$ ($QSGW$)~\cite{brivio_prb_2014,aguilera_prb_2015,clark_cp_2021,aguilera_prb_2021}, even more so if vertex corrections in the screened interaction W are included ($QSG\tilde{W}$)~\cite{kresse_prl_2007,pasquarello_prb_2015,schilf_prb_2023}. Notably, beyond-$G_0W_0$ MBPT calculations often treat SOC with the second-variation approach~\cite{kioupakis_prb_2010,yazyev_prb_2012} or with lower levels of theory such $G_0W_0$~\cite{wiktor_jpcl_2017} or hybrid functionals~\cite{kresse_prb_2007}, which are approximations that might be rather inaccurate in systems where SOC is more relevant~\cite{aguilera_prb_2013,marsili_PRB_2021}. Finally, we note that Hedin's equations~\cite{hedin_pr_1965} have spin dependence, while their extension to spin-dependent electron-electron interactions and the corresponding $GW$ approximation have been developed~\cite{biermann_prl_2008,biermann_jpcm_2009}. Nonetheless, actual $GW$ calculations are performed with a spin-independent interaction at the diagrammatic level~\cite{marsili_PRB_2021,louie_prb_2022}. In other words, spin-dependent screening effects are not usually accounted for.

In this work we develop a theory and implementation to calculate accurate spectral properties of NC electronic structures of materials, such as in presence of strong SOC, based on a functional dependent upon the orbital charge and spin-vector densities. Our approach enforces the Koopmans-compliance condition to each spin-orbital and leads to a spectral functional of four-component orbitals densities, which can be obtained from the charge and spin-vector densities of Wannier functions (WFs). The formalism takes into account spin-dependent screening effects related to spin-spin and spin-charge interactions.
We implement and validate the theory as a one-shot approach that corrects DFT band structures, where screening coefficients are calculated through NC density-functional perturbation theory (DFPT)~\cite{baroni_phonons_2001,gorni_spin-fluctuation_2016,gorni_spin_2018,cao_ab_2018,urru_density_2019, urru_lattice_2020}.

\section{A primer on Koopmans-compliant functionals}
Koopmans-compliant (KC) spectral functionals~\cite{Dabo2010,Borghi2014,colonna_jctc_2018,Linh2018, linscott_koopmans_2023} are orbital-dependent functionals capable of delivering accurate spectral properties for molecular~\cite{nguyen_first-principles_2015,nguyen_first-principles_2016,Elliott2019,colonna_jctc_2019} and extended systems~\cite{Linh2018,de_almeida_electronic_2021,degennaro_prb_2022,colonna_jctc_2022,ingall_accurate_2024} at low computational cost and complexity. Remarkably, the KC approach maintains a simple functional formulation while being typically as accurate as state of the art in Green's function theory~\cite{Linh2018,colonna_jctc_2019,colonna_jctc_2022,ingall_accurate_2024}, at a cost which is broadly comparable to standard DFT. The simplicity and accuracy of the KC framework rests on three fundamental concepts: linearization, screening, and localization. First, a generalized linearization condition is imposed on each charged excitation: the energy of any orbital must be independent of the occupation of the orbital itself. This is a necessary condition for a correct description of an electron addition/removal process and implies that the KC total energy functional is piecewise linear with respect to fractional occupations. 
Second, screening and relaxation effects (due to the electron addition/removal) are taken into account by orbital-dependent screening coefficients, which can be calculated by finite differences~\cite{Linh2018} or linear-response approaches~\cite{colonna_jctc_2018,colonna_jctc_2022}. Finally, the Koopmans compliance is imposed on the variational orbitals---i.e., those that minimize the KC energy functional---which are typically localized in space. 
For periodic systems, the variational orbitals are Wannier like, and typically resemble maximally-localized WFs (MLWFs)~\cite{Marzari2012,Linh2018,colonna_jctc_2022}. This property has allowed the development of a Wannier-interpolation and unfolding scheme to calculate the band structure from a supercell Koopmans-functional calculation~\cite{degennaro_prb_2022} and more recently the development of a convenient Koopmans formulation that operates fully in periodic-boundary conditions (PBC) and it is based on explicit Brillouin-zone (BZ) sampling and DFPT~\cite{colonna_jctc_2022}. 
This Koopmans-Wannier (KCW) formulation~\cite{colonna_jctc_2022} can be deployed as a one-shot correction to DFT and delivers improved scaling with system size, making band-structure calculations with KC functionals much more straightforward. KC functionals resonate with other efforts aimed at calculating excitation energies where the piece-wise linearity (PWL) condition and the use of localized orbitals are often a key ingredient~\cite{Anisimov2005,Anisimov2007,Kronik2007,Galli2014,Weitao2015,Weitao2017,Kronik2021,Ma2016}.
For an exhaustive and detailed description of the Koopmans functionals we refer the reader to Refs.~\cite{linscott_koopmans_2023, colonna_jctc_2022, Linh2018}.

\section{Non-collinear Koopmans-compliant functionals}
As first step and in the spirit of what has been done for collinear systems in Ref.~\cite{Borghi2014}, we introduce a NC KC functional that, once added to the NC DFT energy functional, linearizes the total energy with respect to orbital occupations:
\begin{widetext}
\begin{equation}
  \label{eq:nc_koopfunc}
\Pi_i^{\rm rKI}= - \left\{E^{\rm DFT}[\rho,\mathbf{m}] - E^{\rm DFT}[\rho^{f_i=0},\mathbf{m}^{f_i=0}]\right\} + f_i \left\{E^{\rm DFT}[\rho^{f_i=1},\mathbf{m}^{f_i=1}] - E^{\rm DFT}[\rho^{f_i=0},\mathbf{m}^{f_i=0}]\right\}
\end{equation}
\end{widetext}
where $E^{\rm DFT}$ is the DFT total energy, which is a functional of the total electron charge $\rho$ and spin-vector $\mbf{m}$ densities, and $f_i$ is the occupation of the $i$th orbital. 
This correction removes from the underlying DFT energy functional the contribution that is non-linear in the occupation $f_i$ and replaces it with a linear term that interpolates between integer occupation numbers; this enforces a generalized PWL condition that makes single-particle eigenvalues consistent with the energy differences that define charged excitations.
Evaluating the total energy differences appearing in the curly brackets of Eq.~(\ref{eq:nc_koopfunc})
can be done either by resorting to a frozen orbitals approximation plus a \textit{post hoc} scaling down of the frozen orbitals correction via a screening coefficient, as originally proposed~\cite{Dabo2010,Borghi2014}, or by resorting to a Taylor expansion of the DFT energy with respect to the occupation $f_i$ truncated to second order as discussed in Refs.~\cite{colonna_jctc_2018, colonna_jctc_2022}. 
In this work we follow the latter strategy as it enables an efficient implementation in PBCs using a primitive cell setup and a sampling of the BZ, and ultimately gives direct access to the band structure of periodic solids at reduced computational costs~\cite{colonna_jctc_2022} (a strategy to go beyond the second order approximation is discussed in Sec.~\ref{subsec:beyond_2nd}). 
By applying the second-order approximation to the NC Koopmans functional in Eq.~(\ref{eq:nc_koopfunc}) we get:
\begin{equation}
  \label{eq:2nd-ord_exp}
  \Pi_i^{\rm (2)rKI}  = \frac{1}{2}f_i(1-f_i)\frac{d^2E^{\rm DFT}}{df_i^2}\Big|_{\bar{f}}
\end{equation}
where $\bar{f}$ is the reference ground-state occupation and the superscript (r) marks that orbital relaxation effects are taken into account.

We note that even in the case of TR-invariant systems, the addition or removal of an electron generally breaks TR as the system becomes spin polarized. In fact, derivatives are first computed for a general NC system, possibly with a non-vanishing spin magnetization, and only \emph{later} evaluated for a system with TR symmetry (where $|m(\mathbf{r})|=0$ everywhere). This is a crucial aspect of the theory that we will elaborate more later on: the KC functional framework correctly requires dealing with perturbations that break TR-symmetry, even for TR-invariant systems. 

We exploit Hellmann-Feynman theorem and express Eq.~(\ref{eq:2nd-ord_exp}) as
\begin{eqnarray}
  \frac{d^2E^{\rm DFT}}{df_i^2}\Big|_{\bar{f}}&=& \frac{d\varepsilon_i}{df_i}\Big|_{\bar{f}} \nonumber\\
  &=& \left[\braket{\psi_i| \frac{dV_{\rm Hxc}}{df_i}|\psi_i}+\braket{\psi_i| \frac{d\mbf{W}_{\rm xc}\cdot\bm{\sigma}}{df_i}|\psi_i}\right]_{\bar{f}}\nonumber \\
  &=& \left[\braket{\psi_i| \frac{dV_{\rm Hxc}}{df_i}|\psi_i}+\braket{\psi_i| \frac{d{W}_{\rm xc}\hat{\mbf{m}}}{df_i}\cdot\bm{\sigma}|\psi_i}\right]_{\bar{f}},
  \label{eq:2nd_deriv_HF} 
\end{eqnarray}
where $\varepsilon_i = dE^{\rm DFT}/df_i = \langle \psi_i | \hat{h}^{\rm DFT} | \psi_i \rangle$ is the expectation value of the DFT Hamiltonian on the single-particle spin-orbitals $|\psi_i\rangle$. In the expression above, we separated the Hartree and exchange-correlation potential into a scalar part $V_{\rm Hxc}$ (which includes the Hartree term) and a spin-dependent part $\mbf{W}_{\rm xc}$, the latter is expressed on the basis of Pauli matrices $\bm{\sigma}$. Note that while our theory is very general, common DFT NC exchange-correlation potentials adopt the local spin-density approximation (LSDA)~\cite{hedin_JPCSSP_1972}, hence they always point to the direction of local spin magnetization and do not include any spin torque~\cite{gross_prl_2007}.

We can evaluate the two terms of Eq.~(\ref{eq:nc_koopfunc}) by using the chain rule for functional derivatives:
\begin{eqnarray}
  \braket{\psi_i| \frac{dV_{\rm Hxc}}{df_i}|\psi_i} &=& \int d\mbf{r}d\mbf{r}' n_{i,\rho}(\mbf{r})\left(F_{\rm Hxc}^{\rho,\rho}(\mbf{r},\mbf{r}')\frac{d \rho(\mbf{r})}{d f_i}\right. + \nonumber \\
  && \left.\sum_{\alpha}F_{\rm xc}^{\rho,m_{\alpha}}(\mbf{r},\mbf{r}')\frac{d m_{\alpha}(\mbf{r'})}{d f_i}\right) ,
  \label{eq:hf_scalar}
\end{eqnarray}
  \begin{eqnarray}
  \braket{\psi_i|\frac{d{W}_{\rm xc,\alpha}}{df_i}\sigma_{\alpha}|\psi_i} &=& \int d\mbf{r}d\mbf{r}' n_{i,m_\alpha}(\mbf{r})\nonumber
  \left(F_{\rm xc}^{m_{\alpha},\rho}(\mbf{r},\mbf{r}')\frac{d \rho(\mbf{r})}{d f_i} +\right. \nonumber \\
  &&\left. \sum_{\beta}F_{\rm xc}^{m_{\alpha},m_{\beta}}(\mbf{r},\mbf{r}')\frac{d m_{\beta}(\mbf{r'})}{d f_i}\right)
  \label{eq:hf_spin}
\end{eqnarray}
where $F_{\rm Hxc}^{i,j}$ represents the charge and spin-magnetization components of the Hartree and exchange-correlation (Hxc) kernel. Equations~(\ref{eq:hf_scalar}) and (\ref{eq:hf_spin}) highlight the symmetry between the scalar and spin-dependent components, so we introduce a compact notation based on four-vector quantities for the electron number $\bm{n}_i(\mbf{r})$ and charge $\bm{\rho}_i(\mbf{r})$ densities, and four-by-four matrices for the Hxc kernel $\mbf{F}_{\rm Hxc}$:
\begin{equation}
  \label{eq:compact_2nd_order_deriv_NC}
  \braket{\psi_i| \frac{d\mbf{V}_{\rm Hxc}}{df_i}\cdot\tilde{\bm{\sigma}} |\psi_i}=\int d\mbf{r}d\mbf{r}' \bm{n}_i(\mbf{r}) \mbf{F}_{\rm Hxc}(\mbf{r},\mbf{r}')\frac{d \bm{\rho}(\mbf{r}')}{d f_i}
\end{equation}
where scalar and vector-matrix products are understood, and we introduced an extended set of Pauli matrices $\tilde{\bm{\sigma}}$ including a two-by-two identity matrix $\sigma_0$. Equation~(\ref{eq:compact_2nd_order_deriv_NC}) is perspicuous: the NC case can be recast in the same form of a collinear problem for four-vector densities and promoting the Hxc kernel to four-by-four matrices (compare Eq.~(\ref{eq:compact_2nd_order_deriv_NC}) above with Eq.~(5) in Ref.~\cite{colonna_jctc_2018}). This holds true also for the Dyson equations (see Supplementary Material~\cite{SM} for the derivation) that allow us to write the derivative of the density as
\begin{equation}
 \frac{d \bm{\rho}(\mbf{r})}{d f_i} = \bm{n}_i(\mbf{r}) + \int d\mbf{r}' \bm{\chi}(\mbf{r},\mbf{r}')\int d\mbf{r''}\mbf{F}_{\rm Hxc}(\mbf{r}',\mbf{r}'') \bm{n}_i(\mbf{r}''),
\end{equation}
where the NC interacting response function $ \bm{\chi}$ is calculated from the non-interacting one $ \bm{\chi}_{0}$ as 


\begin{eqnarray}
  \bm{\chi}(\mbf{r},\mbf{r}') &=& \bm{\chi}_0(\mbf{r},\mbf{r}') +   \int d\mbf{r}'' \bm{\chi}_0(\mbf{r},\mbf{r}'')\cdot \nonumber \\
&&\int d\mbf{r}''' \mbf{F}_{\rm Hxc}(\mbf{r}'',\mbf{r}''')  \ \bm{\chi}(\mbf{r}''',\mbf{r}').
\end{eqnarray}

We can use these results in Eq.~(\ref{eq:2nd-ord_exp}) and obtain an expression for the second-order expansion of the NC KC functional:
\begin{equation}
  \label{eq:2nd-KIfunc-dens}
  \Pi_i^{\rm (2)rKI} = \frac{1}{2}f_i(1-f_i) \int d\mbf{r}d\mbf{r}' \bm{n}_i(\mbf{r})  \bm{\mathbb{F}}_{\rm Hxc}(\mbf{r},\mbf{r}') \bm{n}_i(\mbf{r}')
\end{equation}
where we define the screened Hxc kernel as 
\begin{eqnarray}
  \label{eq:screened_kernel}
 \bm{\mathbb{F}}_{\rm Hxc}(\mbf{r},\mbf{r}')  &=&  \mbf{F}_{\rm Hxc}(\mbf{r},\mbf{r}')+ \int d\mbf{r}'' \mbf{F}_{\rm Hxc}(\mbf{r},\mbf{r}'') \nonumber \\
 &&\int d\mbf{r}''' \bm{\chi}(\mbf{r}'',\mbf{r}''')\mbf{F}_{\rm Hxc}(\mbf{r}''',\mbf{r}').
\end{eqnarray}

\subsection{Non-collinear Koopmans potentials}
\begin{figure*}[th]
  \centering
  \includegraphics[width=1\linewidth]{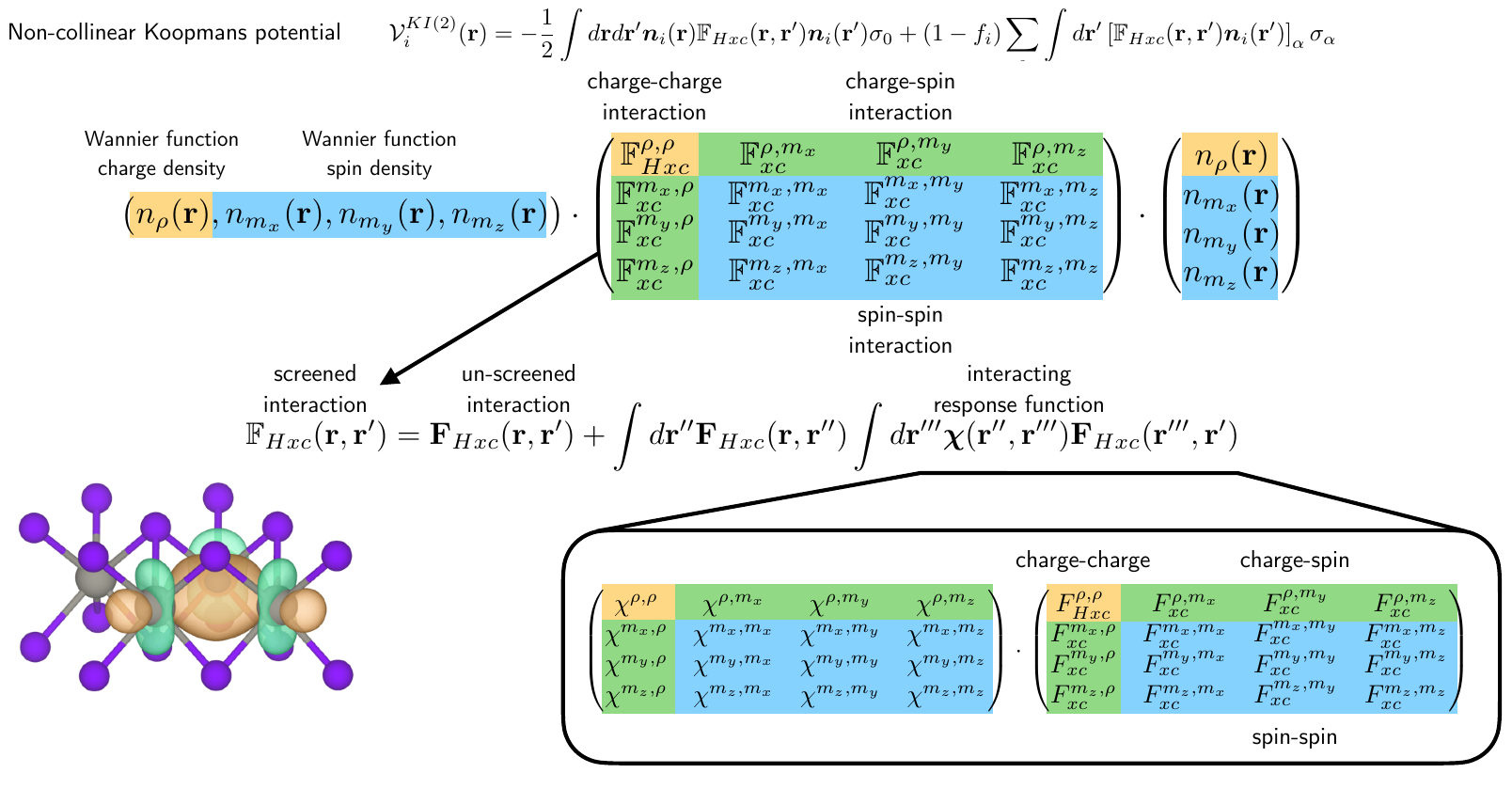}
  \caption{Schematics of the charge-charge (yellow), charge-spin (green) and spin-spin (blue) interactions and screening effects in the non-collinear Koopmans-Wannier functional theory. In the bottom left corner, one of the valence Wannier functions for bulk WSe$_2$ is shown. The non-collinear Koopmans potential is based on an interaction kernel that couples the spin and charge degrees of freedom of the (Wannier) orbital densities, even for non-magnetic systems with time-reversal symmetry. Also, the screening of the interaction is treated as non-collinear and involves spin-dependent terms originating both from the kernel and the response function of the charge and magnetization densities. Note that in the random phase approximation all the spin-charge and spin-spin channels would be absent and the charge-charge interaction would be given simply by the Hartree term alone (as opposed to the Hartree and exchange-correlation kernel, which appear instead in the Koopmans approach).}
  \label{fig:int_screen}
\end{figure*}

From Eq.~(\ref{eq:2nd-KIfunc-dens}) we can derive the corresponding local and orbital-dependent potential by taking the functional derivative with respect to all components of the four-vector orbital density and write the result in a compact form:
\begin{eqnarray}
  \label{eq:nc_kop_potential}
  \mathcal{V}^{\rm KI(2)}_{i}(\mbf{r}) =&& -\frac{1}{2}\int d\mbf{r}d\mbf{r}' \bm{n}_i(\mbf{r}) \mathbb{F}_{\rm Hxc}(\mbf{r},\mbf{r}') \bm{n}_i(\mbf{r}')\sigma_0 + \nonumber \\
  (1-f_i) &&\sum_{\alpha}\int d\mbf{r}' \left[\mathbb{F}_{\rm Hxc}(\mbf{r},\mbf{r}') \bm{n}_i(\mbf{r}')\right]_{\alpha}\sigma_{\alpha}.
\end{eqnarray}
The first term of Eq.~(\ref{eq:nc_kop_potential}) is a scalar shift while the other four terms are local potentials. Even for TR-invariant (non-magnetic) systems, not only the charge component of the orbital density is non-vanishing but necessarily also some of its spin components as each spin-orbital has always a finite spin density. Indeed, TR symmetry only implies that the total spin density is vanishing, not orbital spin densities. In addition, even the orbital charge density alone couples not only with the charge-charge component of the $\mathbb{F}_{\rm Hxc}$ but also with the charge-spin components, once again even for non-magnetic systems with TR symmetry. Fig.~\ref{fig:int_screen} summarizes the charge-charge, charge-spin and spin-spin interactions and screening effects in the non-collinear KCW functional theory, while Sec.~\ref{subsec:spin-dep} contains a more thorough discussion of these spin-dependent effects.

\subsection{Wannier Hamiltonian for the non-collinear Koopmans correction}
At variance with DFT, KC functionals are not invariant under unitary rotations of the occupied manifold, due to their orbital-dependent nature: the energy functional is minimized by the so-called variational orbitals. These variational orbitals have been shown to be very similar to maximally-localized WFs (MLWFs), which can be a good proxy to avoid the minimization procedure~\cite{Linh2018,colonna_jctc_2018,colonna_jctc_2019,colonna_jctc_2022}. We introduce a notation for the most general case, where a set of $J$ WFs is extracted from a higher number of $\nbands_\kk$ entangled bands~\cite{Marzari1997,Souza2001,Marzari2012}:
\begin{subequations}
  \begin{align}
  \ket{{\bf R}j}&= \;\frac{1}{N}\sum_\kk
  \,e^{-i{\bf k\cdot {\bf R}}}\ket{\psi^\text{W}_{j{\bf k}}}\,,
  \label{eq:wannier-dis}\\
  \ket{\psi^\text{W}_{j{\bf k}}}&=
  \sum_{n=1}^{\nbands_\kk}\,\ket{\psi_{n{\bf k}}}V_{\kk,nj}\,,
  \label{eq:psi-smooth}
  \end{align}
  \label{eq:wannier-psi-dis}%
  \end{subequations}
where the $\nbands_\kk \times J$ matrices $V_\kk=\tilde{V_\kk} U_\kk$ represent the net result of the disentanglement ($\tilde V_\kk$, subspace-selection) and
maximal localization ($U_\kk$, gauge-selection) steps. The case of isolated bands~\cite{Marzari1997} can be retrieved by setting $V_\kk=\tilde V_\kk U_\kk$ and replacing $\nbands_\kk$ with $J$.

As already done for collinear KC functionals~\cite{colonna_jctc_2022}, we also adopt MLWFs in place of variational orbitals and express the Koopmans correction in a WF basis:
\begin{equation}
  \Delta H_{ij}^{\rm KI(2)}(\mbf{R}) = -\frac{1}{2}\Delta_{\mbf{0}\mbf{j}}^{\rm KI(2)} \delta_{\mbf{R},\mbf{0}}\delta_{i,j}+ \Delta H_{ij,\mbf{r}}^{\rm KI(2)}(\mbf{R}).
  \label{eq:ki_ham}
\end{equation}
The two terms come from the scalar and local Koopmans potentials of Eq.~(\ref{eq:nc_kop_potential}). 
The first correction is purely on-site and leads mostly to a downward rigid shift of the bands
\begin{equation}
  \Delta_{\mbf{0}\mbf{j}}^{\rm KI(2)} = \frac{1}{N_{\mbf{q}}}\sum_{\mbf{q}} \langle \bm{n}^{\mbf{0}j}_{\mbf{q}} | \mathbb{F}^{\mbf{q}}_{\rm Hxc} | \bm{n}^{\mbf{0}j}_{\mbf{q}} \rangle
  \label{eq:first_term}
\end{equation}
where $\bm{n}_{\mbf{q}}^{\mbf{0}j}(\mbf{r})$ is the $\mathbf{q}-$component of charge-spin four-vector density of the WF $\ket{\mbf{0}j}$. The second correction acts only on empty states:
\begin{align}
  \label{eq:second_term}
  \Delta H_{ij,\mbf{r}}^{\rm KI(2)}(\mbf{R}) = & (1-f_i) \frac{1}{N_{\mbf{k}}}\sum_{\mbf{k}}  e^{i\mbf{k}\cdot\mbf{R}} \nonumber \\ 
  &  \frac{1}{N_{\mbf{q}}}\sum_{\mbf{q}} \langle \bm{n}_{\mbf{k}-\mbf{q},\mbf{k}}^{ji} | \mathbb{F}^{\mbf{q}}_{\rm Hxc} | \bm{n}^{\mbf{0}j}_{\mbf{q}} \rangle,
\end{align}
where we introduce a monochromatic expansion of the densities calculated as the overlap between the periodic part of Bloch states in the Wannier gauge $u^{W}_{i,\mbf{k}}(\mbf{r})$ at different $k$-points:
\begin{equation}
  \bm{n}^{ij}_{\mbf{k},\mbf{k}+\mbf{q}}(\mbf{r}) = \braket{u^{\rm W}_{i,\mbf{k}}(\mbf{r})|\boldsymbol{\tilde{\sigma}}|u^{\rm W}_{j,\mbf{k}+\mbf{q}}(\mbf{r})}.
\end{equation}
At variance with the collinear KCW formulation~\cite{colonna_jctc_2022}, the corresponding NC expression for the non-scalar correction term (Eq.~(\ref{eq:second_term})) leads to a coupling between $\mathbf{q}$ and $\mathbf{k}-\mathbf{q}$ (as opposed to $\mathbf{k}+\mathbf{q}$) if TR symmetry is not assumed.
\subsection{Screening coefficents and DFPT}
\label{sec:screening}
In the spirit of the original formulation of KC functionals~\cite{Dabo2010}, we split the screened Koopmans correction into an un-screened one, obtained from Eq.~(\ref{eq:2nd-KIfunc-dens}) by using the bare Hxc kernel  $\mbf{F}_{\rm Hxc}$ (this would give PWL in the absence of orbital relaxation), and an orbital-dependent screening coefficient defined as the ratio between the screened and un-screened second-order Koopmans correction:
\begin{equation}
\alpha_{\bm{0}i}  
= \frac{\langle \bm{n}_{\mbf{0}i} | \bm{\mathbb{F}}_{\rm Hxc} | \bm{n}_{\mbf{0}i} \rangle}{\langle \bm{n}_{\mbf{0}i} | \mbf{F}_{\rm Hxc} | \bm{n}_{\bm{0}i} \rangle}
= 1+\frac{\langle \bm{n}_{\bm{0}i} | \mbf{F}_{\rm Hxc}\bm{\chi}\mbf{F}_{\rm Hxc} | \bm{n}_{\bm{0}i} \rangle}{\langle \bm{n}_{\bm{0}i} | \mbf{F}_{\rm Hxc} | \bm{n}_{\bm{0}i} \rangle}.
\end{equation}
Within this approximation the KI Hamiltonian in the WF basis reads
\begin{equation}
    H^{\rm KI(2)}_{ij}(\mbf{R}) = H^{\rm DFT}_{ij}(\mbf{R}) + \alpha_{\mbf{0}i} \Delta H^{\rm uKI(2)}_{ij}(\mbf{R})
    \label{eq:ki_ham_}
\end{equation}
where $\Delta H^{\rm uKI(2)}_{ij}(\mbf{R})$ is the un-screened Koopmans correction to the DFT Hamiltonian $H^{\rm DFT}_{ij}(\mbf{R})$, that is the analogous of Eq.~(\ref{eq:ki_ham}) where the screened Hxc kernel $\mathbb{F}_{\rm Hxc}$ is replaced by the bare one $\mbf{F}_{\rm Hxc}$. The final expression for the KI Hamitonian in Eq.~(\ref{eq:ki_ham_}) is an exact reformulation for the diagonal matrix elements, but introduces an approximation for the off-diagonal ones (that can be understood by comparing the second term in the right-hand side of Eq.~(\ref{eq:ki_ham_}) with Eq.~(\ref{eq:ki_ham})). However, because of the localization properties of WFs, off-diagonal matrix elements are often much smaller compared to diagonal ones (and exactly zero for occupied states) and this approximation has negligible effects on the final band structure. 

Introducing the un-screened monochromatic perturbing potential $\mbf{V}_{{\rm pert},\mbf{q}}^{\mbf{0}i}$ 
\begin{equation}
  \mbf{\mbf{V}}_{{\rm pert},\mbf{q}}^{\mbf{0}i}(\mbf{r}) = \int d\mbf{r}' \left[\mbf{F}^{\mbf{q}}_{\rm Hxc}(\mbf{r},\mbf{r}') \bm{n}_{\mbf{q}}^{\mbf{0}i}(\mbf{r}')\right],
\label{eq:kcw_pert_u}
\end{equation}
the expression for the screening coefficient can be recast into a linear response problem:
\begin{equation}
 \alpha_{\bm{0}i}  
= 1 + \frac{ \sum_{\mbf{q}} \langle \mbf{V}_{{\rm pert},\mbf{q}}^{\mbf{0}i} | \Delta_{\mbf{q}}^{\mbf{0}i} \boldsymbol{\rho} \rangle }{ \sum_{\mbf{q}} \langle \mbf{V}_{{\rm pert},\mbf{q}}^{\mbf{0}i} | \bm{n}_{\mbf{q}}^{\mbf{0}i} \rangle}
\label{eq:alpha_lr}
\end{equation}
where $\Delta_{\mbf{q}}^{\mbf{0}i} \boldsymbol{\rho}(\mbf{r}) = \int d\mbf{r}' \boldsymbol{\chi}_{\mbf{q}}(\mbf{r},\mbf{r}')\mbf{V}_{{\rm pert},\mbf{q}}^{\mbf{0}i}(\mbf{r}')$ is the four-density response of the system to the 
perturbation $\mbf{V}_{{\rm pert},\mbf{q}}^{\mbf{0}i} (\mbf{r})$. Formally, this is identical to the expression derived in Ref.~\cite{colonna_jctc_2022} for the collinear case but now using four-vector quantities rather than just the density component. The matrix-vector product between the Hxc kernel and the four-density involves charge-charge, charge-spin and spin-spin couplings even for non-magnetic systems. As already done for the collinear case~\cite{colonna_jctc_2018, colonna_jctc_2022}, we avoid the explicit evaluation of the NC response functions as sum over empty states by resorting to DFPT~\cite{baroni_phonons_2001}, which is based on the Sternheimer equation and requires computing only occupied states. The generalization of DFPT to the magnetic case with broken TR symmetry was first developed for the calculation of spin-fluctuation spectra~\cite{cao_ab_2018,gorni_spin-fluctuation_2016,gorni_spin_2018} and later extended
to the calculation of vibrational spectra~\cite{urru_density_2019, urru_lattice_2020} and Hubbard parameters~\cite{binci_noncollinear_2023}. We follow the same strategy and adapt it to the Koopmans-specific external perturbations $\mbf{V}_{{\rm pert},\mbf{q}}^{\mbf{0}i}(\mbf{r}')$. The  four-density response in Eq.~(\ref{eq:alpha_lr}) can be conveniently written~\cite{gorni_spin_2018} in terms of the (periodic part of the) ground state KS spinors $u_{\mbf{k},v}$ and $\hat{\mathcal{T}}u_{-\mbf{k},v}$ and its linear variations $\Delta^{\mbf{0}i}_{\mbf{q}} u_{\mbf{k},v}$ and $\hat{\mathcal{T}}\Delta^{\mbf{0}i}_{-\mbf{q}} u_{-\mbf{k},v}$:
\begin{widetext}
\begin{align}
    \Delta^{\mbf{0}i}_{\mbf{q}}\rho(\mbf{r}) & = \sum_{\mbf{k}v} \left[ u^{\dag}_{\mbf{k}v}(\mbf{r}) \Delta^{\mbf{0}i}_{\mbf{q}}u_{\mbf{k}v}(\mbf{r}) + (\hat{\mathcal{T}}u_{-\mbf{k}v}(\mbf{r}))^\dag (\hat{\mathcal{T}}\Delta^{\mbf{0}i}_{-\mbf{q}}u_{-\mbf{k}v}(\mbf{r})) \right] \nonumber \\
    \Delta^{\mbf{0}i}_{\mbf{q}}\mbf{m}(\mbf{r}) & = \sum_{\mbf{k}v} \left[ u^{\dag}_{\mbf{k}v}(\mbf{r}) \bm{\sigma} \Delta_{\mbf{q}}u_{\mbf{k}v}(\mbf{r}) - (\hat{\mathcal{T}}u_{-\mbf{k}v}(\mbf{r}))^\dag \bm{\sigma} (\hat{\mathcal{T}}\Delta^{\mbf{0}i}_{-\mbf{q}}u_{-\mbf{k}v}(\mbf{r})) \right] 
    \label{eq:deltarhoq_nc_}
\end{align}
where $\hat{\mathcal{T}}= i\sigma_y \hat{\mathcal{K}}$ is the time-reversal operator with $\hat{\mathcal{K}}$ being the complex-conjugation operator. The linear variations defining the four-density response are given by the solutions of the following Sternheimer equations~\cite{gorni_spin_2018,urru_density_2019}:
\begin{align}
    (h_{\mbf{k}+\mbf{q}}  -\varepsilon_{\mbf{k}v} ) |\Delta^{\mbf{0}i}_{\mbf{q}}u_{\mbf{k}v}\rangle & = -{P}_c^{\mbf{k}+\mbf{q}} \Delta^{\mbf{0}i}_{\mbf{q}}V_{\rm SCF} |u_{\mbf{k}v} \rangle \nonumber \\
    (h^{[-\mbf{B}]}_{\mbf{k}+\mbf{q}} -\varepsilon_{-\mbf{k}v} )  |\hat{\mathcal{T}}\left(\Delta^{\mbf{0}i}_{-\mbf{q}}u_{-\mbf{k}v}\right)\rangle & = -{\Pi}_c^{\mbf{k}+\mbf{q}} \Delta^{\mbf{0}i}_{\mbf{q}}V^{[-\mbf{B}]}_{\rm SCF} | \hat{\mathcal{T}}\left(u_{-\mbf{k}v}\right)\rangle .\nonumber \\
    \label{eq:sternheimer}
\end{align}
\end{widetext}
In the expressions above $h_{\mbf{k}+\mbf{q}}$ is the KS Hamiltonian at wave-vector $\mbf{k}+\mbf{q}$, ${P}_c^{\mbf{k}+\mbf{q}}$ is the projector operator on the empty states with wave-vector ${\mbf{k}+\mbf{q}}$ and it is defined as ${P}_c^{\mbf{k}} = \mathcal{I} - {P}_v^{\mbf{k}}$ with $\mathcal{I}$ the identity and ${P}_v^{\mbf{k}} = \sum_{i=1}^{\rm occ} | u_{\mathbf{k}i}\rangle \langle u_{\mathbf{k}i}|$,  ${\Pi}_c^{\mbf{k}+\mbf{q}} = \hat{\mathcal{T}}{P}_c^{-\mbf{k}-\mbf{q}}\hat{\mathcal{T}}^{-1}$, and $\Delta^{\mbf{0}i}_{\mbf{q}}V_{\rm SCF}$ is the monochromatic $\mbf{q}$-component of the screened perturbing potential, which reads:
\begin{align}
    \Delta^{\mbf{0}i}_{\mbf{q}}\mathbf{V}_{\rm SCF}(\mbf{r}) = \mbf{V}_{{\rm pert},\mbf{q}}^{\mbf{0}i}(\mbf{r}) + \Delta^{\mbf{0}i}_{\mbf{q}} \mathbf{V}_{\rm Hxc}[\Delta^{\mbf{0}i}_{\mbf{q}}\bm{\rho}](\mbf{r}) \nonumber \\
    \Delta^{\mbf{0}i}_{\mbf{q}} \mathbf{V}_{\rm Hxc}[\Delta_{\mbf{q}}\bm{\rho}](\mbf{r}) =  \int d\mbf{r}' \mbf{F}^{\mbf{q}}_{\rm Hxc}(\mbf{r},\mbf{r}') \Delta^{\mbf{0}i}_{\mbf{q}}\bm{\rho}(\mbf{r}')
    \label{eq:DV_scf}
\end{align}
The superscript ``$^{[-\mbf{B}]}$'' indicates that the corresponding operator is evaluated after reversing the direction of the exchange-correlation magnetic field $\mbf{W}_{xc}$, i.e., the vector part of total Hxc potential (see. Eq.~\ref{eq:2nd_deriv_HF}). Equations~(\ref{eq:deltarhoq_nc_})-(\ref{eq:DV_scf}) form a set of coupled equations whose self-consistent solution provides the four-density variation and hence the screening coefficients in Eq.~(\ref{eq:alpha_lr}). 

Note that, at variance with the phonons case and more similarly to the case of magnons, our Koopmans perturbation is magnetic (see Fig.~\ref{fig:int_screen}) and involves the coupling of the spin and charge degrees of freedom through the interaction kernel in both terms appearing in Eq.~(\ref{eq:DV_scf}).

\subsection{\label{subsec:beyond_2nd}Corrections beyond second order}
One can go beyond the second-order approximation used in the derivation above by adding a corrective term to the Taylor expansion of the KI energy correction (Eq.~(\ref{eq:2nd-ord_exp})).  A formally exact correction is given by $\Delta_i^{\rm r} = \Pi_i^{\rm rKI} -  \Pi_i^{\rm (2)rKI}$, and would revert the second-order approximation into the full KI functional in Eq.~(\ref{eq:nc_koopfunc}). If relaxation effects are neglected, the correction can be approximated as $\Delta_i^{\rm r} \simeq \Delta_i^{u} = \Pi_i^{\rm uKI} -  \Pi_i^{\rm (2)uKI}$. For convenience, the effect of this corrective term is essentially incorporated into a renormalized screening coefficient:
\begin{align}
& \Pi_{\mathbf{0}i}^{\rm rKI}  = \Pi_{\mathbf{0}i}^{\rm (2)rKI} + \Delta^{\rm r}_{\mathbf{0}i} \simeq \alpha_{\mathbf{0}i} \Pi_{\mathbf{0}i}^{\rm (2)uKI} + \Delta_{\mathbf{0}i}^{\rm u} = \tilde{\alpha}_{\mathbf{0}i} \Pi_{\mathbf{0}i}^{\rm (2)uKI} \nonumber \\
& \rightarrow \tilde{\alpha}_{\mathbf{0}i} = \alpha_{\mathbf{0}i} + \frac{\Delta_{\mathbf{0}i}^{\rm u}}{\Pi_{\mathbf{0}i}^{\rm (2)uKI}} = \alpha_{\mathbf{0}i} + \Delta \alpha_{\mathbf{0}i}.
\label{eq:alpha_corr}
\end{align}
Being dependent only on un-screened quantities, $\Delta_i^{u}$ can be computed by the sole knowledge of the KS orbitals of the neutral system. Evaluating the corrective term $\Delta\alpha_i$ at the occupation of the reference neutral system we get (see Supplementary Sec.~C. of the SM~\cite{SM} for more details on the derivation):
\begin{equation}
    \Delta \alpha_{\mathbf{0}i} = \frac{-2 \tilde{\Delta}_{\mathbf{0}i}^{u}}{\langle \bm{n}_{\mathbf{0}i} | \mathbf{F}^{\rm xc}[\boldsymbol{\rho}] | \bm{n}_{\mathbf{0}i} \rangle},
\end{equation}
where 
\begin{align}
 \tilde{\Delta}_{\mathbf{0}i}^{u} = & E^{\rm DFT}_{xc}[\boldsymbol{\rho}] - E^{\rm DFT}_{\rm xc}[\boldsymbol{\rho}^{N-1}_{\mathbf{0}i}] +  \nonumber \\ 
                & - \braket{{\mathbf{0}i} |\mathbf{V}^{\rm DFT}_{xc}[\boldsymbol{\rho}] \cdot\bm{\tilde{\sigma}} |  {\mathbf{0}i}} + \frac{1}{2} \braket{\bm{n}_{\mathbf{0}i} | \mathbf{F}^{\rm xc}[\boldsymbol{\rho}] | \bm{n}_{\mathbf{0}i}}
\end{align}
In the expression above $\boldsymbol{\rho}^{N-1}_{\mathbf{0}i}(\mathbf{r}) = \boldsymbol{\rho}(\mathbf{r}) - \boldsymbol{\rho}_{\mathbf{0}i}(\mathbf{r})$
is the density of the system where an electron is removed from the $i$th WF and no orbital relaxation is accounted for (a similar expression holds for empty states, where the energy of the $N+1$ electron system needs to be computed). Only terms involving the xc potential and kernel appear in this expression, as the kinetic, external and Hartree terms are strictly linear or quadratic in the occupation and are therefore already accounted for in  $\Pi_i^{\rm (2)rKI}$. 

\section{Methods}
We implement NC KCW functionals in the \code{KCW} code of the \code{Quantum ESPRESSO} (QE) distribution~\cite{giannozzi_qe_2009,giannozzi_qe_2017,carnimeo_qe_2023}. The NC screening coefficents are obtained by solving the two NC Sternheimer equations of Eq.~(\ref{eq:sternheimer}) through the corresponding solvers in the \code{PHonon} code of \code{Quantum ESPRESSO}, originally developed for calculating phonons~\cite{baroni_phonons_2001}. Notably, and at variance with phonons, in the NC KCW theory the perturbation is always magnetic and a magnetic calculation is performed also for TR-invariant systems.
The starting DFT calculations are done in the LSDA approximation~\cite{hedin_JPCSSP_1972} and norm-conserving ONCV pseudopotentials~\cite{Hamann2013,Hamann2017} are used. NC calculations have been performed with fully-relativistic pseudopotentials generated from the scalar-relativistic LDA PseudoDojo library version 0.4.1~\cite{vanSetten2018}. We use a plane-wave kinetic energy cutoff of 80 Ry for the wavefunctions and 320 Ry for the electronic density. 
We employ MLWFs~\cite{Marzari1997,Souza2001,Marzari2012} in the calculation of the Koopmans correction and perform separate Wannierization for the valence and conduction bands. MLWFs for the valence bands are computed first for each isolated manifold separately and then merged together, to reduce the mixing between core and valence bands and obtain a block diagonal unitary matrix. We use the disentanglement approach~\cite{Souza2001} to construct MLWFs for the low-lying conduction bands.
In all the following numerical calculations, results without SOC are obtained with collinear spin-polarized DFT and collinear KCW~\cite{colonna_jctc_2022}. Results with SOC are obtained with NC DFT and NC KCW. In all cases, and for the reasons discussed above, the full Hxc kernel is calculated, including spin-spin and charge-spin components. The renormalized screening coefficients designed to include a correction beyond the second-order Taylor expansion and described in Sec.~\ref{subsec:beyond_2nd} are used for all the system except CrI$_3$, where corrections beyond second order are negligible and are not included for computational efficiency.
Hybrid-functional calculations are performed by using the Heyd-Scuseria-Ernzerhof (HSE) functional~\cite{HSE} with the acceleration provided by the Adaptively Compressed Exchange Operator~\cite{lin_ace_2016}. For the evaluation of the Fock operator we use a reduced cutoff of 80 Ry for WSe$_2$ and CsPbBr$_3$ and 120 Ry for CrI$_3$ and a $q$-point grid ($6\times6\times2$ for WSe$_2$,  $4\times4\times4$ for CsPbBr$_3$, and $3\times3\times3$ for CrI$_3$) that is coarser by a factor of two with respect to the $k$-point mesh. PBE PseudoDojo pseudopotentials are used for the HSE calculations. For the case of CrI$_3$ we used pseudopotentials with semicore electrons in the valence and an increased energy cutoff of 120 Ry and 480 Ry for the wavefunctions and electron density, respectively.

The data used to produce the results of this work are available at the Materials Cloud Archive~\cite{MC_archive}.
\section{Numerical results}

\subsection{GaAs}

\begin{figure}[tb]
  \centering
  \includegraphics[width=1\linewidth]{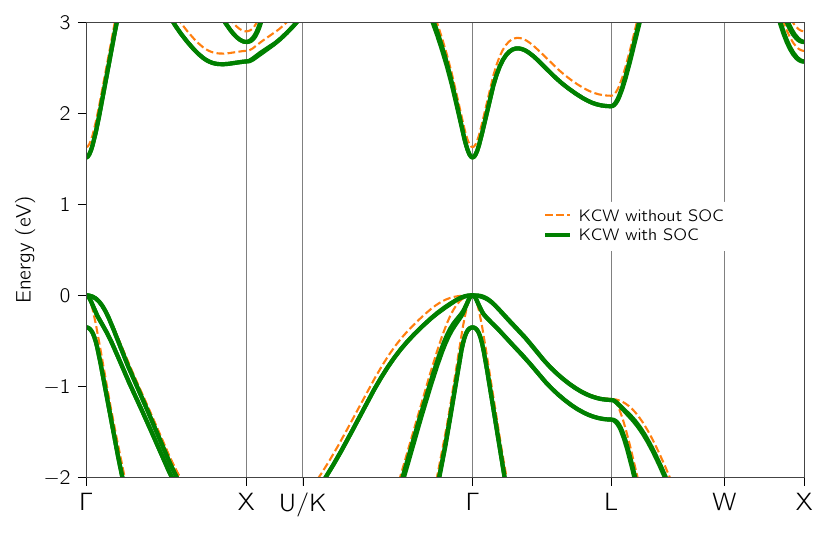}
  \caption{Band structure of GaAs obtained with Koopmans-Wannier (KCW) functionals both with and without spin-orbit coupling (SOC). Calculations with SOC (green solid line) have been performed with the non-collinear framework, including screening coefficients from non-collinear density-functional perturbation theory. KCW corrections are calculated on top of LSDA simulations, lines are the result of Wannier interpolation and energy zero is set at the top of the valence bands.}
  \label{fig:gaas_bands}
\end{figure}

First, we test our theory and implementation on GaAs where the effect of SOC on the band structure is moderate and has been well characterized experimentally~\cite{semicon_book,cardona_prb_1974}. We calculate the KCW band structure with and without SOC, results are shown in Fig.~\ref{fig:gaas_bands}. The KCW Hamiltonian and screening coefficents are calculated respectively with $12\times12\times12$ $k$-point and $6\times6\times6$ $q$-point grids. In Table~\ref{tab:gaas}, we report the band gap and two SOC-driven energy splitting, the $\Delta E_{SOC}^{15}$ at the top of the valence bands and the $\Delta E_{SOC}^{d}$ between the low-lying $J=5/2$ and $J=3/2$ $d$-orbitals of Ga. The effect of SOC on the band gap is small, about 0.1 eV of reduction at the KCW and LDA levels, in agreement with previous studies~\cite{kresse_prb_2007}, but contributes to the overall accuracy: our KCW band gap with SOC is about $1.51$ eV and in good agreement with the experimental value ($1.52$ eV). It is well known that LDA generally underestimates the band gap, here we obtain $0.17$ eV with SOC. However, at least for GaAs, LDA yields SOC energy splittings that match experiments rather accurately. Notably, not only KCW is capable to substantially correct the LDA band gap (from $0.17$ to $1.51$ eV) but at the same time KCW preserves the SOC energy splittings of the valence band that are already accurate at the LDA level (see Tab.~\ref{tab:gaas}).

\begin{table}[t]
  \begin{tabular}{|c|c|c|c|} 
   \hline
    with SOC & LDA & KCW & Exp. \\ 
   \hline\hline
   $E_g$ (eV) & 0.17 & 1.51  & 1.52 eV\\ 
   \hline
   $\Delta E_{SOC}^{15}$ (eV) & 0.35 & 0.35 &  0.3464(5)~\cite{semicon_book} \\
   \hline
   $\Delta E_{SOC}^{d}$  (eV) & 0.45 & 0.46 &  $0.4\pm1$~\cite{cardona_prb_1974} \\
   \hline
  \end{tabular}
  \caption{Comparison of spin-orbit coupling (SOC) effects on the band structure of GaAs between the local-density approximations (LDA), Koopmans-Wannier (KCW) functionals and experiments. The band gap ($E_g$) and spin-orbit driven splittings at the top of the valence bands $\Delta E_{SOC}^{15}$ and between low-lying Ga $d$-orbitals ($J=5/2$ and $J=3/2$) are considered. KCW corrections preserve the already-accurate LDA splittings due to SOC but substantially correct the band gap and bring it on top of experiments.}
  \label{tab:gaas}
  \end{table}
%

\subsection{WSe$_2$}
Now we consider a transition metal dichalcogenide, 2H-WSe$_2$, where SOC is stronger due to the presence of tungsten, and the spin texture is strongly modulated in both real and momentum space~\cite{riley_natphys_2014,kis_natrevmat_2017}. Although being globally centrosymmetric, the material exhibits a large spin polarization and spin-valley coupling of its bulk electronic states, due to the in-plane net dipole moment of each of the  two layers in the unit cell (local inversion asymmetry)~\cite{riley_natphys_2014,zhang_hidden_2014}.

\begin{figure}[t]
  \centering
  \includegraphics[width=1\linewidth]{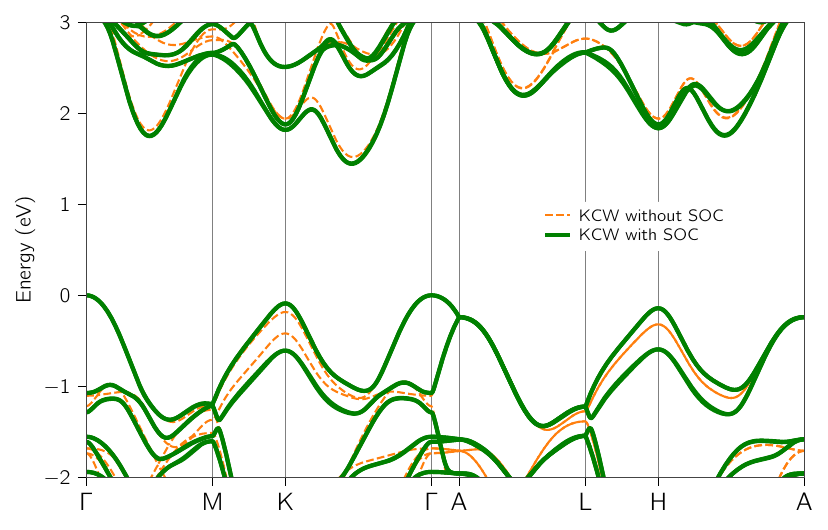}
  \caption{Band structure of WSe$_2$ obtained with Koopmans-Wannier (KCW) functionals both with and without spin-orbit coupling (SOC). Calculations with SOC (green solid line) have been performed with the non-collinear framework, including screening coefficients from non-collinear density-functional perturbation theory. KCW corrections are calculated on top of LSDA simulations, lines are the result of Wannier interpolation and the energy zero is set at the top of the valence bands.}
  \label{fig:wse2_bands}
\end{figure}

To facilitate comparison, we consider the experimental structural parameters of Ref.~\cite{schutte_jssc_1987}. We report in Fig.~\ref{fig:wse2_bands} the band structure calculated at the KCW level both with and without SOC. The KCW Hamiltonian and screening coefficents are calculated with $6\times6\times2$ $k/q$-point grids. The top valence bands are characterized by a splitting at the special point K of about half eV (in agreement with experiments~\cite{riley_natphys_2014}). The band gap is indirect and reduced by the inclusion of SOC. We calculate the band gap with LDA, HSE hybrid functionals and KCW both with and without SOC, and compare with experiments~\cite{bourzeg_prb_1992,kam_jpcssp_1984,traving_prb_1997,finteis_prb_1997} and $GW_0$ calculations~\cite{jiang_JPCC_2012} from the literature; all results are reported in Tab.~\ref{tab:ws2}. The $GW_0$ calculations of Ref.~\cite{jiang_JPCC_2012} are partially self-consistent for the Green's function $G$ through updated quasi-particle energies at each iteration, but with fixed screening $W_0$ in the random phase approximation (RPA); at variance with our work, their starting point is the Perdew-Burke-Ernzerhof (PBE) functional~\cite{perdew_prl_1996} and SOC is included only perturbatively. 

The experimental results are spread by about 0.1 eV around the value 1.3 eV and, as expected, LDA underestimates the band gap by about one third, even by including SOC (0.9 eV). All the beyond-DFT methods considered here perform reasonably well if SOC is included: HSE (1.38 eV) and KCW (1.44 eV) are on the upper side of the experimental range while self-consistent $GW_0$ (1.19 eV) on the lower side. Notably the effect of SOC on the band gap is about the same (around 80-90 meV) among different methods (LDA, HSE and KCW). 

\subsection{CsPbBr$_3$}
\begin{figure}[t]
  \centering
  \includegraphics[width=1\linewidth]{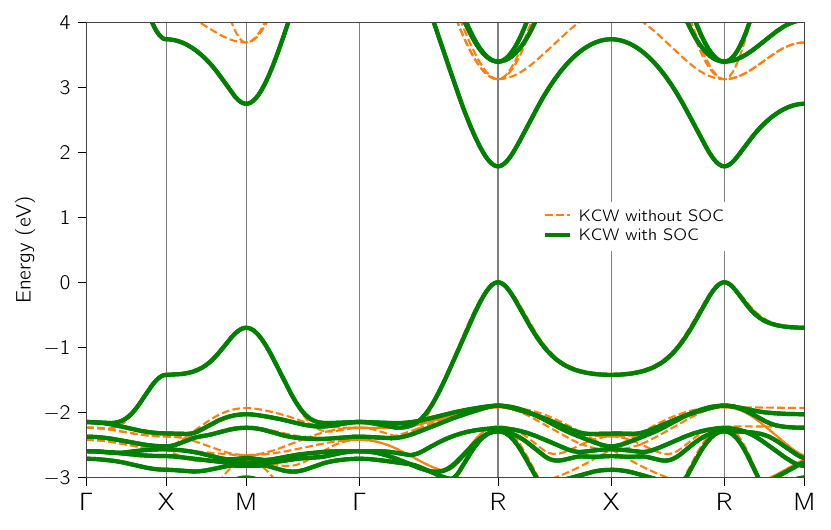}
  \caption{Band structure of CsPbBr$_3$ obtained with Koopmans-Wannier (KCW) functionals both with and without spin-orbit coupling (SOC). Calculations with SOC (green solid line) have been performed with the non-collinear framework, including screening coefficients from non-collinear density-functional perturbation theory. KCW corrections are calculated on top of LSDA simulations, lines are the result of Wannier interpolation and energy zero is set at the top of the valence bands. The main effect of SOC is to reduce the band gap, which is direct, by 1.34 eV. The KCW band gap with SOC (1.78 eV) compares well with $QSG\tilde{W}+\Delta_{\rm SOC}^{G_0W_0}$ (1.53 eV) and even experiments after removing temperature effects (1.85 eV).}
  \label{fig:perov_bands}
\end{figure}

Next, we consider a lead-halide perovskite, CsPbBr$_3$, where SOC is known to have a dramatic effect on the band gap magnitude~\cite{wiktor_jpcl_2017}. We consider the system in its high-temperature cubic phase and set the lattice constant to the experimental one ~\cite{stoumpos_crystal_2013} (5.87 \AA). We calculate the KCW band structure
with and without SOC, results are shown in Fig.~\ref{fig:perov_bands}. The KCW Hamiltonian and screening coefficients are calculated on a $4 \times 4 \times 4$ $k/q$-mesh. The band gap is direct and located at the high symmetry point R. The bottom of the conduction band is mainly from lead $6p$-states. At the R point, SOC splits the 3 times-degenerate band into $6p_{1/2}$ and $6p_{3/2}$ bands, leading to a reduction of the band gap by more than 1 eV regardless of the level of theory. In Tab.~\ref{tab:perov} we compare the KCW band gap with results from MBPT~\cite{wiktor_jpcl_2017}. Calculations without SOC show that the KCW band gap is in close agreement with that obtained with quasi-particle self-consistent $GW$ plus vertex corrections ($QSG\tilde{W}$) in the screened coulomb interaction, and significantly larger than the  $G_0W_0$ one. By including SOC the KCW band gap reduces by 1.34 eV (a reduction comparable to that observed for LDA and HSE: 1.22 eV and 1.31 eV, respectively) leading to a zero-temperature band gap of 1.78 eV which compares favorably with high-temperature experimental results~\cite{hoffman_transformation_2016} (2.36 eV) once the effects of temperature are accounted for~\cite{wiktor_jpcl_2017} (0.51 eV). The band gap from $G_0W_0$ with SOC is about 1 eV too small compared to experiments, while a fully consistent comparison with state-of-the-art $GW$ calculations ($QSG\tilde{W}$) with SOC is actually not possible as, to the best of our knowledge, no such simulation has been reported in the literature yet. Still, Ref.~\cite{wiktor_jpcl_2017} included the effect of SOC at the $G_0W_0$ level and add that on top of $QSG\tilde{W}$ calculations without SOC, leading to a predicted band gap of about 1.53 eV, which slightly underestimates the experimental value. 

\begin{table}[t]
  \begin{tabular}{|c|c|c|} 
    \hline
    & Method & Band gap (eV) \\
    \hline
    \hline
     \multirow{3}{*}{without SOC }&LDA & 0.9 \\ 
                  & HSE & 1.47 \\
                & KCW & 1.52  \\
    \hline
    \multirow{4}{*}{with SOC }&LDA & 0.82 \\ 
    & HSE & 1.38 \\
  & $GW_0$ & 1.19~\cite{jiang_JPCC_2012} \\
  & KCW & 1.44  \\
  \hline
  & Exp  & 1.22~\cite{kam_jpcssp_1984},1.27~\cite{bourzeg_prb_1992}, \\
  && 1.2~\cite{traving_prb_1997},1.3$\pm$0.1~\cite{finteis_prb_1997} \\
  \hline
   \end{tabular}
   \caption{Indirect band gap of bulk WSe$_2$ calculated with different methods, with and without including spin-orbit coupling, and compared with experimental results~\cite{bourzeg_prb_1992,kam_jpcssp_1984,traving_prb_1997,finteis_prb_1997}. Koopmans-Wannier (KCW) calculations are performed on top of LSDA, $GW_0$ results are taken from Ref.~\cite{jiang_JPCC_2012}.}
   \label{tab:ws2}
   \end{table}

\begin{figure}[th]
  \centering
  \includegraphics[width=1\linewidth]{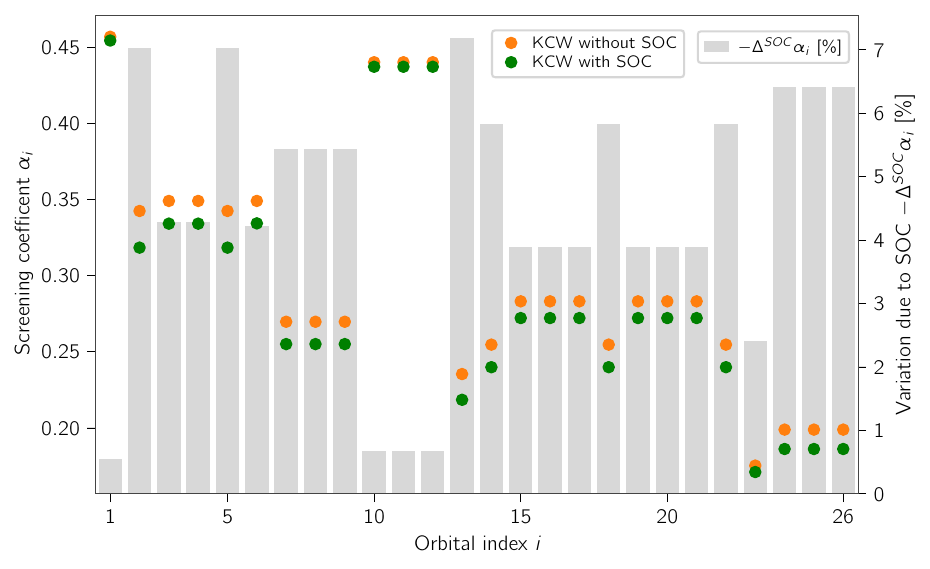}
  \caption{Screening coefficients for the MLWFs of CsPbBr$_3$ calculated with and without SOC. Results including SOC (green dots) have been obtained with the non-collinear linear-response formalism introduced in this work starting from a non-collinear LSDA ground state.  Results without SOC have been performed with the collinear formalism introduced in Refs.~\cite{colonna_jctc_2018, colonna_jctc_2022}, starting from a collinear LSDA ground state. Relative variation between results with and without SOC are shown with gray bars. Changes up to~7\% signify the importance of SOC effects in the response properties of the system. }
  \label{fig:perov_alphas}
\end{figure}

Finally, we mention that in this case the quality of the final result is also due to the consistent evaluation of the screening coefficients through the non-collinear linear response formalism detailed in Sec.~\ref{sec:screening}. At variance with the other systems presented above, the screening coefficients are significantly affected by the inclusion of SOC when solving the LR equations. This is ultimately due to the fact that SOC drastically modifies the band structure of CsPbBr$_3$, leading to a significant change in the response function of the system. 
In Fig.~\ref{fig:perov_alphas}, we compare the screening coefficients evaluated with and without SOC for the MLWFs of CsPbBr$_3$.
We observe relative variations as large as~7 \%; these are much larger than those observed for GaAs and WSe$_2$ (always $<$ 1 \%) and signify the importance of consistently including SOC in the calculation of the response function of this system.

\begin{table}[b]
\begin{tabular}{|c|c|c|} 
  \hline
  & Method & Band gap (eV) \\
  \hline
  \hline
   \multirow{4}{*}{without SOC } 
            & LDA      & 1.40 \\ 
            & HSE      & 2.09 \\ 
            & $G_0W_0$ & 2.56~\cite{wiktor_jpcl_2017} \\
            & $QSG\tilde{W}$& 3.15~\cite{wiktor_jpcl_2017}  \\
            & KCW      & 3.12 \\
  \hline
  \multirow{4}{*}{with SOC }
           & LDA      & 0.18 \\ 
           & HSE      & 0.78 \\ 
           & $G_0W_0$ & 0.94~\cite{wiktor_jpcl_2017} \\
           & $QSG\tilde{W} + \Delta_{\rm SOC}^{G_0W_0}$ & 1.53~\cite{wiktor_jpcl_2017} \\
           & KCW      & 1.78 \\
\hline
           & Exp $ + \Delta E_T $ & 1.85~\cite{hoffman_transformation_2016, wiktor_jpcl_2017}  \\
\hline
 \end{tabular}
 \caption{Direct band gap of CsPbBr$_3$ at the high-symmetry point R. Many-body perturbation theory results are from  Ref.~\cite{wiktor_jpcl_2017}. For a meaningful comparison with zero-temperature simulations (both $GW$ and KCW), the experimental band gap~\cite{hoffman_transformation_2016} (2.36 eV) is corrected by removing the temperature effects evaluated at the hybrid-DFT PBE0 level of theory in Ref.~\cite{wiktor_jpcl_2017} ($\Delta E_T^{\rm PBE0} = 0.51$ eV).} 
 \label{tab:perov}
 \end{table}

\subsection{CrI$_3$}
\begin{figure}[ht]
  \centering
  \includegraphics[width=1\linewidth]{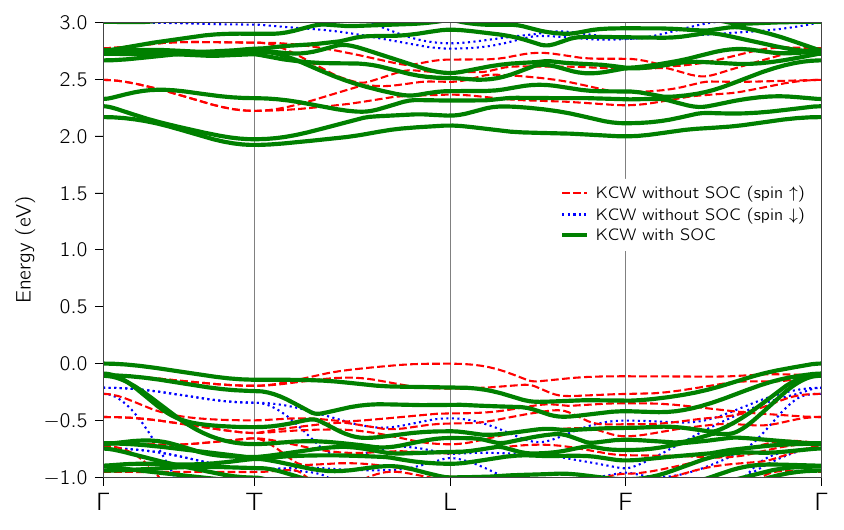}
  \caption{Band structure of CrI$_3$ obtained with Koopmans-Wannier (KCW) functionals both with and without spin-orbit coupling (SOC). Calculations with SOC (green solid line) have been performed with the non-collinear framework, including screening coefficients from non-collinear density-functional perturbation theory. Spin-polarized collinear calculations without SOC are shown for the spin-up (red dashed line) and spin-down (blue dotted line) channels. KCW corrections are calculated on top of LSDA simulations, lines are the result of Wannier interpolation and energy zero is set at the top of the valence bands. SOC reduces the band gap by 0.3 eV and push the $\Gamma$ point of the top valence band above the $L$ point. The KCW band gap with SOC (1.92 eV) compares well with $QSG\tilde{W}$ calculations ($2.2$--$2.5$ eV~\cite{acharya_prb_2021,acharya_npj2D_2022}) and even optical experiments (1.2 eV~\cite{dillon_jap_1965}) after the large exciton binding energy (1 eV) is accounted for~\cite{acharya_npj2D_2022}.}
  \label{fig:cri3_bands}
\end{figure}

Finally, we consider the ferromagnetic semiconductor CrI$_3$, where both magnetism and SOC play a significant role in the electronic band structure. Bulk CrI$_3$ is a layered material where each layer consists of edge-sharing CrI$_6$ octahedra with Cr atoms arranged in a honeycomb lattice. We consider the system in its low-temperature rhombohedral phase and adopt experimental lattice constant and atomic positions~\cite{mcguire_coupling_2015}. The KCW band structures with and without SOC are shown in Fig.~\ref{fig:cri3_bands}. The KCW Hamiltonian and screening coefficients are calculated on a $4 \times 4 \times 4$ $k/q$-mesh. 

\begin{table}[hb]
\begin{tabular}{|c|c|c|} 
  \hline
  & Method & Band gap (eV) \\
  \hline
  \hline
   \multirow{4}{*}{without SOC } 
            & LDA      & 0.89\\ 
            & HSE      & 1.85\\
            & $G_0W_0$ & 2.07~\cite{kutepov_prm_2021} \\
            & $QSGW80$ &  2.23~\cite{lee_prb_2020}\\
            & $QSGW$ & 3.11~\cite{kutepov_prm_2021} \\
            & KCW      & 2.22 \\
  \hline
  \multirow{4}{*}{with SOC }
  & LDA      & 0.63\\ 
  & HSE      & 1.45\\ 
  & $G_0W_0$ & 1.99~\cite{kutepov_prm_2021} \\
  & $QSGW80$ & 1.68~\cite{lee_prb_2020} \\
  & $QSGW$ &  2.64~\cite{kutepov_prm_2021},3.0~\cite{acharya_prb_2021,acharya_npj2D_2022} \\
  & $QSG\tilde{W}$&   2.5~\cite{acharya_prb_2021}, 2.2~\cite{acharya_npj2D_2022}\\
  & KCW      & 1.92 \\
\hline
           & Exp~\cite{dillon_jap_1965}$ + E_b$ ($1$ eV) & 2.2  \\
\hline
 \end{tabular}
 \caption{Indirect band gap of CrI$_3$. Many-body perturbation theory results are from  Refs.~\cite{lee_prb_2020,kutepov_prm_2021,acharya_prb_2021,acharya_npj2D_2022}. For a meaningful comparison, the experimental optical band gap~\cite{dillon_jap_1965} (1.2 eV) is corrected by the exciton binding energy as calculated with the Bethe-Salpeter equation in Ref.~\cite{acharya_npj2D_2022} ($E_b = 1 $ eV).} 
 \label{tab:cri3}
 \end{table}

In the following we focus the discussion on calculations with SOC, that decrease the band gap by about $0.3$ eV both at the LDA and KCW level, and push the $\Gamma$ point of the top valence band above the $L$ point. Indeed, the gap is indirect both with and without SOC, but the transition shifts from $L-T$ without SOC to $\Gamma-T$ with SOC. 

In Tab.~\ref{tab:cri3} we compare the KCW band gap with results obtained with LDA and HSE hybrids functionals, as well as with MBPT calculations from the literature~\cite{lee_prb_2020,kutepov_prm_2021,acharya_prb_2021,acharya_npj2D_2022} that cover different flavors of $QSGW$ with and without vertex corrections. From Ref.~\cite{kutepov_prm_2021}, we also report the results for $G_0W_0$ calculations on top of PBE. In addition to the $QSG\tilde{W}$ results, where the polarizability includes vertex corrections (ladder diagrams) by solving a Bethe-Salpeter equation (BSE) for the two-particle Hamiltonian, we also report $QSGW80$ calculations from Ref.~\cite{lee_prb_2020}, which uses an empirical mixing of $80\%$ $QSGW$ and $20\%$ LDA to generate the xc potential~\cite{deguchi_JJAP_2016,chantis_prl_2006}. 

The band gap of CrI$_3$ gradually increases in going from LDA to QSGW through HSE and $G_0W_0$: with SOC, the gap is $0.64$ eV with LDA, $1.45$ eV with HSE, $1.99$ eV with $G_0W_0$~\cite{kutepov_prm_2021} and between $2.64$~\cite{kutepov_prm_2021} and $3.0$~\cite{acharya_prb_2021,acharya_npj2D_2022} eV with $QSGW$. Including vertex corrections within the $QSG\tilde{W}$ approach reduces the band gap from $3.0$ eV~\cite{acharya_prb_2021,acharya_npj2D_2022} to $2.2$--$2.5$ eV~\cite{acharya_npj2D_2022, acharya_prb_2021}.
The KCW band gap (calculated on top of LSDA) is $1.92$ eV, which compares well with $G_0W_0$ ($1.99$ eV~\cite{kutepov_prm_2021}) and is not far from $QSG\tilde{W}$ results ($2.2$--$2.5$~\cite{acharya_npj2D_2022, acharya_prb_2021} eV). 
The comparison with the experimental optical gap (1.2 eV) is not straightforward due to the strong excitonic effects that are expected in 2D and layered materials. If we consider $1$ eV of exciton binding energy as calculated in Ref.~\cite{acharya_npj2D_2022} through the BSE, the estimated electronic gap should be around $2.2$ eV, which compares well with KCW ($1.92$ eV), $G_0W_0$ ($1.99$ eV~\cite{kutepov_prm_2021}) and $QSG\tilde{W}$ ($2.2$--$2.5$ eV~\cite{acharya_prb_2021,acharya_npj2D_2022}). On the other hand, $QSGW$ band gaps ($2.64$~\cite{kutepov_prm_2021}, $3.0$~\cite{acharya_prb_2021,acharya_npj2D_2022} eV) are overestimated with respect to experiments (even if considering the effect of large exciton binding energies), while  HSE (1.45 eV) underestimates, although less severely than LDA (0.63 eV). This resonates with the common knowledge that $G_0W_0$ calculations can often be remarkably accurate, despite their simplified self-energy, due to a compensation of errors between the lack of self-consistency and vertex corrections~\cite{Martin2016}. 

Similar trends hold also for spin-polarized calculations without SOC (see Tab.~\ref{tab:cri3}), although the $QSGW80$ (2.23 eV~\cite{lee_prb_2020}) and also the KCW band gaps ($2.22$ eV) are now slightly larger than in $G_0W_0$ (2.07~\cite{kutepov_prm_2021}); we have not found in the literature $QSG\tilde{W}$ results without SOC for a comparison.

To further elucidate the role of spin-dependent screening, we also calculate the KCW corrections at the RPA level, i.e., neglecting xc effects in the response function, and obtain larger values for the screening coefficients (that signals a reduction in electronic screening) such that the band gap increases by approximately $0.6$ eV. This is consistent with the trend observed in going from $QSGW$ to $QSG\tilde{W}$~\cite{acharya_prb_2021,acharya_npj2D_2022} (see also Tab.~\ref{tab:cri3}) and support the importance of going beyond the RPA to describe electronic screening and predict more accurately band structure properties.

\section{\label{sec:disc}Discussion}
The numerical results presented for GaAs, WSe$_2$, CsPbBr$_3$ and CrI$_3$ indicate that NC KCW calculations can provide accurate band structures for semiconductors in presence of SOC interactions. Remarkably, the theory is inherently spin-dependent and treats charge-charge, spin-charge, and spin-spin interactions on the same footing. This aspect has a few important implications we want to briefly discuss here.

\subsection{\label{subsec:spin-dep}Spin-dependent interaction and screening}
The NC Koopmans-functional theory includes spin-dependent effects in two different ways: by considering a spin-dependent effective potential for each spin-orbital (Eq.~(\ref{eq:nc_kop_potential})) and by considering spin-dependent screening effects (see e.g., Eq.~(\ref{eq:alpha_lr})). As argued in Ref.~\cite{colonna_jctc_2019}, the KI effective potential (Eq.~(\ref{eq:nc_kop_potential})) transforms the xc potential of the underlying DFT into a local, orbital- and spin-dependent approximation of the electronic self-energy beyond the RPA, i.e., including approximate (DFT-based) vertex corrections both in the effective potential and in the screening. 
These two effects can be parsed by considering the relaxed Koopmans functional introduced in Eq.~(\ref{eq:2nd-KIfunc-dens}), the relaxed Koopmans potential in Eq.~(\ref{eq:nc_kop_potential}), and their schematic representation given in Fig.~(\ref{fig:int_screen}): 
The unscreened KI effective potential involves the product of the bare Hxc kernel with the WF four-density, where there are always not only charge-charge terms but also charge-spin and spin-spin terms. In fact, even for a non-magnetic system, each WF has a finite magnetization density that sums up to one and the four-vector has always at least one non-vanishing spin-component (more than one if, e.g., Rashba SOC is present). 

All these degrees of freedom are coupled by the four-by-four matrix representing the NC Hxc kernel. The inclusion of screening effects transforms the bare Hxc kernel into the screened one via the Dyson equation in Eq.~(\ref{eq:screened_kernel}), which involves matrix-matrix products with the spin-dependent response function. This differs from the standard RPA approximation where only the classical test-charge-test-charge interaction is considered~\cite{Vignale_2005, del_sole_gwg_1994, hybertsen_ab_1987}. A RPA in the Koopmans framework essentially amounts to neglecting all the spin-charge (green) and spin-spin (blue) channels in Fig.~(\ref{fig:int_screen}) or, equivalently, to set to zero all the xc components of the Hxc kernel in Eq.~(\ref{eq:nc_kop_potential}) both in the effective interaction and in the response function, and ultimately makes the KI potential a local approximation of the static $GW$ self-energy (COHSEX)~\cite{colonna_jctc_2019}.  
The inclusion of xc effects via the $\mathbf{F}_{\rm xc}$ kernel in the Dyson equation~(\ref{eq:screened_kernel}) transforms the test charge-test charge (RPA) response function in the test charge-test electron response~\cite{hybertsen_ab_1987,del_sole_gwg_1994,bruneval_many-body_2005}. The electrons contributing to the screening are now dressed by an approximate xc-hole and the potential induced by the additional electron or hole includes spin-dependent xc interactions beyond the classical Hartree term. Moreover, this improved response function is used to screen the bare Hxc kernel which by itself already includes non-classical terms.

From the perspective of photoemission experiments, an accurate theory of band structures should indeed take into account that an electron removal/addition perturbs not only the charge distribution of the orbitals but also their spin densities, and the two can be coupled by spin-dependent interactions such as SOC.
We therefor argue that the additional physics borrowed from that of the homogeneous electron gas via the screened xc kernel of the underlying density functional might explain the quality of the results presented here and in previous benchmarks~\cite{Linh2018, colonna_jctc_2019, de_almeida_electronic_2021} and makes Koopmans functionals typically as accurate as state-of-the-art Green's function methods.

The crucial role of orbital- and spin-dependent interactions emerges also in MBPT, where it can be taken into account either by adding second-order exchange diagrams through vertex corrections or by implementing self-screening corrections to the $GW$ self-energy~\cite{self-screen_prb_2012,self-screen_prb_2023}. Indeed, the absence of spin-dependent interactions in standard $GW$ manifests through the presence of self-screening in the RPA, where an electron screens itself. Aryasetiawan, Sakuma and Karlsson have shown~\cite{self-screen_prb_2012} that removing the self-screening terms in the $GW$ self-energy is partially equivalent to adding vertex corrections in the form of exchange diagrams, that leads to bands gaps of semiconductors that better agree with experiments~\cite{self-screen_prb_2023}. 
While this physics is not captured by standard $GW$ and RPA, and requires self-screening or vertex corrections to be treated in MBPT, it is naturally built-in for Koopmans functionals and, likely, for other orbital-density-dependent functionals that implement a generalized PWL. 

\subsection{Spin-torque}
Common density functionals assume the magnetization $\mathbf{m}(\mathbf{r})$ to be locally collinear to the exchange-correlation ``magnetic field'' $\mathbf{W}_{xc}(\mathbf{r})$~\cite{kubler_jpf_1988}. Strictly speaking, this is allowed only within purely local functionals like LSDA, but it is also used in the context of the generalized gradient approximation (GGA)~\cite{nordstrom_prb_2002}. This approximation implies a vanishing spin torque $\mathbf{m}(\mathbf{r})\times \mathbf{W}_{xc}(\mathbf{r})$ everywhere in space. While including the spin-torque within a pure density-functional framework is a non-trivial task, approaches based on the optimized effective potential (OEP)~\cite{gross_prl_2007,ullrich_prb_2018}, which involve the knowledge of the KS orbitals, and source-free functionals~\cite{sharma_jctc_2018,dewhurst_eplb_2018} have been developed for that purpose.
At variance with standard DFT, the NC Koopmans potential (Eq.~\ref{eq:nc_kop_potential}) can naturally lead to spin-orbit torque: the WF spin-densities are not forced to be collinear among themselves and all aligned to the total spin magnetization. This is also in line with recent numerical evidences that the NC extension of Perdew-Zunger self-interaction correction (PZ-SIC) functional~\cite{perdew_self-interaction_1981}, which is also an orbital-density dependent functional, produces non vanishing spin-torque~\cite{tancogne-dejean_self-interaction_2023}. We therefor expect similar conclusions to apply to the NC Koopmans functional discussed here and even more to the KIPZ Koopmans functional~\cite{Borghi2014}, which adds to the basic KI functional an extra orbital-dependent term inspired by the PZ-SIC scheme. 
 
\section{Summary and Conclusions}
We have developed NC Koopmans-compliant spectral functionals and discussed the role of spin-dependent interactions and screening in the Koopmans theory. In particular, we propose a NC perturbative formulation based on MLWFs and DFPT, that allows calculating accurate band structures in presence of SOC with a one-shot correction on top of LSDA calculations. 
The approach has been validated on GaAs, WSe$_2$, CsPbBr$_3$ and CrI$_3$, where the predicted band gaps compare well with experiments and match in accuracy state-of-the-art MBPT. These results reiterate the message that Koopmans functionals i) are able to deliver charged excitations that are as accurate as those obtained from diagrammatic approaches, and ii) thus provide a reliable and efficient alternative when such approaches become unfeasible due to the higher computational complexity and cost.  

We have argued that the NC KCW theory presented here includes spin-dependent interactions and screening effects that are missing in standard diagrammatic approaches based on the RPA. 
Indeed, the NC KCW approach involves simple functionals of the charge and spin-vector densities of WFs, where the interaction kernel couples the spin and charge degrees of freedom, even for non-magnetic systems with TR symmetry. Also, the screening of the interaction is treated as non-collinear and involves spin-dependent terms originating both from the kernel and the response function of the charge and spin magnetization densities. 

The method is computationally efficient and simple, requiring essentially the same resources and convergence tests of a NC phonon calculation in DFPT. That, together with automated Wannierization protocols~\cite{Vitale2020,Qiao2023,Qiao2023a}, makes it particularly suited for high-throughput computational screening of materials databases and for studying complex materials, also at finite temperature~\cite{Bart_PRB_2016,wiktor_jpcl_2017,ZG_PRR_2022}.

Finally, we note that accurate spin-resolved band structures are very relevant not only for spectral properties such as band gaps or effective masses, but also for calculating magnetic exchange constants through the magnetic force theorem~\cite{rmp_attila_2023} and other response properties; the impact of NC KCW corrections beyond band structures will be the subject of future work.

\section{Acknowledgements}

The authors acknowledge useful discussions with Andrea Dal Corso and Andrea Ferretti. 
A.M. acknowledges that this study was funded by the University of the Trieste through the “Microgrant 2023” program and by the European Union - NextGenerationEU, through the ICSC -- Centro Nazionale di Ricerca in High Performance Computing, Big Data and Quantum Computing -- (CUP J93C22000540006, PNRR Investimento M4.C2.1.4), and in the framework of the PRIN Project ``Simultaneous electrical control of spin and valley polarization in van der Waals magnetic materials'' (SECSY - CUP J53D23001400001, PNRR Investimento M4.C2.1.1). The views and opinions expressed are solely those of the authors and do not necessarily reflect those of the European Union, nor can the European Union be held responsible for them.
N.C. acknowledges support from the NCCR MARVEL, a National Centre of Competence in Research, funded by the Swiss National Science Foundation (grant number 205602).
A.M. acknowledges CINECA for simulation time on Galileo100, under the ISCRA initiative and CINECA-UniTS agreement, for the availability of high-performance computing resources and support. N.C. acknowledges CSCS for high-performance computing resources under the CSCS-PSI agreement. 
\bibliography{biblio}

\begin{thebibliography}{123}%
\makeatletter
\providecommand \@ifxundefined [1]{%
 \@ifx{#1\undefined}
}%
\providecommand \@ifnum [1]{%
 \ifnum #1\expandafter \@firstoftwo
 \else \expandafter \@secondoftwo
 \fi
}%
\providecommand \@ifx [1]{%
 \ifx #1\expandafter \@firstoftwo
 \else \expandafter \@secondoftwo
 \fi
}%
\providecommand \natexlab [1]{#1}%
\providecommand \enquote  [1]{``#1''}%
\providecommand \bibnamefont  [1]{#1}%
\providecommand \bibfnamefont [1]{#1}%
\providecommand \citenamefont [1]{#1}%
\providecommand \href@noop [0]{\@secondoftwo}%
\providecommand \href [0]{\begingroup \@sanitize@url \@href}%
\providecommand \@href[1]{\@@startlink{#1}\@@href}%
\providecommand \@@href[1]{\endgroup#1\@@endlink}%
\providecommand \@sanitize@url [0]{\catcode `\\12\catcode `\$12\catcode
  `\&12\catcode `\#12\catcode `\^12\catcode `\_12\catcode `\%12\relax}%
\providecommand \@@startlink[1]{}%
\providecommand \@@endlink[0]{}%
\providecommand \url  [0]{\begingroup\@sanitize@url \@url }%
\providecommand \@url [1]{\endgroup\@href {#1}{\urlprefix }}%
\providecommand \urlprefix  [0]{URL }%
\providecommand \Eprint [0]{\href }%
\providecommand \doibase [0]{https://doi.org/}%
\providecommand \selectlanguage [0]{\@gobble}%
\providecommand \bibinfo  [0]{\@secondoftwo}%
\providecommand \bibfield  [0]{\@secondoftwo}%
\providecommand \translation [1]{[#1]}%
\providecommand \BibitemOpen [0]{}%
\providecommand \bibitemStop [0]{}%
\providecommand \bibitemNoStop [0]{.\EOS\space}%
\providecommand \EOS [0]{\spacefactor3000\relax}%
\providecommand \BibitemShut  [1]{\csname bibitem#1\endcsname}%
\let\auto@bib@innerbib\@empty
\bibitem [{\citenamefont {Kohn}\ and\ \citenamefont {Sham}(1965)}]{KS_pr_1965}%
  \BibitemOpen
  \bibfield  {author} {\bibinfo {author} {\bibfnamefont {W.}~\bibnamefont
  {Kohn}}\ and\ \bibinfo {author} {\bibfnamefont {L.~J.}\ \bibnamefont
  {Sham}},\ }\bibfield  {title} {\bibinfo {title} {{Self-Consistent Equations
  Including Exchange and Correlation Effects}},\ }\href
  {https://doi.org/10.1103/PhysRev.140.A1133} {\bibfield  {journal} {\bibinfo
  {journal} {Phys. Rev.}\ }\textbf {\bibinfo {volume} {140}},\ \bibinfo {pages}
  {A1133} (\bibinfo {year} {1965})}\BibitemShut {NoStop}%
\bibitem [{\citenamefont {Martin}(2020)}]{martin_book_2020}%
  \BibitemOpen
  \bibfield  {author} {\bibinfo {author} {\bibfnamefont {R.~M.}\ \bibnamefont
  {Martin}},\ }\href {https://doi.org/10.1017/9781108555586} {\emph {\bibinfo
  {title} {{Electronic Structure: Basic Theory and Practical Methods}}}},\
  \bibinfo {edition} {2nd}\ ed.\ (\bibinfo  {publisher} {Cambridge University
  Press},\ \bibinfo {year} {2020})\BibitemShut {NoStop}%
\bibitem [{\citenamefont {Kubler}(2017)}]{kubler_book_2017}%
  \BibitemOpen
  \bibfield  {author} {\bibinfo {author} {\bibfnamefont {J.}~\bibnamefont
  {Kubler}},\ }\href@noop {} {\emph {\bibinfo {title} {{ Theory of Itinerant
  Electron Magnetism, International Series of Monographs on Physics Vol. 106
  }}}}\ (\bibinfo  {publisher} {Oxford University Press},\ \bibinfo {year}
  {2017})\BibitemShut {NoStop}%
\bibitem [{\citenamefont {Nagaosa}\ and\ \citenamefont
  {Tokura}(2013)}]{nagaosa_naturenanotech_2013}%
  \BibitemOpen
  \bibfield  {author} {\bibinfo {author} {\bibfnamefont {N.}~\bibnamefont
  {Nagaosa}}\ and\ \bibinfo {author} {\bibfnamefont {Y.}~\bibnamefont
  {Tokura}},\ }\bibfield  {title} {\bibinfo {title} {{Topological properties
  and dynamics of magnetic skyrmions}},\ }\href
  {https://doi.org/10.1038/nnano.2013.243} {\bibfield  {journal} {\bibinfo
  {journal} {Nature Nanotechnology}\ }\textbf {\bibinfo {volume} {8}},\
  \bibinfo {pages} {899} (\bibinfo {year} {2013})}\BibitemShut {NoStop}%
\bibitem [{\citenamefont {Fert}\ \emph {et~al.}(2017)\citenamefont {Fert},
  \citenamefont {Reyren},\ and\ \citenamefont {Cros}}]{fer_natrevmat_2017}%
  \BibitemOpen
  \bibfield  {author} {\bibinfo {author} {\bibfnamefont {A.}~\bibnamefont
  {Fert}}, \bibinfo {author} {\bibfnamefont {N.}~\bibnamefont {Reyren}},\ and\
  \bibinfo {author} {\bibfnamefont {V.}~\bibnamefont {Cros}},\ }\bibfield
  {title} {\bibinfo {title} {{Magnetic skyrmions: advances in physics and
  potential applications}},\ }\href
  {https://doi.org/10.1038/natrevmats.2017.31} {\bibfield  {journal} {\bibinfo
  {journal} {Nature Reviews Materials}\ }\textbf {\bibinfo {volume} {2}},\
  \bibinfo {pages} {17031} (\bibinfo {year} {2017})}\BibitemShut {NoStop}%
\bibitem [{\citenamefont {Kamber}\ \emph {et~al.}(2020)\citenamefont {Kamber},
  \citenamefont {Bergman}, \citenamefont {Eich}, \citenamefont {Iuşan},
  \citenamefont {Steinbrecher}, \citenamefont {Hauptmann}, \citenamefont
  {Nordström}, \citenamefont {Katsnelson}, \citenamefont {Wegner},
  \citenamefont {Eriksson},\ and\ \citenamefont
  {Khajetoorians}}]{kamber_science_2020}%
  \BibitemOpen
  \bibfield  {author} {\bibinfo {author} {\bibfnamefont {U.}~\bibnamefont
  {Kamber}}, \bibinfo {author} {\bibfnamefont {A.}~\bibnamefont {Bergman}},
  \bibinfo {author} {\bibfnamefont {A.}~\bibnamefont {Eich}}, \bibinfo {author}
  {\bibfnamefont {D.}~\bibnamefont {Iuşan}}, \bibinfo {author} {\bibfnamefont
  {M.}~\bibnamefont {Steinbrecher}}, \bibinfo {author} {\bibfnamefont
  {N.}~\bibnamefont {Hauptmann}}, \bibinfo {author} {\bibfnamefont
  {L.}~\bibnamefont {Nordström}}, \bibinfo {author} {\bibfnamefont {M.~I.}\
  \bibnamefont {Katsnelson}}, \bibinfo {author} {\bibfnamefont
  {D.}~\bibnamefont {Wegner}}, \bibinfo {author} {\bibfnamefont
  {O.}~\bibnamefont {Eriksson}},\ and\ \bibinfo {author} {\bibfnamefont
  {A.~A.}\ \bibnamefont {Khajetoorians}},\ }\bibfield  {title} {\bibinfo
  {title} {{Self-induced spin glass state in elemental and crystalline
  neodymium}},\ }\href {https://doi.org/10.1126/science.aay6757} {\bibfield
  {journal} {\bibinfo  {journal} {Science}\ }\textbf {\bibinfo {volume}
  {368}},\ \bibinfo {pages} {eaay6757} (\bibinfo {year} {2020})}\BibitemShut
  {NoStop}%
\bibitem [{\citenamefont {Verlhac}\ \emph {et~al.}(2022)\citenamefont
  {Verlhac}, \citenamefont {Niggli}, \citenamefont {Bergman}, \citenamefont
  {Kamber}, \citenamefont {Bagrov}, \citenamefont {Iu{\c{s}}an}, \citenamefont
  {Nordstr{\"o}m}, \citenamefont {Katsnelson}, \citenamefont {Wegner},
  \citenamefont {Eriksson},\ and\ \citenamefont
  {Khajetoorians}}]{verlhac_natphys_2022}%
  \BibitemOpen
  \bibfield  {author} {\bibinfo {author} {\bibfnamefont {B.}~\bibnamefont
  {Verlhac}}, \bibinfo {author} {\bibfnamefont {L.}~\bibnamefont {Niggli}},
  \bibinfo {author} {\bibfnamefont {A.}~\bibnamefont {Bergman}}, \bibinfo
  {author} {\bibfnamefont {U.}~\bibnamefont {Kamber}}, \bibinfo {author}
  {\bibfnamefont {A.}~\bibnamefont {Bagrov}}, \bibinfo {author} {\bibfnamefont
  {D.}~\bibnamefont {Iu{\c{s}}an}}, \bibinfo {author} {\bibfnamefont
  {L.}~\bibnamefont {Nordstr{\"o}m}}, \bibinfo {author} {\bibfnamefont {M.~I.}\
  \bibnamefont {Katsnelson}}, \bibinfo {author} {\bibfnamefont
  {D.}~\bibnamefont {Wegner}}, \bibinfo {author} {\bibfnamefont
  {O.}~\bibnamefont {Eriksson}},\ and\ \bibinfo {author} {\bibfnamefont
  {A.~A.}\ \bibnamefont {Khajetoorians}},\ }\bibfield  {title} {\bibinfo
  {title} {{Thermally induced magnetic order from glassiness in elemental
  neodymium}},\ }\href {https://doi.org/10.1038/s41567-022-01633-9} {\bibfield
  {journal} {\bibinfo  {journal} {Nature Physics}\ }\textbf {\bibinfo {volume}
  {18}},\ \bibinfo {pages} {905} (\bibinfo {year} {2022})}\BibitemShut
  {NoStop}%
\bibitem [{\citenamefont {Szilva}\ \emph
  {et~al.}(2023{\natexlab{a}})\citenamefont {Szilva}, \citenamefont {Kvashnin},
  \citenamefont {Stepanov}, \citenamefont {Nordstr\"om}, \citenamefont
  {Eriksson}, \citenamefont {Lichtenstein},\ and\ \citenamefont
  {Katsnelson}}]{szilva_rmp_2023}%
  \BibitemOpen
  \bibfield  {author} {\bibinfo {author} {\bibfnamefont {A.}~\bibnamefont
  {Szilva}}, \bibinfo {author} {\bibfnamefont {Y.}~\bibnamefont {Kvashnin}},
  \bibinfo {author} {\bibfnamefont {E.~A.}\ \bibnamefont {Stepanov}}, \bibinfo
  {author} {\bibfnamefont {L.}~\bibnamefont {Nordstr\"om}}, \bibinfo {author}
  {\bibfnamefont {O.}~\bibnamefont {Eriksson}}, \bibinfo {author}
  {\bibfnamefont {A.~I.}\ \bibnamefont {Lichtenstein}},\ and\ \bibinfo {author}
  {\bibfnamefont {M.~I.}\ \bibnamefont {Katsnelson}},\ }\bibfield  {title}
  {\bibinfo {title} {{Quantitative theory of magnetic interactions in
  solids}},\ }\href {https://doi.org/10.1103/RevModPhys.95.035004} {\bibfield
  {journal} {\bibinfo  {journal} {Rev. Mod. Phys.}\ }\textbf {\bibinfo {volume}
  {95}},\ \bibinfo {pages} {035004} (\bibinfo {year}
  {2023}{\natexlab{a}})}\BibitemShut {NoStop}%
\bibitem [{\citenamefont {Mosconi}\ \emph {et~al.}(2016)\citenamefont
  {Mosconi}, \citenamefont {Umari},\ and\ \citenamefont
  {De~Angelis}}]{Mosconi_PCCP_2016}%
  \BibitemOpen
  \bibfield  {author} {\bibinfo {author} {\bibfnamefont {E.}~\bibnamefont
  {Mosconi}}, \bibinfo {author} {\bibfnamefont {P.}~\bibnamefont {Umari}},\
  and\ \bibinfo {author} {\bibfnamefont {F.}~\bibnamefont {De~Angelis}},\
  }\bibfield  {title} {\bibinfo {title} {{Electronic and optical properties of
  MAPbX$_3$ perovskites (X = I, Br, Cl): a unified DFT and GW theoretical
  analysis}},\ }\href {https://doi.org/10.1039/C6CP03969C} {\bibfield
  {journal} {\bibinfo  {journal} {Phys. Chem. Chem. Phys.}\ }\textbf {\bibinfo
  {volume} {18}},\ \bibinfo {pages} {27158} (\bibinfo {year}
  {2016})}\BibitemShut {NoStop}%
\bibitem [{\citenamefont {Scherpelz}\ \emph {et~al.}(2016)\citenamefont
  {Scherpelz}, \citenamefont {Govoni}, \citenamefont {Hamada},\ and\
  \citenamefont {Galli}}]{galli_jctc_2016}%
  \BibitemOpen
  \bibfield  {author} {\bibinfo {author} {\bibfnamefont {P.}~\bibnamefont
  {Scherpelz}}, \bibinfo {author} {\bibfnamefont {M.}~\bibnamefont {Govoni}},
  \bibinfo {author} {\bibfnamefont {I.}~\bibnamefont {Hamada}},\ and\ \bibinfo
  {author} {\bibfnamefont {G.}~\bibnamefont {Galli}},\ }\bibfield  {title}
  {\bibinfo {title} {{Implementation and Validation of Fully Relativistic GW
  Calculations: Spin-Orbit Coupling in Molecules, Nanocrystals, and Solids}},\
  }\href {https://doi.org/10.1021/acs.jctc.6b00114} {\bibfield  {journal}
  {\bibinfo  {journal} {Journal of Chemical Theory and Computation}\ }\textbf
  {\bibinfo {volume} {12}},\ \bibinfo {pages} {3523} (\bibinfo {year}
  {2016})}\BibitemShut {NoStop}%
\bibitem [{\citenamefont {Corso}(2013)}]{dalcorso_JPCM_2013}%
  \BibitemOpen
  \bibfield  {author} {\bibinfo {author} {\bibfnamefont {A.~D.}\ \bibnamefont
  {Corso}},\ }\bibfield  {title} {\bibinfo {title} {{Ab initio phonon
  dispersions of transition and noble metals: effects of the exchange and
  correlation functional}},\ }\href
  {https://doi.org/10.1088/0953-8984/25/14/145401} {\bibfield  {journal}
  {\bibinfo  {journal} {Journal of Physics: Condensed Matter}\ }\textbf
  {\bibinfo {volume} {25}},\ \bibinfo {pages} {145401} (\bibinfo {year}
  {2013})}\BibitemShut {NoStop}%
\bibitem [{\citenamefont {Marrazzo}\ \emph {et~al.}(2018)\citenamefont
  {Marrazzo}, \citenamefont {Gibertini}, \citenamefont {Campi}, \citenamefont
  {Mounet},\ and\ \citenamefont {Marzari}}]{marrazzo_PRL_2018}%
  \BibitemOpen
  \bibfield  {author} {\bibinfo {author} {\bibfnamefont {A.}~\bibnamefont
  {Marrazzo}}, \bibinfo {author} {\bibfnamefont {M.}~\bibnamefont {Gibertini}},
  \bibinfo {author} {\bibfnamefont {D.}~\bibnamefont {Campi}}, \bibinfo
  {author} {\bibfnamefont {N.}~\bibnamefont {Mounet}},\ and\ \bibinfo {author}
  {\bibfnamefont {N.}~\bibnamefont {Marzari}},\ }\bibfield  {title} {\bibinfo
  {title} {{Prediction of a Large-Gap and Switchable Kane-Mele Quantum Spin
  Hall Insulator}},\ }\href {https://doi.org/10.1103/PhysRevLett.120.117701}
  {\bibfield  {journal} {\bibinfo  {journal} {Phys. Rev. Lett.}\ }\textbf
  {\bibinfo {volume} {120}},\ \bibinfo {pages} {117701} (\bibinfo {year}
  {2018})}\BibitemShut {NoStop}%
\bibitem [{\citenamefont {Chen}\ \emph {et~al.}(2019)\citenamefont {Chen},
  \citenamefont {Maezono}, \citenamefont {Chen}, \citenamefont {Grosche},
  \citenamefont {Pickard},\ and\ \citenamefont
  {Monserrat}}]{monserrat_JPM_2019}%
  \BibitemOpen
  \bibfield  {author} {\bibinfo {author} {\bibfnamefont {S.}~\bibnamefont
  {Chen}}, \bibinfo {author} {\bibfnamefont {R.}~\bibnamefont {Maezono}},
  \bibinfo {author} {\bibfnamefont {J.}~\bibnamefont {Chen}}, \bibinfo {author}
  {\bibfnamefont {F.~M.}\ \bibnamefont {Grosche}}, \bibinfo {author}
  {\bibfnamefont {C.~J.}\ \bibnamefont {Pickard}},\ and\ \bibinfo {author}
  {\bibfnamefont {B.}~\bibnamefont {Monserrat}},\ }\bibfield  {title} {\bibinfo
  {title} {{Chemical and structural stability of superconducting In$_5$Bi$_3$
  driven by spin-orbit coupling}},\ }\href
  {https://doi.org/10.1088/2515-7639/ab4c2b} {\bibfield  {journal} {\bibinfo
  {journal} {Journal of Physics: Materials}\ }\textbf {\bibinfo {volume} {3}},\
  \bibinfo {pages} {015007} (\bibinfo {year} {2019})}\BibitemShut {NoStop}%
\bibitem [{\citenamefont {von Barth}\ and\ \citenamefont
  {Hedin}(1972)}]{hedin_JPCSSP_1972}%
  \BibitemOpen
  \bibfield  {author} {\bibinfo {author} {\bibfnamefont {U.}~\bibnamefont {von
  Barth}}\ and\ \bibinfo {author} {\bibfnamefont {L.}~\bibnamefont {Hedin}},\
  }\bibfield  {title} {\bibinfo {title} {{A local exchange-correlation
  potential for the spin polarized case. i}},\ }\href
  {https://doi.org/10.1088/0022-3719/5/13/012} {\bibfield  {journal} {\bibinfo
  {journal} {Journal of Physics C: Solid State Physics}\ }\textbf {\bibinfo
  {volume} {5}},\ \bibinfo {pages} {1629} (\bibinfo {year} {1972})}\BibitemShut
  {NoStop}%
\bibitem [{\citenamefont {Kubler}\ \emph {et~al.}(1988)\citenamefont {Kubler},
  \citenamefont {Hock}, \citenamefont {Sticht},\ and\ \citenamefont
  {Williams}}]{kubler_jpf_1988}%
  \BibitemOpen
  \bibfield  {author} {\bibinfo {author} {\bibfnamefont {J.}~\bibnamefont
  {Kubler}}, \bibinfo {author} {\bibfnamefont {K.~H.}\ \bibnamefont {Hock}},
  \bibinfo {author} {\bibfnamefont {J.}~\bibnamefont {Sticht}},\ and\ \bibinfo
  {author} {\bibfnamefont {A.~R.}\ \bibnamefont {Williams}},\ }\bibfield
  {title} {\bibinfo {title} {{Density functional theory of non-collinear
  magnetism}},\ }\href {https://doi.org/10.1088/0305-4608/18/3/018} {\bibfield
  {journal} {\bibinfo  {journal} {Journal of Physics F: Metal Physics}\
  }\textbf {\bibinfo {volume} {18}},\ \bibinfo {pages} {469} (\bibinfo {year}
  {1988})}\BibitemShut {NoStop}%
\bibitem [{\citenamefont {Sakuma}\ \emph {et~al.}(2011)\citenamefont {Sakuma},
  \citenamefont {Friedrich}, \citenamefont {Miyake}, \citenamefont {Bl\"ugel},\
  and\ \citenamefont {Aryasetiawan}}]{sakuma_prb_2011}%
  \BibitemOpen
  \bibfield  {author} {\bibinfo {author} {\bibfnamefont {R.}~\bibnamefont
  {Sakuma}}, \bibinfo {author} {\bibfnamefont {C.}~\bibnamefont {Friedrich}},
  \bibinfo {author} {\bibfnamefont {T.}~\bibnamefont {Miyake}}, \bibinfo
  {author} {\bibfnamefont {S.}~\bibnamefont {Bl\"ugel}},\ and\ \bibinfo
  {author} {\bibfnamefont {F.}~\bibnamefont {Aryasetiawan}},\ }\bibfield
  {title} {\bibinfo {title} {${GW}$ calculations including spin-orbit coupling:
  {Application} to {Hg} chalcogenides},\ }\href
  {https://doi.org/10.1103/PhysRevB.84.085144} {\bibfield  {journal} {\bibinfo
  {journal} {Phys. Rev. B}\ }\textbf {\bibinfo {volume} {84}},\ \bibinfo
  {pages} {085144} (\bibinfo {year} {2011})}\BibitemShut {NoStop}%
\bibitem [{\citenamefont {Aguilera}\ \emph
  {et~al.}(2013{\natexlab{a}})\citenamefont {Aguilera}, \citenamefont
  {Friedrich},\ and\ \citenamefont {Bl\"ugel}}]{aguilera_prb_2013}%
  \BibitemOpen
  \bibfield  {author} {\bibinfo {author} {\bibfnamefont {I.}~\bibnamefont
  {Aguilera}}, \bibinfo {author} {\bibfnamefont {C.}~\bibnamefont
  {Friedrich}},\ and\ \bibinfo {author} {\bibfnamefont {S.}~\bibnamefont
  {Bl\"ugel}},\ }\bibfield  {title} {\bibinfo {title} {{Spin}-orbit coupling in
  quasiparticle studies of topological insulators},\ }\href
  {https://doi.org/10.1103/PhysRevB.88.165136} {\bibfield  {journal} {\bibinfo
  {journal} {Phys. Rev. B}\ }\textbf {\bibinfo {volume} {88}},\ \bibinfo
  {pages} {165136} (\bibinfo {year} {2013}{\natexlab{a}})}\BibitemShut
  {NoStop}%
\bibitem [{\citenamefont {Aguilera}\ \emph
  {et~al.}(2013{\natexlab{b}})\citenamefont {Aguilera}, \citenamefont
  {Friedrich}, \citenamefont {Bihlmayer},\ and\ \citenamefont
  {Bl\"ugel}}]{aguilera_prb_2013bis}%
  \BibitemOpen
  \bibfield  {author} {\bibinfo {author} {\bibfnamefont {I.}~\bibnamefont
  {Aguilera}}, \bibinfo {author} {\bibfnamefont {C.}~\bibnamefont {Friedrich}},
  \bibinfo {author} {\bibfnamefont {G.}~\bibnamefont {Bihlmayer}},\ and\
  \bibinfo {author} {\bibfnamefont {S.}~\bibnamefont {Bl\"ugel}},\ }\bibfield
  {title} {\bibinfo {title} {${GW}$ study of topological insulators
  {Bi}$_2${Se}$_3$, {Bi}$_2${Te}$_3$, and {Sb}$_2${Te}$_3$: {Beyond} the
  perturbative one-shot approach},\ }\href
  {https://doi.org/10.1103/PhysRevB.88.045206} {\bibfield  {journal} {\bibinfo
  {journal} {Phys. Rev. B}\ }\textbf {\bibinfo {volume} {88}},\ \bibinfo
  {pages} {045206} (\bibinfo {year} {2013}{\natexlab{b}})}\BibitemShut
  {NoStop}%
\bibitem [{\citenamefont {Molina-S\'anchez}\ \emph {et~al.}(2013)\citenamefont
  {Molina-S\'anchez}, \citenamefont {Sangalli}, \citenamefont {Hummer},
  \citenamefont {Marini},\ and\ \citenamefont {Wirtz}}]{molina_prb_2013}%
  \BibitemOpen
  \bibfield  {author} {\bibinfo {author} {\bibfnamefont {A.}~\bibnamefont
  {Molina-S\'anchez}}, \bibinfo {author} {\bibfnamefont {D.}~\bibnamefont
  {Sangalli}}, \bibinfo {author} {\bibfnamefont {K.}~\bibnamefont {Hummer}},
  \bibinfo {author} {\bibfnamefont {A.}~\bibnamefont {Marini}},\ and\ \bibinfo
  {author} {\bibfnamefont {L.}~\bibnamefont {Wirtz}},\ }\bibfield  {title}
  {\bibinfo {title} {{Effect} of spin-orbit interaction on the optical spectra
  of single-layer, double-layer, and bulk {MoS}$_{2}$},\ }\href
  {https://doi.org/10.1103/PhysRevB.88.045412} {\bibfield  {journal} {\bibinfo
  {journal} {Phys. Rev. B}\ }\textbf {\bibinfo {volume} {88}},\ \bibinfo
  {pages} {045412} (\bibinfo {year} {2013})}\BibitemShut {NoStop}%
\bibitem [{\citenamefont {Nechaev}\ \emph {et~al.}(2013)\citenamefont
  {Nechaev}, \citenamefont {Hatch}, \citenamefont {Bianchi}, \citenamefont
  {Guan}, \citenamefont {Friedrich}, \citenamefont {Aguilera}, \citenamefont
  {Mi}, \citenamefont {Iversen}, \citenamefont {Bl\"ugel}, \citenamefont
  {Hofmann},\ and\ \citenamefont {Chulkov}}]{chulkov_prb_2013}%
  \BibitemOpen
  \bibfield  {author} {\bibinfo {author} {\bibfnamefont {I.~A.}\ \bibnamefont
  {Nechaev}}, \bibinfo {author} {\bibfnamefont {R.~C.}\ \bibnamefont {Hatch}},
  \bibinfo {author} {\bibfnamefont {M.}~\bibnamefont {Bianchi}}, \bibinfo
  {author} {\bibfnamefont {D.}~\bibnamefont {Guan}}, \bibinfo {author}
  {\bibfnamefont {C.}~\bibnamefont {Friedrich}}, \bibinfo {author}
  {\bibfnamefont {I.}~\bibnamefont {Aguilera}}, \bibinfo {author}
  {\bibfnamefont {J.~L.}\ \bibnamefont {Mi}}, \bibinfo {author} {\bibfnamefont
  {B.~B.}\ \bibnamefont {Iversen}}, \bibinfo {author} {\bibfnamefont
  {S.}~\bibnamefont {Bl\"ugel}}, \bibinfo {author} {\bibfnamefont
  {P.}~\bibnamefont {Hofmann}},\ and\ \bibinfo {author} {\bibfnamefont {E.~V.}\
  \bibnamefont {Chulkov}},\ }\bibfield  {title} {\bibinfo {title} {Evidence for
  a direct band gap in the topological insulator {Bi}$_2${Se}$_3$ from theory
  and experiment},\ }\href {https://doi.org/10.1103/PhysRevB.87.121111}
  {\bibfield  {journal} {\bibinfo  {journal} {Phys. Rev. B}\ }\textbf {\bibinfo
  {volume} {87}},\ \bibinfo {pages} {121111(R)} (\bibinfo {year}
  {2013})}\BibitemShut {NoStop}%
\bibitem [{\citenamefont {Marrazzo}\ \emph {et~al.}(2019)\citenamefont
  {Marrazzo}, \citenamefont {Gibertini}, \citenamefont {Campi}, \citenamefont
  {Mounet},\ and\ \citenamefont {Marzari}}]{marrazzo_nanolett_2019}%
  \BibitemOpen
  \bibfield  {author} {\bibinfo {author} {\bibfnamefont {A.}~\bibnamefont
  {Marrazzo}}, \bibinfo {author} {\bibfnamefont {M.}~\bibnamefont {Gibertini}},
  \bibinfo {author} {\bibfnamefont {D.}~\bibnamefont {Campi}}, \bibinfo
  {author} {\bibfnamefont {N.}~\bibnamefont {Mounet}},\ and\ \bibinfo {author}
  {\bibfnamefont {N.}~\bibnamefont {Marzari}},\ }\bibfield  {title} {\bibinfo
  {title} {{Relative} {Abundance} of $\mathbb{{Z}}_2$ {Topological} {Order} in
  {Exfoliable} {Two-Dimensional} {Insulators}},\ }\href
  {https://doi.org/10.1021/acs.nanolett.9b02689} {\bibfield  {journal}
  {\bibinfo  {journal} {Nano Letters}\ }\textbf {\bibinfo {volume} {19}},\
  \bibinfo {pages} {8431} (\bibinfo {year} {2019})}\BibitemShut {NoStop}%
\bibitem [{\citenamefont {Marsili}\ \emph {et~al.}(2021)\citenamefont
  {Marsili}, \citenamefont {Molina-S\'anchez}, \citenamefont {Palummo},
  \citenamefont {Sangalli},\ and\ \citenamefont {Marini}}]{marsili_PRB_2021}%
  \BibitemOpen
  \bibfield  {author} {\bibinfo {author} {\bibfnamefont {M.}~\bibnamefont
  {Marsili}}, \bibinfo {author} {\bibfnamefont {A.}~\bibnamefont
  {Molina-S\'anchez}}, \bibinfo {author} {\bibfnamefont {M.}~\bibnamefont
  {Palummo}}, \bibinfo {author} {\bibfnamefont {D.}~\bibnamefont {Sangalli}},\
  and\ \bibinfo {author} {\bibfnamefont {A.}~\bibnamefont {Marini}},\
  }\bibfield  {title} {\bibinfo {title} {{Spinorial} formulation of the
  ${GW}$-{BSE} equations and spin properties of excitons in two-dimensional
  transition metal dichalcogenides},\ }\href
  {https://doi.org/10.1103/PhysRevB.103.155152} {\bibfield  {journal} {\bibinfo
   {journal} {Phys. Rev. B}\ }\textbf {\bibinfo {volume} {103}},\ \bibinfo
  {pages} {155152} (\bibinfo {year} {2021})}\BibitemShut {NoStop}%
\bibitem [{\citenamefont {Nabok}\ \emph {et~al.}(2022)\citenamefont {Nabok},
  \citenamefont {Tas}, \citenamefont {Kusaka}, \citenamefont {Durgun},
  \citenamefont {Friedrich}, \citenamefont {Bihlmayer}, \citenamefont
  {Bl\"ugel}, \citenamefont {Hirahara},\ and\ \citenamefont
  {Aguilera}}]{aguilera_prb_2022}%
  \BibitemOpen
  \bibfield  {author} {\bibinfo {author} {\bibfnamefont {D.}~\bibnamefont
  {Nabok}}, \bibinfo {author} {\bibfnamefont {M.}~\bibnamefont {Tas}}, \bibinfo
  {author} {\bibfnamefont {S.}~\bibnamefont {Kusaka}}, \bibinfo {author}
  {\bibfnamefont {E.}~\bibnamefont {Durgun}}, \bibinfo {author} {\bibfnamefont
  {C.}~\bibnamefont {Friedrich}}, \bibinfo {author} {\bibfnamefont
  {G.}~\bibnamefont {Bihlmayer}}, \bibinfo {author} {\bibfnamefont
  {S.}~\bibnamefont {Bl\"ugel}}, \bibinfo {author} {\bibfnamefont
  {T.}~\bibnamefont {Hirahara}},\ and\ \bibinfo {author} {\bibfnamefont
  {I.}~\bibnamefont {Aguilera}},\ }\bibfield  {title} {\bibinfo {title} {Bulk
  and surface electronic structure of {Bi}$_4${Te}$_3$ from ${GW}$ calculations
  and photoemission experiments},\ }\href
  {https://doi.org/10.1103/PhysRevMaterials.6.034204} {\bibfield  {journal}
  {\bibinfo  {journal} {Phys. Rev. Mater.}\ }\textbf {\bibinfo {volume} {6}},\
  \bibinfo {pages} {034204} (\bibinfo {year} {2022})}\BibitemShut {NoStop}%
\bibitem [{\citenamefont {Yeh}\ \emph {et~al.}(2022)\citenamefont {Yeh},
  \citenamefont {Shee}, \citenamefont {Sun}, \citenamefont {Gull},\ and\
  \citenamefont {Zgid}}]{zgid_prb_2022}%
  \BibitemOpen
  \bibfield  {author} {\bibinfo {author} {\bibfnamefont {C.-N.}\ \bibnamefont
  {Yeh}}, \bibinfo {author} {\bibfnamefont {A.}~\bibnamefont {Shee}}, \bibinfo
  {author} {\bibfnamefont {Q.}~\bibnamefont {Sun}}, \bibinfo {author}
  {\bibfnamefont {E.}~\bibnamefont {Gull}},\ and\ \bibinfo {author}
  {\bibfnamefont {D.}~\bibnamefont {Zgid}},\ }\bibfield  {title} {\bibinfo
  {title} {Relativistic self-consistent ${GW}$: Exact two-component formalism
  with one-electron approximation for solids},\ }\href
  {https://doi.org/10.1103/PhysRevB.106.085121} {\bibfield  {journal} {\bibinfo
   {journal} {Phys. Rev. B}\ }\textbf {\bibinfo {volume} {106}},\ \bibinfo
  {pages} {085121} (\bibinfo {year} {2022})}\BibitemShut {NoStop}%
\bibitem [{\citenamefont {Gaurav}\ \emph {et~al.}(2024)\citenamefont {Gaurav},
  \citenamefont {Vibin},\ and\ \citenamefont {Zgid}}]{zgid_fd_2024}%
  \BibitemOpen
  \bibfield  {author} {\bibinfo {author} {\bibfnamefont {H.}~\bibnamefont
  {Gaurav}}, \bibinfo {author} {\bibfnamefont {A.}~\bibnamefont {Vibin}},\ and\
  \bibinfo {author} {\bibfnamefont {D.}~\bibnamefont {Zgid}},\ }\bibfield
  {title} {\bibinfo {title} {Challenges with relativistic {GW} calculations in
  solids and molecules},\ }\bibfield  {journal} {\bibinfo  {journal} {Faraday
  Discussions}\ }\href {https://doi.org/10.1039/d4fd00043a}
  {10.1039/d4fd00043a} (\bibinfo {year} {2024})\BibitemShut {NoStop}%
\bibitem [{\citenamefont {Brivio}\ \emph {et~al.}(2014)\citenamefont {Brivio},
  \citenamefont {Butler}, \citenamefont {Walsh},\ and\ \citenamefont {van
  Schilfgaarde}}]{brivio_prb_2014}%
  \BibitemOpen
  \bibfield  {author} {\bibinfo {author} {\bibfnamefont {F.}~\bibnamefont
  {Brivio}}, \bibinfo {author} {\bibfnamefont {K.~T.}\ \bibnamefont {Butler}},
  \bibinfo {author} {\bibfnamefont {A.}~\bibnamefont {Walsh}},\ and\ \bibinfo
  {author} {\bibfnamefont {M.}~\bibnamefont {van Schilfgaarde}},\ }\bibfield
  {title} {\bibinfo {title} {Relativistic quasiparticle self-consistent
  electronic structure of hybrid halide perovskite photovoltaic absorbers},\
  }\href {https://doi.org/10.1103/PhysRevB.89.155204} {\bibfield  {journal}
  {\bibinfo  {journal} {Phys. Rev. B}\ }\textbf {\bibinfo {volume} {89}},\
  \bibinfo {pages} {155204} (\bibinfo {year} {2014})}\BibitemShut {NoStop}%
\bibitem [{\citenamefont {Aguilera}\ \emph {et~al.}(2015)\citenamefont
  {Aguilera}, \citenamefont {Friedrich},\ and\ \citenamefont
  {Bl\"ugel}}]{aguilera_prb_2015}%
  \BibitemOpen
  \bibfield  {author} {\bibinfo {author} {\bibfnamefont {I.}~\bibnamefont
  {Aguilera}}, \bibinfo {author} {\bibfnamefont {C.}~\bibnamefont
  {Friedrich}},\ and\ \bibinfo {author} {\bibfnamefont {S.}~\bibnamefont
  {Bl\"ugel}},\ }\bibfield  {title} {\bibinfo {title} {Electronic phase
  transitions of bismuth under strain from relativistic self-consistent ${GW}$
  calculations},\ }\href {https://doi.org/10.1103/PhysRevB.91.125129}
  {\bibfield  {journal} {\bibinfo  {journal} {Phys. Rev. B}\ }\textbf {\bibinfo
  {volume} {91}},\ \bibinfo {pages} {125129} (\bibinfo {year}
  {2015})}\BibitemShut {NoStop}%
\bibitem [{\citenamefont {Clark}\ \emph {et~al.}(2021)\citenamefont {Clark},
  \citenamefont {Freyse}, \citenamefont {Aguilera}, \citenamefont {Frolov},
  \citenamefont {Ionov}, \citenamefont {Bozhko}, \citenamefont {Yashina},\ and\
  \citenamefont {Sánchez-Barriga}}]{clark_cp_2021}%
  \BibitemOpen
  \bibfield  {author} {\bibinfo {author} {\bibfnamefont {O.~J.}\ \bibnamefont
  {Clark}}, \bibinfo {author} {\bibfnamefont {F.}~\bibnamefont {Freyse}},
  \bibinfo {author} {\bibfnamefont {I.}~\bibnamefont {Aguilera}}, \bibinfo
  {author} {\bibfnamefont {A.~S.}\ \bibnamefont {Frolov}}, \bibinfo {author}
  {\bibfnamefont {A.~M.}\ \bibnamefont {Ionov}}, \bibinfo {author}
  {\bibfnamefont {S.~I.}\ \bibnamefont {Bozhko}}, \bibinfo {author}
  {\bibfnamefont {L.~V.}\ \bibnamefont {Yashina}},\ and\ \bibinfo {author}
  {\bibfnamefont {J.}~\bibnamefont {Sánchez-Barriga}},\ }\bibfield  {title}
  {\bibinfo {title} {Observation of a giant mass enhancement in the ultrafast
  electron dynamics of a topological semimetal},\ }\bibfield  {journal}
  {\bibinfo  {journal} {Communications Physics}\ }\textbf {\bibinfo {volume}
  {4}},\ \href {https://doi.org/10.1038/s42005-021-00657-6}
  {10.1038/s42005-021-00657-6} (\bibinfo {year} {2021})\BibitemShut {NoStop}%
\bibitem [{\citenamefont {Aguilera}\ \emph {et~al.}(2021)\citenamefont
  {Aguilera}, \citenamefont {Kim}, \citenamefont {Friedrich}, \citenamefont
  {Bihlmayer},\ and\ \citenamefont {Bl\"ugel}}]{aguilera_prb_2021}%
  \BibitemOpen
  \bibfield  {author} {\bibinfo {author} {\bibfnamefont {I.}~\bibnamefont
  {Aguilera}}, \bibinfo {author} {\bibfnamefont {H.-J.}\ \bibnamefont {Kim}},
  \bibinfo {author} {\bibfnamefont {C.}~\bibnamefont {Friedrich}}, \bibinfo
  {author} {\bibfnamefont {G.}~\bibnamefont {Bihlmayer}},\ and\ \bibinfo
  {author} {\bibfnamefont {S.}~\bibnamefont {Bl\"ugel}},\ }\bibfield  {title}
  {\bibinfo {title} {$\mathbb{Z}_2$ topology of bismuth},\ }\href
  {https://doi.org/10.1103/PhysRevMaterials.5.L091201} {\bibfield  {journal}
  {\bibinfo  {journal} {Phys. Rev. Mater.}\ }\textbf {\bibinfo {volume} {5}},\
  \bibinfo {pages} {L091201} (\bibinfo {year} {2021})}\BibitemShut {NoStop}%
\bibitem [{\citenamefont {Shishkin}\ \emph {et~al.}(2007)\citenamefont
  {Shishkin}, \citenamefont {Marsman},\ and\ \citenamefont
  {Kresse}}]{kresse_prl_2007}%
  \BibitemOpen
  \bibfield  {author} {\bibinfo {author} {\bibfnamefont {M.}~\bibnamefont
  {Shishkin}}, \bibinfo {author} {\bibfnamefont {M.}~\bibnamefont {Marsman}},\
  and\ \bibinfo {author} {\bibfnamefont {G.}~\bibnamefont {Kresse}},\
  }\bibfield  {title} {\bibinfo {title} {{Accurate Quasiparticle Spectra from
  Self-Consistent GW Calculations with Vertex Corrections}},\ }\href
  {https://doi.org/10.1103/PhysRevLett.99.246403} {\bibfield  {journal}
  {\bibinfo  {journal} {Phys. Rev. Lett.}\ }\textbf {\bibinfo {volume} {99}},\
  \bibinfo {pages} {246403} (\bibinfo {year} {2007})}\BibitemShut {NoStop}%
\bibitem [{\citenamefont {Chen}\ and\ \citenamefont
  {Pasquarello}(2015)}]{pasquarello_prb_2015}%
  \BibitemOpen
  \bibfield  {author} {\bibinfo {author} {\bibfnamefont {W.}~\bibnamefont
  {Chen}}\ and\ \bibinfo {author} {\bibfnamefont {A.}~\bibnamefont
  {Pasquarello}},\ }\bibfield  {title} {\bibinfo {title} {{Accurate} band gaps
  of extended systems via efficient vertex corrections in ${GW}$},\ }\href
  {https://doi.org/10.1103/PhysRevB.92.041115} {\bibfield  {journal} {\bibinfo
  {journal} {Phys. Rev. B}\ }\textbf {\bibinfo {volume} {92}},\ \bibinfo
  {pages} {041115(R)} (\bibinfo {year} {2015})}\BibitemShut {NoStop}%
\bibitem [{\citenamefont {Cunningham}\ \emph {et~al.}(2023)\citenamefont
  {Cunningham}, \citenamefont {Gr\"uning}, \citenamefont {Pashov},\ and\
  \citenamefont {van Schilfgaarde}}]{schilf_prb_2023}%
  \BibitemOpen
  \bibfield  {author} {\bibinfo {author} {\bibfnamefont {B.}~\bibnamefont
  {Cunningham}}, \bibinfo {author} {\bibfnamefont {M.}~\bibnamefont
  {Gr\"uning}}, \bibinfo {author} {\bibfnamefont {D.}~\bibnamefont {Pashov}},\
  and\ \bibinfo {author} {\bibfnamefont {M.}~\bibnamefont {van Schilfgaarde}},\
  }\bibfield  {title} {\bibinfo {title} {{QS}${G\hat{W}}$: {Quasiparticle}
  self-consistent ${GW}$ with ladder diagrams in ${W}$},\ }\href
  {https://doi.org/10.1103/PhysRevB.108.165104} {\bibfield  {journal} {\bibinfo
   {journal} {Phys. Rev. B}\ }\textbf {\bibinfo {volume} {108}},\ \bibinfo
  {pages} {165104} (\bibinfo {year} {2023})}\BibitemShut {NoStop}%
\bibitem [{\citenamefont {Kioupakis}\ \emph {et~al.}(2010)\citenamefont
  {Kioupakis}, \citenamefont {Tiago},\ and\ \citenamefont
  {Louie}}]{kioupakis_prb_2010}%
  \BibitemOpen
  \bibfield  {author} {\bibinfo {author} {\bibfnamefont {E.}~\bibnamefont
  {Kioupakis}}, \bibinfo {author} {\bibfnamefont {M.~L.}\ \bibnamefont
  {Tiago}},\ and\ \bibinfo {author} {\bibfnamefont {S.~G.}\ \bibnamefont
  {Louie}},\ }\bibfield  {title} {\bibinfo {title} {{Quasiparticle} electronic
  structure of bismuth telluride in the ${GW}$ approximation},\ }\href
  {https://doi.org/10.1103/PhysRevB.82.245203} {\bibfield  {journal} {\bibinfo
  {journal} {Phys. Rev. B}\ }\textbf {\bibinfo {volume} {82}},\ \bibinfo
  {pages} {245203} (\bibinfo {year} {2010})}\BibitemShut {NoStop}%
\bibitem [{\citenamefont {Yazyev}\ \emph {et~al.}(2012)\citenamefont {Yazyev},
  \citenamefont {Kioupakis}, \citenamefont {Moore},\ and\ \citenamefont
  {Louie}}]{yazyev_prb_2012}%
  \BibitemOpen
  \bibfield  {author} {\bibinfo {author} {\bibfnamefont {O.~V.}\ \bibnamefont
  {Yazyev}}, \bibinfo {author} {\bibfnamefont {E.}~\bibnamefont {Kioupakis}},
  \bibinfo {author} {\bibfnamefont {J.~E.}\ \bibnamefont {Moore}},\ and\
  \bibinfo {author} {\bibfnamefont {S.~G.}\ \bibnamefont {Louie}},\ }\bibfield
  {title} {\bibinfo {title} {{Quasiparticle} effects in the bulk and
  surface-state bands of {Bi}$_{2}${Se}$_{3}$ and {Bi}$_{2}${Te}$_{3}$
  topological insulators},\ }\href {https://doi.org/10.1103/PhysRevB.85.161101}
  {\bibfield  {journal} {\bibinfo  {journal} {Phys. Rev. B}\ }\textbf {\bibinfo
  {volume} {85}},\ \bibinfo {pages} {161101(R)} (\bibinfo {year}
  {2012})}\BibitemShut {NoStop}%
\bibitem [{\citenamefont {Wiktor}\ \emph {et~al.}(2017)\citenamefont {Wiktor},
  \citenamefont {Rothlisberger},\ and\ \citenamefont
  {Pasquarello}}]{wiktor_jpcl_2017}%
  \BibitemOpen
  \bibfield  {author} {\bibinfo {author} {\bibfnamefont {J.}~\bibnamefont
  {Wiktor}}, \bibinfo {author} {\bibfnamefont {U.}~\bibnamefont
  {Rothlisberger}},\ and\ \bibinfo {author} {\bibfnamefont {A.}~\bibnamefont
  {Pasquarello}},\ }\bibfield  {title} {\bibinfo {title} {{Predictive
  Determination of Band Gaps of Inorganic Halide Perovskites}},\ }\href
  {https://doi.org/10.1021/acs.jpclett.7b02648} {\bibfield  {journal} {\bibinfo
   {journal} {The Journal of Physical Chemistry Letters}\ }\textbf {\bibinfo
  {volume} {8}},\ \bibinfo {pages} {5507} (\bibinfo {year} {2017})}\BibitemShut
  {NoStop}%
\bibitem [{\citenamefont {Shishkin}\ and\ \citenamefont
  {Kresse}(2007)}]{kresse_prb_2007}%
  \BibitemOpen
  \bibfield  {author} {\bibinfo {author} {\bibfnamefont {M.}~\bibnamefont
  {Shishkin}}\ and\ \bibinfo {author} {\bibfnamefont {G.}~\bibnamefont
  {Kresse}},\ }\bibfield  {title} {\bibinfo {title} {{Self}-consistent ${GW}$
  calculations for semiconductors and insulators},\ }\href
  {https://doi.org/10.1103/PhysRevB.75.235102} {\bibfield  {journal} {\bibinfo
  {journal} {Phys. Rev. B}\ }\textbf {\bibinfo {volume} {75}},\ \bibinfo
  {pages} {235102} (\bibinfo {year} {2007})}\BibitemShut {NoStop}%
\bibitem [{\citenamefont {Hedin}(1965)}]{hedin_pr_1965}%
  \BibitemOpen
  \bibfield  {author} {\bibinfo {author} {\bibfnamefont {L.}~\bibnamefont
  {Hedin}},\ }\bibfield  {title} {\bibinfo {title} {{New Method for Calculating
  the One-Particle Green's Function with Application to the Electron-Gas
  Problem}},\ }\href {https://doi.org/10.1103/PhysRev.139.A796} {\bibfield
  {journal} {\bibinfo  {journal} {Phys. Rev.}\ }\textbf {\bibinfo {volume}
  {139}},\ \bibinfo {pages} {A796} (\bibinfo {year} {1965})}\BibitemShut
  {NoStop}%
\bibitem [{\citenamefont {Aryasetiawan}\ and\ \citenamefont
  {Biermann}(2008)}]{biermann_prl_2008}%
  \BibitemOpen
  \bibfield  {author} {\bibinfo {author} {\bibfnamefont {F.}~\bibnamefont
  {Aryasetiawan}}\ and\ \bibinfo {author} {\bibfnamefont {S.}~\bibnamefont
  {Biermann}},\ }\bibfield  {title} {\bibinfo {title} {{Generalized Hedin's
  Equations for Quantum Many-Body Systems with Spin-Dependent Interactions}},\
  }\href {https://doi.org/10.1103/PhysRevLett.100.116402} {\bibfield  {journal}
  {\bibinfo  {journal} {Phys. Rev. Lett.}\ }\textbf {\bibinfo {volume} {100}},\
  \bibinfo {pages} {116402} (\bibinfo {year} {2008})}\BibitemShut {NoStop}%
\bibitem [{\citenamefont {Aryasetiawan}\ and\ \citenamefont
  {Biermann}(2009)}]{biermann_jpcm_2009}%
  \BibitemOpen
  \bibfield  {author} {\bibinfo {author} {\bibfnamefont {F.}~\bibnamefont
  {Aryasetiawan}}\ and\ \bibinfo {author} {\bibfnamefont {S.}~\bibnamefont
  {Biermann}},\ }\bibfield  {title} {\bibinfo {title} {{Generalized Hedin}
  equations and $\sigma {G} \sigma {W}$ approximation for quantum many-body
  systems with spin-dependent interactions},\ }\href
  {https://doi.org/10.1088/0953-8984/21/6/064232} {\bibfield  {journal}
  {\bibinfo  {journal} {Journal of Physics: Condensed Matter}\ }\textbf
  {\bibinfo {volume} {21}},\ \bibinfo {pages} {064232} (\bibinfo {year}
  {2009})}\BibitemShut {NoStop}%
\bibitem [{\citenamefont {Barker}\ \emph {et~al.}(2022)\citenamefont {Barker},
  \citenamefont {Deslippe}, \citenamefont {Lischner}, \citenamefont {Jain},
  \citenamefont {Yazyev}, \citenamefont {Strubbe},\ and\ \citenamefont
  {Louie}}]{louie_prb_2022}%
  \BibitemOpen
  \bibfield  {author} {\bibinfo {author} {\bibfnamefont {B.~A.}\ \bibnamefont
  {Barker}}, \bibinfo {author} {\bibfnamefont {J.}~\bibnamefont {Deslippe}},
  \bibinfo {author} {\bibfnamefont {J.}~\bibnamefont {Lischner}}, \bibinfo
  {author} {\bibfnamefont {M.}~\bibnamefont {Jain}}, \bibinfo {author}
  {\bibfnamefont {O.~V.}\ \bibnamefont {Yazyev}}, \bibinfo {author}
  {\bibfnamefont {D.~A.}\ \bibnamefont {Strubbe}},\ and\ \bibinfo {author}
  {\bibfnamefont {S.~G.}\ \bibnamefont {Louie}},\ }\bibfield  {title} {\bibinfo
  {title} {{Spinor} ${GW}$/{Bethe-Salpeter} calculations in {BerkeleyGW}:
  Implementation, symmetries, benchmarking, and performance},\ }\href
  {https://doi.org/10.1103/PhysRevB.106.115127} {\bibfield  {journal} {\bibinfo
   {journal} {Phys. Rev. B}\ }\textbf {\bibinfo {volume} {106}},\ \bibinfo
  {pages} {115127} (\bibinfo {year} {2022})}\BibitemShut {NoStop}%
\bibitem [{\citenamefont {Baroni}\ \emph {et~al.}(2001)\citenamefont {Baroni},
  \citenamefont {de~Gironcoli}, \citenamefont {Dal~Corso},\ and\ \citenamefont
  {Giannozzi}}]{baroni_phonons_2001}%
  \BibitemOpen
  \bibfield  {author} {\bibinfo {author} {\bibfnamefont {S.}~\bibnamefont
  {Baroni}}, \bibinfo {author} {\bibfnamefont {S.}~\bibnamefont
  {de~Gironcoli}}, \bibinfo {author} {\bibfnamefont {A.}~\bibnamefont
  {Dal~Corso}},\ and\ \bibinfo {author} {\bibfnamefont {P.}~\bibnamefont
  {Giannozzi}},\ }\bibfield  {title} {\bibinfo {title} {{Phonons and related
  crystal properties from density-functional perturbation theory}},\ }\href
  {https://doi.org/10.1103/RevModPhys.73.515} {\bibfield  {journal} {\bibinfo
  {journal} {Rev. Mod. Phys.}\ }\textbf {\bibinfo {volume} {73}},\ \bibinfo
  {pages} {515} (\bibinfo {year} {2001})}\BibitemShut {NoStop}%
\bibitem [{\citenamefont {Gorni}(2016)}]{gorni_spin-fluctuation_2016}%
  \BibitemOpen
  \bibfield  {author} {\bibinfo {author} {\bibfnamefont {T.}~\bibnamefont
  {Gorni}},\ }\emph {\bibinfo {title} {Spin-fluctuation spectra in magnetic
  systems: a novel approach based on TDDFT}},\ \href
  {https://hdl.handle.net/20.500.11767/43342} {Ph.D. thesis},\ \bibinfo
  {school} {SISSA} (\bibinfo {year} {2016})\BibitemShut {NoStop}%
\bibitem [{\citenamefont {Gorni}\ \emph {et~al.}(2018)\citenamefont {Gorni},
  \citenamefont {Timrov},\ and\ \citenamefont {Baroni}}]{gorni_spin_2018}%
  \BibitemOpen
  \bibfield  {author} {\bibinfo {author} {\bibfnamefont {T.}~\bibnamefont
  {Gorni}}, \bibinfo {author} {\bibfnamefont {I.}~\bibnamefont {Timrov}},\ and\
  \bibinfo {author} {\bibfnamefont {S.}~\bibnamefont {Baroni}},\ }\bibfield
  {title} {\bibinfo {title} {{Spin dynamics from time-dependent density
  functional perturbation theory}},\ }\href
  {https://doi.org/10.1140/epjb/e2018-90247-9} {\bibfield  {journal} {\bibinfo
  {journal} {The European Physical Journal B}\ }\textbf {\bibinfo {volume}
  {91}},\ \bibinfo {pages} {249} (\bibinfo {year} {2018})}\BibitemShut
  {NoStop}%
\bibitem [{\citenamefont {Cao}\ \emph {et~al.}(2018)\citenamefont {Cao},
  \citenamefont {Lambert}, \citenamefont {Radaelli},\ and\ \citenamefont
  {Giustino}}]{cao_ab_2018}%
  \BibitemOpen
  \bibfield  {author} {\bibinfo {author} {\bibfnamefont {K.}~\bibnamefont
  {Cao}}, \bibinfo {author} {\bibfnamefont {H.}~\bibnamefont {Lambert}},
  \bibinfo {author} {\bibfnamefont {P.~G.}\ \bibnamefont {Radaelli}},\ and\
  \bibinfo {author} {\bibfnamefont {F.}~\bibnamefont {Giustino}},\ }\bibfield
  {title} {\bibinfo {title} {{Ab initio calculation of spin fluctuation spectra
  using time-dependent density functional perturbation theory, plane waves, and
  pseudopotentials}},\ }\href {https://doi.org/10.1103/PhysRevB.97.024420}
  {\bibfield  {journal} {\bibinfo  {journal} {Physical Review B}\ }\textbf
  {\bibinfo {volume} {97}},\ \bibinfo {pages} {024420} (\bibinfo {year}
  {2018})}\BibitemShut {NoStop}%
\bibitem [{\citenamefont {Urru}\ and\ \citenamefont
  {Dal~Corso}(2019)}]{urru_density_2019}%
  \BibitemOpen
  \bibfield  {author} {\bibinfo {author} {\bibfnamefont {A.}~\bibnamefont
  {Urru}}\ and\ \bibinfo {author} {\bibfnamefont {A.}~\bibnamefont
  {Dal~Corso}},\ }\bibfield  {title} {\bibinfo {title} {{Density functional
  perturbation theory for lattice dynamics with fully relativistic ultrasoft
  pseudopotentials: The magnetic case}},\ }\href
  {https://doi.org/10.1103/PhysRevB.100.045115} {\bibfield  {journal} {\bibinfo
   {journal} {Physical Review B}\ }\textbf {\bibinfo {volume} {100}},\ \bibinfo
  {pages} {045115} (\bibinfo {year} {2019})}\BibitemShut {NoStop}%
\bibitem [{\citenamefont {Urru}(2020)}]{urru_lattice_2020}%
  \BibitemOpen
  \bibfield  {author} {\bibinfo {author} {\bibfnamefont {A.}~\bibnamefont
  {Urru}},\ }\emph {\bibinfo {title} {Lattice dynamics with Fully Relativistic
  Pseudopotentials for magnetic systems, with selected applications}},\ \href
  {https://hdl.handle.net/20.500.11767/115671} {Ph.D. thesis},\ \bibinfo
  {school} {SISSA} (\bibinfo {year} {2020})\BibitemShut {NoStop}%
\bibitem [{\citenamefont {Dabo}\ \emph {et~al.}(2010)\citenamefont {Dabo},
  \citenamefont {Ferretti}, \citenamefont {Poilvert}, \citenamefont {Li},
  \citenamefont {Marzari},\ and\ \citenamefont {Cococcioni}}]{Dabo2010}%
  \BibitemOpen
  \bibfield  {author} {\bibinfo {author} {\bibfnamefont {I.}~\bibnamefont
  {Dabo}}, \bibinfo {author} {\bibfnamefont {A.}~\bibnamefont {Ferretti}},
  \bibinfo {author} {\bibfnamefont {N.}~\bibnamefont {Poilvert}}, \bibinfo
  {author} {\bibfnamefont {Y.}~\bibnamefont {Li}}, \bibinfo {author}
  {\bibfnamefont {N.}~\bibnamefont {Marzari}},\ and\ \bibinfo {author}
  {\bibfnamefont {M.}~\bibnamefont {Cococcioni}},\ }\bibfield  {title}
  {\bibinfo {title} {{Koopmans}' condition for density-functional theory},\
  }\href {https://doi.org/10.1103/PhysRevB.82.115121} {\bibfield  {journal}
  {\bibinfo  {journal} {Phys. Rev. B}\ }\textbf {\bibinfo {volume} {82}},\
  \bibinfo {pages} {115121} (\bibinfo {year} {2010})}\BibitemShut {NoStop}%
\bibitem [{\citenamefont {Borghi}\ \emph {et~al.}(2014)\citenamefont {Borghi},
  \citenamefont {Ferretti}, \citenamefont {Nguyen}, \citenamefont {Dabo},\ and\
  \citenamefont {Marzari}}]{Borghi2014}%
  \BibitemOpen
  \bibfield  {author} {\bibinfo {author} {\bibfnamefont {G.}~\bibnamefont
  {Borghi}}, \bibinfo {author} {\bibfnamefont {A.}~\bibnamefont {Ferretti}},
  \bibinfo {author} {\bibfnamefont {N.~L.}\ \bibnamefont {Nguyen}}, \bibinfo
  {author} {\bibfnamefont {I.}~\bibnamefont {Dabo}},\ and\ \bibinfo {author}
  {\bibfnamefont {N.}~\bibnamefont {Marzari}},\ }\bibfield  {title} {\bibinfo
  {title} {{Koopmans-compliant functionals and their performance against
  reference molecular data}},\ }\href
  {https://doi.org/10.1103/PhysRevB.90.075135} {\bibfield  {journal} {\bibinfo
  {journal} {Phys. Rev. B}\ }\textbf {\bibinfo {volume} {90}},\ \bibinfo
  {pages} {075135} (\bibinfo {year} {2014})}\BibitemShut {NoStop}%
\bibitem [{\citenamefont {Colonna}\ \emph {et~al.}(2018)\citenamefont
  {Colonna}, \citenamefont {Nguyen}, \citenamefont {Ferretti},\ and\
  \citenamefont {Marzari}}]{colonna_jctc_2018}%
  \BibitemOpen
  \bibfield  {author} {\bibinfo {author} {\bibfnamefont {N.}~\bibnamefont
  {Colonna}}, \bibinfo {author} {\bibfnamefont {N.~L.}\ \bibnamefont {Nguyen}},
  \bibinfo {author} {\bibfnamefont {A.}~\bibnamefont {Ferretti}},\ and\
  \bibinfo {author} {\bibfnamefont {N.}~\bibnamefont {Marzari}},\ }\bibfield
  {title} {\bibinfo {title} {{Screening in Orbital-Density-Dependent
  Functionals}},\ }\href {https://doi.org/10.1021/acs.jctc.7b01116} {\bibfield
  {journal} {\bibinfo  {journal} {Journal of Chemical Theory and Computation}\
  }\textbf {\bibinfo {volume} {14}},\ \bibinfo {pages} {2549} (\bibinfo {year}
  {2018})}\BibitemShut {NoStop}%
\bibitem [{\citenamefont {Nguyen}\ \emph {et~al.}(2018)\citenamefont {Nguyen},
  \citenamefont {Colonna}, \citenamefont {Ferretti},\ and\ \citenamefont
  {Marzari}}]{Linh2018}%
  \BibitemOpen
  \bibfield  {author} {\bibinfo {author} {\bibfnamefont {N.~L.}\ \bibnamefont
  {Nguyen}}, \bibinfo {author} {\bibfnamefont {N.}~\bibnamefont {Colonna}},
  \bibinfo {author} {\bibfnamefont {A.}~\bibnamefont {Ferretti}},\ and\
  \bibinfo {author} {\bibfnamefont {N.}~\bibnamefont {Marzari}},\ }\bibfield
  {title} {\bibinfo {title} {{Koopmans-Compliant Spectral Functionals for
  Extended Systems}},\ }\href {https://doi.org/10.1103/PhysRevX.8.021051}
  {\bibfield  {journal} {\bibinfo  {journal} {Phys. Rev. X}\ }\textbf {\bibinfo
  {volume} {8}},\ \bibinfo {pages} {021051} (\bibinfo {year}
  {2018})}\BibitemShut {NoStop}%
\bibitem [{\citenamefont {Linscott}\ \emph {et~al.}(2023)\citenamefont
  {Linscott}, \citenamefont {Colonna}, \citenamefont {De~Gennaro},
  \citenamefont {Nguyen}, \citenamefont {Borghi}, \citenamefont {Ferretti},
  \citenamefont {Dabo},\ and\ \citenamefont
  {Marzari}}]{linscott_koopmans_2023}%
  \BibitemOpen
  \bibfield  {author} {\bibinfo {author} {\bibfnamefont {E.~B.}\ \bibnamefont
  {Linscott}}, \bibinfo {author} {\bibfnamefont {N.}~\bibnamefont {Colonna}},
  \bibinfo {author} {\bibfnamefont {R.}~\bibnamefont {De~Gennaro}}, \bibinfo
  {author} {\bibfnamefont {N.~L.}\ \bibnamefont {Nguyen}}, \bibinfo {author}
  {\bibfnamefont {G.}~\bibnamefont {Borghi}}, \bibinfo {author} {\bibfnamefont
  {A.}~\bibnamefont {Ferretti}}, \bibinfo {author} {\bibfnamefont
  {I.}~\bibnamefont {Dabo}},\ and\ \bibinfo {author} {\bibfnamefont
  {N.}~\bibnamefont {Marzari}},\ }\bibfield  {title} {\bibinfo {title}
  {{koopmans: An Open-Source Package for Accurately and Efficiently Predicting
  Spectral Properties with Koopmans Functionals}},\ }\href
  {https://doi.org/10.1021/acs.jctc.3c00652} {\bibfield  {journal} {\bibinfo
  {journal} {Journal of Chemical Theory and Computation}\ }\textbf {\bibinfo
  {volume} {19}},\ \bibinfo {pages} {7097} (\bibinfo {year}
  {2023})}\BibitemShut {NoStop}%
\bibitem [{\citenamefont {Nguyen}\ \emph {et~al.}(2015)\citenamefont {Nguyen},
  \citenamefont {Borghi}, \citenamefont {Ferretti}, \citenamefont {Dabo},\ and\
  \citenamefont {Marzari}}]{nguyen_first-principles_2015}%
  \BibitemOpen
  \bibfield  {author} {\bibinfo {author} {\bibfnamefont {N.~L.}\ \bibnamefont
  {Nguyen}}, \bibinfo {author} {\bibfnamefont {G.}~\bibnamefont {Borghi}},
  \bibinfo {author} {\bibfnamefont {A.}~\bibnamefont {Ferretti}}, \bibinfo
  {author} {\bibfnamefont {I.}~\bibnamefont {Dabo}},\ and\ \bibinfo {author}
  {\bibfnamefont {N.}~\bibnamefont {Marzari}},\ }\bibfield  {title} {\bibinfo
  {title} {{First-Principles Photoemission Spectroscopy and Orbital Tomography
  in Molecules from Koopmans-Compliant Functionals}},\ }\href
  {https://doi.org/10.1103/PhysRevLett.114.166405} {\bibfield  {journal}
  {\bibinfo  {journal} {Physical Review Letters}\ }\textbf {\bibinfo {volume}
  {114}},\ \bibinfo {pages} {166405} (\bibinfo {year} {2015})}\BibitemShut
  {NoStop}%
\bibitem [{\citenamefont {Nguyen}\ \emph {et~al.}(2016)\citenamefont {Nguyen},
  \citenamefont {Borghi}, \citenamefont {Ferretti},\ and\ \citenamefont
  {Marzari}}]{nguyen_first-principles_2016}%
  \BibitemOpen
  \bibfield  {author} {\bibinfo {author} {\bibfnamefont {N.~L.}\ \bibnamefont
  {Nguyen}}, \bibinfo {author} {\bibfnamefont {G.}~\bibnamefont {Borghi}},
  \bibinfo {author} {\bibfnamefont {A.}~\bibnamefont {Ferretti}},\ and\
  \bibinfo {author} {\bibfnamefont {N.}~\bibnamefont {Marzari}},\ }\bibfield
  {title} {\bibinfo {title} {{First-Principles Photoemission Spectroscopy of}
  {DNA} and {RNA} {Nucleobases} from {Koopmans-Compliant Functionals}},\ }\href
  {https://doi.org/10.1021/acs.jctc.6b00145} {\bibfield  {journal} {\bibinfo
  {journal} {Journal of Chemical Theory and Computation}\ }\textbf {\bibinfo
  {volume} {12}},\ \bibinfo {pages} {3948} (\bibinfo {year}
  {2016})}\BibitemShut {NoStop}%
\bibitem [{\citenamefont {Elliott}\ \emph {et~al.}(2019)\citenamefont
  {Elliott}, \citenamefont {Colonna}, \citenamefont {Marsili}, \citenamefont
  {Marzari},\ and\ \citenamefont {Umari}}]{Elliott2019}%
  \BibitemOpen
  \bibfield  {author} {\bibinfo {author} {\bibfnamefont {J.~D.}\ \bibnamefont
  {Elliott}}, \bibinfo {author} {\bibfnamefont {N.}~\bibnamefont {Colonna}},
  \bibinfo {author} {\bibfnamefont {M.}~\bibnamefont {Marsili}}, \bibinfo
  {author} {\bibfnamefont {N.}~\bibnamefont {Marzari}},\ and\ \bibinfo {author}
  {\bibfnamefont {P.}~\bibnamefont {Umari}},\ }\bibfield  {title} {\bibinfo
  {title} {{Koopmans Meets Bethe--Salpeter: Excitonic Optical Spectra without
  GW}},\ }\href {https://doi.org/10.1021/acs.jctc.8b01271} {\bibfield
  {journal} {\bibinfo  {journal} {Journal of Chemical Theory and Computation}\
  }\textbf {\bibinfo {volume} {15}},\ \bibinfo {pages} {3710} (\bibinfo {year}
  {2019})}\BibitemShut {NoStop}%
\bibitem [{\citenamefont {Colonna}\ \emph {et~al.}(2019)\citenamefont
  {Colonna}, \citenamefont {Nguyen}, \citenamefont {Ferretti},\ and\
  \citenamefont {Marzari}}]{colonna_jctc_2019}%
  \BibitemOpen
  \bibfield  {author} {\bibinfo {author} {\bibfnamefont {N.}~\bibnamefont
  {Colonna}}, \bibinfo {author} {\bibfnamefont {N.~L.}\ \bibnamefont {Nguyen}},
  \bibinfo {author} {\bibfnamefont {A.}~\bibnamefont {Ferretti}},\ and\
  \bibinfo {author} {\bibfnamefont {N.}~\bibnamefont {Marzari}},\ }\bibfield
  {title} {\bibinfo {title} {{Koopmans-Compliant Functionals and Potentials and
  Their Application to the GW100 Test Set}},\ }\href
  {https://doi.org/10.1021/acs.jctc.8b00976} {\bibfield  {journal} {\bibinfo
  {journal} {Journal of Chemical Theory and Computation}\ }\textbf {\bibinfo
  {volume} {15}},\ \bibinfo {pages} {1905} (\bibinfo {year}
  {2019})}\BibitemShut {NoStop}%
\bibitem [{\citenamefont {de~Almeida}\ \emph {et~al.}(2021)\citenamefont
  {de~Almeida}, \citenamefont {Nguyen}, \citenamefont {Colonna}, \citenamefont
  {Chen}, \citenamefont {Rodrigues~Miranda}, \citenamefont {Pasquarello},\ and\
  \citenamefont {Marzari}}]{de_almeida_electronic_2021}%
  \BibitemOpen
  \bibfield  {author} {\bibinfo {author} {\bibfnamefont {J.~M.}\ \bibnamefont
  {de~Almeida}}, \bibinfo {author} {\bibfnamefont {N.~L.}\ \bibnamefont
  {Nguyen}}, \bibinfo {author} {\bibfnamefont {N.}~\bibnamefont {Colonna}},
  \bibinfo {author} {\bibfnamefont {W.}~\bibnamefont {Chen}}, \bibinfo {author}
  {\bibfnamefont {C.}~\bibnamefont {Rodrigues~Miranda}}, \bibinfo {author}
  {\bibfnamefont {A.}~\bibnamefont {Pasquarello}},\ and\ \bibinfo {author}
  {\bibfnamefont {N.}~\bibnamefont {Marzari}},\ }\bibfield  {title} {\bibinfo
  {title} {{Electronic Structure of Water from Koopmans-Compliant
  Functionals}},\ }\href {https://doi.org/10.1021/acs.jctc.1c00063} {\bibfield
  {journal} {\bibinfo  {journal} {Journal of Chemical Theory and Computation}\
  }\textbf {\bibinfo {volume} {17}},\ \bibinfo {pages} {3923} (\bibinfo {year}
  {2021})}\BibitemShut {NoStop}%
\bibitem [{\citenamefont {De~Gennaro}\ \emph {et~al.}(2022)\citenamefont
  {De~Gennaro}, \citenamefont {Colonna}, \citenamefont {Linscott},\ and\
  \citenamefont {Marzari}}]{degennaro_prb_2022}%
  \BibitemOpen
  \bibfield  {author} {\bibinfo {author} {\bibfnamefont {R.}~\bibnamefont
  {De~Gennaro}}, \bibinfo {author} {\bibfnamefont {N.}~\bibnamefont {Colonna}},
  \bibinfo {author} {\bibfnamefont {E.}~\bibnamefont {Linscott}},\ and\
  \bibinfo {author} {\bibfnamefont {N.}~\bibnamefont {Marzari}},\ }\bibfield
  {title} {\bibinfo {title} {{Bloch's theorem in orbital-density-dependent
  functionals: Band structures from Koopmans spectral functionals}},\ }\href
  {https://doi.org/10.1103/PhysRevB.106.035106} {\bibfield  {journal} {\bibinfo
   {journal} {Phys. Rev. B}\ }\textbf {\bibinfo {volume} {106}},\ \bibinfo
  {pages} {035106} (\bibinfo {year} {2022})}\BibitemShut {NoStop}%
\bibitem [{\citenamefont {Colonna}\ \emph {et~al.}(2022)\citenamefont
  {Colonna}, \citenamefont {De~Gennaro}, \citenamefont {Linscott},\ and\
  \citenamefont {Marzari}}]{colonna_jctc_2022}%
  \BibitemOpen
  \bibfield  {author} {\bibinfo {author} {\bibfnamefont {N.}~\bibnamefont
  {Colonna}}, \bibinfo {author} {\bibfnamefont {R.}~\bibnamefont {De~Gennaro}},
  \bibinfo {author} {\bibfnamefont {E.}~\bibnamefont {Linscott}},\ and\
  \bibinfo {author} {\bibfnamefont {N.}~\bibnamefont {Marzari}},\ }\bibfield
  {title} {\bibinfo {title} {{Koopmans Spectral Functionals in Periodic
  Boundary Conditions}},\ }\href {https://doi.org/10.1021/acs.jctc.2c00161}
  {\bibfield  {journal} {\bibinfo  {journal} {Journal of Chemical Theory and
  Computation}\ }\textbf {\bibinfo {volume} {18}},\ \bibinfo {pages} {5435}
  (\bibinfo {year} {2022})}\BibitemShut {NoStop}%
\bibitem [{\citenamefont {Ingall}\ \emph {et~al.}(2024)\citenamefont {Ingall},
  \citenamefont {Linscott}, \citenamefont {Colonna}, \citenamefont {Page},\
  and\ \citenamefont {Keast}}]{ingall_accurate_2024}%
  \BibitemOpen
  \bibfield  {author} {\bibinfo {author} {\bibfnamefont {J.~E.}\ \bibnamefont
  {Ingall}}, \bibinfo {author} {\bibfnamefont {E.}~\bibnamefont {Linscott}},
  \bibinfo {author} {\bibfnamefont {N.}~\bibnamefont {Colonna}}, \bibinfo
  {author} {\bibfnamefont {A.~J.}\ \bibnamefont {Page}},\ and\ \bibinfo
  {author} {\bibfnamefont {V.~J.}\ \bibnamefont {Keast}},\ }\bibfield  {title}
  {\bibinfo {title} {Accurate and {Efficient} {Computation} of the
  {Fundamental} {Bandgap} of the {Vacancy}-{Ordered} {Double} {Perovskite}
  {Cs$_2$TiBr$_6$}},\ }\bibfield  {journal} {\bibinfo  {journal} {The Journal
  of Physical Chemistry C}\ }\href {https://doi.org/10.1021/acs.jpcc.3c07957}
  {10.1021/acs.jpcc.3c07957} (\bibinfo {year} {2024})\BibitemShut {NoStop}%
\bibitem [{\citenamefont {Marzari}\ \emph {et~al.}(2012)\citenamefont
  {Marzari}, \citenamefont {Mostofi}, \citenamefont {Yates}, \citenamefont
  {Souza},\ and\ \citenamefont {Vanderbilt}}]{Marzari2012}%
  \BibitemOpen
  \bibfield  {author} {\bibinfo {author} {\bibfnamefont {N.}~\bibnamefont
  {Marzari}}, \bibinfo {author} {\bibfnamefont {A.~A.}\ \bibnamefont
  {Mostofi}}, \bibinfo {author} {\bibfnamefont {J.~R.}\ \bibnamefont {Yates}},
  \bibinfo {author} {\bibfnamefont {I.}~\bibnamefont {Souza}},\ and\ \bibinfo
  {author} {\bibfnamefont {D.}~\bibnamefont {Vanderbilt}},\ }\bibfield  {title}
  {\bibinfo {title} {{Maximally localized Wannier functions: Theory and
  applications}},\ }\href {https://doi.org/10.1103/RevModPhys.84.1419}
  {\bibfield  {journal} {\bibinfo  {journal} {Rev. Mod. Phys.}\ }\textbf
  {\bibinfo {volume} {84}},\ \bibinfo {pages} {1419} (\bibinfo {year}
  {2012})}\BibitemShut {NoStop}%
\bibitem [{\citenamefont {Anisimov}\ and\ \citenamefont
  {Kozhevnikov}(2005)}]{Anisimov2005}%
  \BibitemOpen
  \bibfield  {author} {\bibinfo {author} {\bibfnamefont {V.~I.}\ \bibnamefont
  {Anisimov}}\ and\ \bibinfo {author} {\bibfnamefont {A.~V.}\ \bibnamefont
  {Kozhevnikov}},\ }\bibfield  {title} {\bibinfo {title} {{Transition state
  method and Wannier functions}},\ }\href
  {https://doi.org/10.1103/PhysRevB.72.075125} {\bibfield  {journal} {\bibinfo
  {journal} {Phys. Rev. B}\ }\textbf {\bibinfo {volume} {72}},\ \bibinfo
  {pages} {075125} (\bibinfo {year} {2005})}\BibitemShut {NoStop}%
\bibitem [{\citenamefont {Anisimov}\ \emph {et~al.}(2007)\citenamefont
  {Anisimov}, \citenamefont {Kozhevnikov}, \citenamefont {Korotin},
  \citenamefont {Lukoyanov},\ and\ \citenamefont {Khafizullin}}]{Anisimov2007}%
  \BibitemOpen
  \bibfield  {author} {\bibinfo {author} {\bibfnamefont {V.~I.}\ \bibnamefont
  {Anisimov}}, \bibinfo {author} {\bibfnamefont {A.~V.}\ \bibnamefont
  {Kozhevnikov}}, \bibinfo {author} {\bibfnamefont {M.~A.}\ \bibnamefont
  {Korotin}}, \bibinfo {author} {\bibfnamefont {A.~V.}\ \bibnamefont
  {Lukoyanov}},\ and\ \bibinfo {author} {\bibfnamefont {D.~A.}\ \bibnamefont
  {Khafizullin}},\ }\bibfield  {title} {\bibinfo {title} {{Orbital density
  functional as a means to restore the discontinuities in the total-energy
  derivative and the exchange-correlation potential}},\ }\href
  {https://doi.org/10.1088/0953-8984/19/10/106206} {\bibfield  {journal}
  {\bibinfo  {journal} {Journal of Physics: Condensed Matter}\ }\textbf
  {\bibinfo {volume} {19}},\ \bibinfo {pages} {106206} (\bibinfo {year}
  {2007})}\BibitemShut {NoStop}%
\bibitem [{\citenamefont {Kraisler}\ and\ \citenamefont
  {Kronik}(2013)}]{Kronik2007}%
  \BibitemOpen
  \bibfield  {author} {\bibinfo {author} {\bibfnamefont {E.}~\bibnamefont
  {Kraisler}}\ and\ \bibinfo {author} {\bibfnamefont {L.}~\bibnamefont
  {Kronik}},\ }\bibfield  {title} {\bibinfo {title} {{Piecewise Linearity of
  Approximate Density Functionals Revisited: Implications for Frontier Orbital
  Energies}},\ }\href {https://doi.org/10.1103/PhysRevLett.110.126403}
  {\bibfield  {journal} {\bibinfo  {journal} {Phys. Rev. Lett.}\ }\textbf
  {\bibinfo {volume} {110}},\ \bibinfo {pages} {126403} (\bibinfo {year}
  {2013})}\BibitemShut {NoStop}%
\bibitem [{\citenamefont {Skone}\ \emph {et~al.}(2014)\citenamefont {Skone},
  \citenamefont {Govoni},\ and\ \citenamefont {Galli}}]{Galli2014}%
  \BibitemOpen
  \bibfield  {author} {\bibinfo {author} {\bibfnamefont {J.~H.}\ \bibnamefont
  {Skone}}, \bibinfo {author} {\bibfnamefont {M.}~\bibnamefont {Govoni}},\ and\
  \bibinfo {author} {\bibfnamefont {G.}~\bibnamefont {Galli}},\ }\bibfield
  {title} {\bibinfo {title} {{Self}-consistent hybrid functional for condensed
  systems},\ }\href {https://doi.org/10.1103/PhysRevB.89.195112} {\bibfield
  {journal} {\bibinfo  {journal} {Phys. Rev. B}\ }\textbf {\bibinfo {volume}
  {89}},\ \bibinfo {pages} {195112} (\bibinfo {year} {2014})}\BibitemShut
  {NoStop}%
\bibitem [{\citenamefont {Li}\ \emph {et~al.}(2015)\citenamefont {Li},
  \citenamefont {Zheng}, \citenamefont {Cohen}, \citenamefont
  {Mori-S\'anchez},\ and\ \citenamefont {Yang}}]{Weitao2015}%
  \BibitemOpen
  \bibfield  {author} {\bibinfo {author} {\bibfnamefont {C.}~\bibnamefont
  {Li}}, \bibinfo {author} {\bibfnamefont {X.}~\bibnamefont {Zheng}}, \bibinfo
  {author} {\bibfnamefont {A.~J.}\ \bibnamefont {Cohen}}, \bibinfo {author}
  {\bibfnamefont {P.}~\bibnamefont {Mori-S\'anchez}},\ and\ \bibinfo {author}
  {\bibfnamefont {W.}~\bibnamefont {Yang}},\ }\bibfield  {title} {\bibinfo
  {title} {{Local Scaling Correction for Reducing Delocalization Error in
  Density Functional Approximations}},\ }\href
  {https://doi.org/10.1103/PhysRevLett.114.053001} {\bibfield  {journal}
  {\bibinfo  {journal} {Phys. Rev. Lett.}\ }\textbf {\bibinfo {volume} {114}},\
  \bibinfo {pages} {053001} (\bibinfo {year} {2015})}\BibitemShut {NoStop}%
\bibitem [{\citenamefont {Li}\ \emph {et~al.}(2017)\citenamefont {Li},
  \citenamefont {Zheng}, \citenamefont {Su},\ and\ \citenamefont
  {Yang}}]{Weitao2017}%
  \BibitemOpen
  \bibfield  {author} {\bibinfo {author} {\bibfnamefont {C.}~\bibnamefont
  {Li}}, \bibinfo {author} {\bibfnamefont {X.}~\bibnamefont {Zheng}}, \bibinfo
  {author} {\bibfnamefont {N.~Q.}\ \bibnamefont {Su}},\ and\ \bibinfo {author}
  {\bibfnamefont {W.}~\bibnamefont {Yang}},\ }\bibfield  {title} {\bibinfo
  {title} {{Localized orbital scaling correction for systematic elimination of
  delocalization error in density functional approximations}},\ }\href
  {https://doi.org/10.1093/nsr/nwx111} {\bibfield  {journal} {\bibinfo
  {journal} {National Science Review}\ }\textbf {\bibinfo {volume} {5}},\
  \bibinfo {pages} {203} (\bibinfo {year} {2017})}\BibitemShut {NoStop}%
\bibitem [{\citenamefont {Wing}\ \emph {et~al.}(2021)\citenamefont {Wing},
  \citenamefont {Ohad}, \citenamefont {Haber}, \citenamefont {Filip},
  \citenamefont {Gant}, \citenamefont {Neaton},\ and\ \citenamefont
  {Kronik}}]{Kronik2021}%
  \BibitemOpen
  \bibfield  {author} {\bibinfo {author} {\bibfnamefont {D.}~\bibnamefont
  {Wing}}, \bibinfo {author} {\bibfnamefont {G.}~\bibnamefont {Ohad}}, \bibinfo
  {author} {\bibfnamefont {J.~B.}\ \bibnamefont {Haber}}, \bibinfo {author}
  {\bibfnamefont {M.~R.}\ \bibnamefont {Filip}}, \bibinfo {author}
  {\bibfnamefont {S.~E.}\ \bibnamefont {Gant}}, \bibinfo {author}
  {\bibfnamefont {J.~B.}\ \bibnamefont {Neaton}},\ and\ \bibinfo {author}
  {\bibfnamefont {L.}~\bibnamefont {Kronik}},\ }\bibfield  {title} {\bibinfo
  {title} {{Band} gaps of crystalline solids from {Wannier}-localization-based
  optimal tuning of a screened range-separated hybrid functional},\ }\href
  {https://doi.org/10.1073/pnas.2104556118} {\bibfield  {journal} {\bibinfo
  {journal} {Proceedings of the National Academy of Sciences}\ }\textbf
  {\bibinfo {volume} {118}},\ \bibinfo {pages} {e2104556118} (\bibinfo {year}
  {2021})}\BibitemShut {NoStop}%
\bibitem [{\citenamefont {Ma}\ and\ \citenamefont {Wang}(2016)}]{Ma2016}%
  \BibitemOpen
  \bibfield  {author} {\bibinfo {author} {\bibfnamefont {J.}~\bibnamefont
  {Ma}}\ and\ \bibinfo {author} {\bibfnamefont {L.-W.}\ \bibnamefont {Wang}},\
  }\bibfield  {title} {\bibinfo {title} {{Using Wannier functions to improve
  solid band gap predictions in density functional theory}},\ }\href
  {https://doi.org/10.1038/srep24924} {\bibfield  {journal} {\bibinfo
  {journal} {Scientific Reports}\ }\textbf {\bibinfo {volume} {6}},\ \bibinfo
  {pages} {24924} (\bibinfo {year} {2016})}\BibitemShut {NoStop}%
\bibitem [{\citenamefont {Sharma}\ \emph {et~al.}(2007)\citenamefont {Sharma},
  \citenamefont {Dewhurst}, \citenamefont {Ambrosch-Draxl}, \citenamefont
  {Kurth}, \citenamefont {Helbig}, \citenamefont {Pittalis}, \citenamefont
  {Shallcross}, \citenamefont {Nordstr\"om},\ and\ \citenamefont
  {Gross}}]{gross_prl_2007}%
  \BibitemOpen
  \bibfield  {author} {\bibinfo {author} {\bibfnamefont {S.}~\bibnamefont
  {Sharma}}, \bibinfo {author} {\bibfnamefont {J.~K.}\ \bibnamefont
  {Dewhurst}}, \bibinfo {author} {\bibfnamefont {C.}~\bibnamefont
  {Ambrosch-Draxl}}, \bibinfo {author} {\bibfnamefont {S.}~\bibnamefont
  {Kurth}}, \bibinfo {author} {\bibfnamefont {N.}~\bibnamefont {Helbig}},
  \bibinfo {author} {\bibfnamefont {S.}~\bibnamefont {Pittalis}}, \bibinfo
  {author} {\bibfnamefont {S.}~\bibnamefont {Shallcross}}, \bibinfo {author}
  {\bibfnamefont {L.}~\bibnamefont {Nordstr\"om}},\ and\ \bibinfo {author}
  {\bibfnamefont {E.~K.~U.}\ \bibnamefont {Gross}},\ }\bibfield  {title}
  {\bibinfo {title} {{First-Principles Approach to Noncollinear Magnetism:
  Towards Spin Dynamics}},\ }\href
  {https://doi.org/10.1103/PhysRevLett.98.196405} {\bibfield  {journal}
  {\bibinfo  {journal} {Phys. Rev. Lett.}\ }\textbf {\bibinfo {volume} {98}},\
  \bibinfo {pages} {196405} (\bibinfo {year} {2007})}\BibitemShut {NoStop}%
\bibitem [{SM()}]{SM}%
  \BibitemOpen
  \href@noop {} {}\bibinfo {howpublished} {See the Supplemental Material for
  more details on the mathematical derivation of the second-order non-collinear
  Koopmans functional, the Koopmans-Wannier Hamiltonian and the corrections
  beyond second order, which includes Refs.~\cite{colonna_jctc_2018,
  colonna_jctc_2022}.}\BibitemShut {Stop}%
\bibitem [{\citenamefont {Marzari}\ and\ \citenamefont
  {Vanderbilt}(1997)}]{Marzari1997}%
  \BibitemOpen
  \bibfield  {author} {\bibinfo {author} {\bibfnamefont {N.}~\bibnamefont
  {Marzari}}\ and\ \bibinfo {author} {\bibfnamefont {D.}~\bibnamefont
  {Vanderbilt}},\ }\bibfield  {title} {\bibinfo {title} {{Maximally localized
  generalized Wannier functions for composite energy bands}},\ }\href
  {https://doi.org/10.1103/PhysRevB.56.12847} {\bibfield  {journal} {\bibinfo
  {journal} {Phys. Rev. B}\ }\textbf {\bibinfo {volume} {56}},\ \bibinfo
  {pages} {12847} (\bibinfo {year} {1997})}\BibitemShut {NoStop}%
\bibitem [{\citenamefont {Souza}\ \emph {et~al.}(2001)\citenamefont {Souza},
  \citenamefont {Marzari},\ and\ \citenamefont {Vanderbilt}}]{Souza2001}%
  \BibitemOpen
  \bibfield  {author} {\bibinfo {author} {\bibfnamefont {I.}~\bibnamefont
  {Souza}}, \bibinfo {author} {\bibfnamefont {N.}~\bibnamefont {Marzari}},\
  and\ \bibinfo {author} {\bibfnamefont {D.}~\bibnamefont {Vanderbilt}},\
  }\bibfield  {title} {\bibinfo {title} {{Maximally localized Wannier functions
  for entangled energy bands}},\ }\href
  {https://doi.org/10.1103/PhysRevB.65.035109} {\bibfield  {journal} {\bibinfo
  {journal} {Phys. Rev. B}\ }\textbf {\bibinfo {volume} {65}},\ \bibinfo
  {pages} {035109} (\bibinfo {year} {2001})}\BibitemShut {NoStop}%
\bibitem [{\citenamefont {Binci}\ and\ \citenamefont
  {Marzari}(2023)}]{binci_noncollinear_2023}%
  \BibitemOpen
  \bibfield  {author} {\bibinfo {author} {\bibfnamefont {L.}~\bibnamefont
  {Binci}}\ and\ \bibinfo {author} {\bibfnamefont {N.}~\bibnamefont
  {Marzari}},\ }\bibfield  {title} {\bibinfo {title} {{Noncollinear}
  {DFT}$+${U} and {Hubbard} parameters with fully relativistic ultrasoft
  pseudopotentials},\ }\href {https://doi.org/10.1103/PhysRevB.108.115157}
  {\bibfield  {journal} {\bibinfo  {journal} {Physical Review B}\ }\textbf
  {\bibinfo {volume} {108}},\ \bibinfo {pages} {115157} (\bibinfo {year}
  {2023})}\BibitemShut {NoStop}%
\bibitem [{\citenamefont {Giannozzi}\ \emph {et~al.}(2009)\citenamefont
  {Giannozzi}, \citenamefont {Baroni}, \citenamefont {Bonini}, \citenamefont
  {Calandra}, \citenamefont {Car}, \citenamefont {Cavazzoni}, \citenamefont
  {{Davide Ceresoli}}, \citenamefont {Chiarotti}, \citenamefont {Cococcioni},
  \citenamefont {Dabo}, \citenamefont {Corso}, \citenamefont {Gironcoli},
  \citenamefont {Fabris}, \citenamefont {Fratesi}, \citenamefont {Gebauer},
  \citenamefont {Gerstmann}, \citenamefont {Gougoussis}, \citenamefont {{Anton
  Kokalj}}, \citenamefont {Lazzeri}, \citenamefont {Martin-Samos},
  \citenamefont {Marzari}, \citenamefont {Mauri}, \citenamefont {Mazzarello},
  \citenamefont {{Stefano Paolini}}, \citenamefont {Pasquarello}, \citenamefont
  {Paulatto}, \citenamefont {Sbraccia}, \citenamefont {Scandolo}, \citenamefont
  {Sclauzero}, \citenamefont {Seitsonen}, \citenamefont {Smogunov},
  \citenamefont {Umari},\ and\ \citenamefont
  {Wentzcovitch}}]{giannozzi_qe_2009}%
  \BibitemOpen
  \bibfield  {author} {\bibinfo {author} {\bibfnamefont {P.}~\bibnamefont
  {Giannozzi}}, \bibinfo {author} {\bibfnamefont {S.}~\bibnamefont {Baroni}},
  \bibinfo {author} {\bibfnamefont {N.}~\bibnamefont {Bonini}}, \bibinfo
  {author} {\bibfnamefont {M.}~\bibnamefont {Calandra}}, \bibinfo {author}
  {\bibfnamefont {R.}~\bibnamefont {Car}}, \bibinfo {author} {\bibfnamefont
  {C.}~\bibnamefont {Cavazzoni}}, \bibinfo {author} {\bibnamefont {{Davide
  Ceresoli}}}, \bibinfo {author} {\bibfnamefont {G.~L.}\ \bibnamefont
  {Chiarotti}}, \bibinfo {author} {\bibfnamefont {M.}~\bibnamefont
  {Cococcioni}}, \bibinfo {author} {\bibfnamefont {I.}~\bibnamefont {Dabo}},
  \bibinfo {author} {\bibfnamefont {A.~D.}\ \bibnamefont {Corso}}, \bibinfo
  {author} {\bibfnamefont {S.~d.}\ \bibnamefont {Gironcoli}}, \bibinfo {author}
  {\bibfnamefont {S.}~\bibnamefont {Fabris}}, \bibinfo {author} {\bibfnamefont
  {G.}~\bibnamefont {Fratesi}}, \bibinfo {author} {\bibfnamefont
  {R.}~\bibnamefont {Gebauer}}, \bibinfo {author} {\bibfnamefont
  {U.}~\bibnamefont {Gerstmann}}, \bibinfo {author} {\bibfnamefont
  {C.}~\bibnamefont {Gougoussis}}, \bibinfo {author} {\bibnamefont {{Anton
  Kokalj}}}, \bibinfo {author} {\bibfnamefont {M.}~\bibnamefont {Lazzeri}},
  \bibinfo {author} {\bibfnamefont {L.}~\bibnamefont {Martin-Samos}}, \bibinfo
  {author} {\bibfnamefont {N.}~\bibnamefont {Marzari}}, \bibinfo {author}
  {\bibfnamefont {F.}~\bibnamefont {Mauri}}, \bibinfo {author} {\bibfnamefont
  {R.}~\bibnamefont {Mazzarello}}, \bibinfo {author} {\bibnamefont {{Stefano
  Paolini}}}, \bibinfo {author} {\bibfnamefont {A.}~\bibnamefont
  {Pasquarello}}, \bibinfo {author} {\bibfnamefont {L.}~\bibnamefont
  {Paulatto}}, \bibinfo {author} {\bibfnamefont {C.}~\bibnamefont {Sbraccia}},
  \bibinfo {author} {\bibfnamefont {S.}~\bibnamefont {Scandolo}}, \bibinfo
  {author} {\bibfnamefont {G.}~\bibnamefont {Sclauzero}}, \bibinfo {author}
  {\bibfnamefont {A.~P.}\ \bibnamefont {Seitsonen}}, \bibinfo {author}
  {\bibfnamefont {A.}~\bibnamefont {Smogunov}}, \bibinfo {author}
  {\bibfnamefont {P.}~\bibnamefont {Umari}},\ and\ \bibinfo {author}
  {\bibfnamefont {R.~M.}\ \bibnamefont {Wentzcovitch}},\ }\bibfield  {title}
  {\bibinfo {title} {{QUANTUM} {ESPRESSO}: a modular and open-source software
  project for quantum simulations of materials},\ }\href
  {https://doi.org/10.1088/0953-8984/21/39/395502} {\bibfield  {journal}
  {\bibinfo  {journal} {J. Phys. Condens. Matter}\ }\textbf {\bibinfo {volume}
  {21}},\ \bibinfo {pages} {395502} (\bibinfo {year} {2009})}\BibitemShut
  {NoStop}%
\bibitem [{\citenamefont {Giannozzi}\ \emph {et~al.}(2017)\citenamefont
  {Giannozzi}, \citenamefont {Andreussi}, \citenamefont {Brumme}, \citenamefont
  {Bunau}, \citenamefont {Buongiorno~Nardelli}, \citenamefont {Calandra},
  \citenamefont {Car}, \citenamefont {Cavazzoni}, \citenamefont {Ceresoli},
  \citenamefont {Cococcioni}, \citenamefont {Colonna}, \citenamefont
  {Carnimeo}, \citenamefont {Dal~Corso}, \citenamefont {De~Gironcoli},
  \citenamefont {Delugas}, \citenamefont {Distasio}, \citenamefont {Ferretti},
  \citenamefont {Floris}, \citenamefont {Fratesi}, \citenamefont {Fugallo},
  \citenamefont {Gebauer}, \citenamefont {Gerstmann}, \citenamefont {Giustino},
  \citenamefont {Gorni}, \citenamefont {Jia}, \citenamefont {Kawamura},
  \citenamefont {Ko}, \citenamefont {Kokalj}, \citenamefont
  {K{\"u}c{\"u}kbenli}, \citenamefont {Lazzeri}, \citenamefont {Marsili},
  \citenamefont {Marzari}, \citenamefont {Mauri}, \citenamefont {Nguyen},
  \citenamefont {Nguyen}, \citenamefont {{Otero-De-La-Roza}}, \citenamefont
  {Paulatto}, \citenamefont {Ponc{\'e}}, \citenamefont {Rocca}, \citenamefont
  {Sabatini}, \citenamefont {Santra}, \citenamefont {Schlipf}, \citenamefont
  {Seitsonen}, \citenamefont {Smogunov}, \citenamefont {Timrov}, \citenamefont
  {Thonhauser}, \citenamefont {Umari}, \citenamefont {Vast}, \citenamefont
  {Wu},\ and\ \citenamefont {Baroni}}]{giannozzi_qe_2017}%
  \BibitemOpen
  \bibfield  {author} {\bibinfo {author} {\bibfnamefont {P.}~\bibnamefont
  {Giannozzi}}, \bibinfo {author} {\bibfnamefont {O.}~\bibnamefont
  {Andreussi}}, \bibinfo {author} {\bibfnamefont {T.}~\bibnamefont {Brumme}},
  \bibinfo {author} {\bibfnamefont {O.}~\bibnamefont {Bunau}}, \bibinfo
  {author} {\bibfnamefont {M.}~\bibnamefont {Buongiorno~Nardelli}}, \bibinfo
  {author} {\bibfnamefont {M.}~\bibnamefont {Calandra}}, \bibinfo {author}
  {\bibfnamefont {R.}~\bibnamefont {Car}}, \bibinfo {author} {\bibfnamefont
  {C.}~\bibnamefont {Cavazzoni}}, \bibinfo {author} {\bibfnamefont
  {D.}~\bibnamefont {Ceresoli}}, \bibinfo {author} {\bibfnamefont
  {M.}~\bibnamefont {Cococcioni}}, \bibinfo {author} {\bibfnamefont
  {N.}~\bibnamefont {Colonna}}, \bibinfo {author} {\bibfnamefont
  {I.}~\bibnamefont {Carnimeo}}, \bibinfo {author} {\bibfnamefont
  {A.}~\bibnamefont {Dal~Corso}}, \bibinfo {author} {\bibfnamefont
  {S.}~\bibnamefont {De~Gironcoli}}, \bibinfo {author} {\bibfnamefont
  {P.}~\bibnamefont {Delugas}}, \bibinfo {author} {\bibfnamefont {R.~A.}\
  \bibnamefont {Distasio}}, \bibinfo {author} {\bibfnamefont {A.}~\bibnamefont
  {Ferretti}}, \bibinfo {author} {\bibfnamefont {A.}~\bibnamefont {Floris}},
  \bibinfo {author} {\bibfnamefont {G.}~\bibnamefont {Fratesi}}, \bibinfo
  {author} {\bibfnamefont {G.}~\bibnamefont {Fugallo}}, \bibinfo {author}
  {\bibfnamefont {R.}~\bibnamefont {Gebauer}}, \bibinfo {author} {\bibfnamefont
  {U.}~\bibnamefont {Gerstmann}}, \bibinfo {author} {\bibfnamefont
  {F.}~\bibnamefont {Giustino}}, \bibinfo {author} {\bibfnamefont
  {T.}~\bibnamefont {Gorni}}, \bibinfo {author} {\bibfnamefont
  {J.}~\bibnamefont {Jia}}, \bibinfo {author} {\bibfnamefont {M.}~\bibnamefont
  {Kawamura}}, \bibinfo {author} {\bibfnamefont {H.~Y.}\ \bibnamefont {Ko}},
  \bibinfo {author} {\bibfnamefont {A.}~\bibnamefont {Kokalj}}, \bibinfo
  {author} {\bibfnamefont {E.}~\bibnamefont {K{\"u}c{\"u}kbenli}}, \bibinfo
  {author} {\bibfnamefont {M.}~\bibnamefont {Lazzeri}}, \bibinfo {author}
  {\bibfnamefont {M.}~\bibnamefont {Marsili}}, \bibinfo {author} {\bibfnamefont
  {N.}~\bibnamefont {Marzari}}, \bibinfo {author} {\bibfnamefont
  {F.}~\bibnamefont {Mauri}}, \bibinfo {author} {\bibfnamefont {N.~L.}\
  \bibnamefont {Nguyen}}, \bibinfo {author} {\bibfnamefont {H.~V.}\
  \bibnamefont {Nguyen}}, \bibinfo {author} {\bibfnamefont {A.}~\bibnamefont
  {{Otero-De-La-Roza}}}, \bibinfo {author} {\bibfnamefont {L.}~\bibnamefont
  {Paulatto}}, \bibinfo {author} {\bibfnamefont {S.}~\bibnamefont {Ponc{\'e}}},
  \bibinfo {author} {\bibfnamefont {D.}~\bibnamefont {Rocca}}, \bibinfo
  {author} {\bibfnamefont {R.}~\bibnamefont {Sabatini}}, \bibinfo {author}
  {\bibfnamefont {B.}~\bibnamefont {Santra}}, \bibinfo {author} {\bibfnamefont
  {M.}~\bibnamefont {Schlipf}}, \bibinfo {author} {\bibfnamefont {A.~P.}\
  \bibnamefont {Seitsonen}}, \bibinfo {author} {\bibfnamefont {A.}~\bibnamefont
  {Smogunov}}, \bibinfo {author} {\bibfnamefont {I.}~\bibnamefont {Timrov}},
  \bibinfo {author} {\bibfnamefont {T.}~\bibnamefont {Thonhauser}}, \bibinfo
  {author} {\bibfnamefont {P.}~\bibnamefont {Umari}}, \bibinfo {author}
  {\bibfnamefont {N.}~\bibnamefont {Vast}}, \bibinfo {author} {\bibfnamefont
  {X.}~\bibnamefont {Wu}},\ and\ \bibinfo {author} {\bibfnamefont
  {S.}~\bibnamefont {Baroni}},\ }\bibfield  {title} {\bibinfo {title} {Advanced
  capabilities for materials modelling with {{Quantum ESPRESSO}}},\ }\href
  {https://doi.org/10.1088/1361-648X/aa8f79} {\bibfield  {journal} {\bibinfo
  {journal} {Journal of Physics Condensed Matter}\ }\textbf {\bibinfo {volume}
  {29}},\ \bibinfo {pages} {465901} (\bibinfo {year} {2017})}\BibitemShut
  {NoStop}%
\bibitem [{\citenamefont {Carnimeo}\ \emph {et~al.}(2023)\citenamefont
  {Carnimeo}, \citenamefont {Affinito}, \citenamefont {Baroni}, \citenamefont
  {Baseggio}, \citenamefont {Bellentani}, \citenamefont {Bertossa},
  \citenamefont {Delugas}, \citenamefont {Ruffino}, \citenamefont {Orlandini},
  \citenamefont {Spiga},\ and\ \citenamefont {Giannozzi}}]{carnimeo_qe_2023}%
  \BibitemOpen
  \bibfield  {author} {\bibinfo {author} {\bibfnamefont {I.}~\bibnamefont
  {Carnimeo}}, \bibinfo {author} {\bibfnamefont {F.}~\bibnamefont {Affinito}},
  \bibinfo {author} {\bibfnamefont {S.}~\bibnamefont {Baroni}}, \bibinfo
  {author} {\bibfnamefont {O.}~\bibnamefont {Baseggio}}, \bibinfo {author}
  {\bibfnamefont {L.}~\bibnamefont {Bellentani}}, \bibinfo {author}
  {\bibfnamefont {R.}~\bibnamefont {Bertossa}}, \bibinfo {author}
  {\bibfnamefont {P.~D.}\ \bibnamefont {Delugas}}, \bibinfo {author}
  {\bibfnamefont {F.~F.}\ \bibnamefont {Ruffino}}, \bibinfo {author}
  {\bibfnamefont {S.}~\bibnamefont {Orlandini}}, \bibinfo {author}
  {\bibfnamefont {F.}~\bibnamefont {Spiga}},\ and\ \bibinfo {author}
  {\bibfnamefont {P.}~\bibnamefont {Giannozzi}},\ }\bibfield  {title} {\bibinfo
  {title} {{Quantum ESPRESSO: One Further Step toward the Exascale}},\ }\href
  {https://doi.org/10.1021/acs.jctc.3c00249} {\bibfield  {journal} {\bibinfo
  {journal} {Journal of Chemical Theory and Computation}\ }\textbf {\bibinfo
  {volume} {19}},\ \bibinfo {pages} {6992} (\bibinfo {year}
  {2023})}\BibitemShut {NoStop}%
\bibitem [{\citenamefont {Hamann}(2013)}]{Hamann2013}%
  \BibitemOpen
  \bibfield  {author} {\bibinfo {author} {\bibfnamefont {D.~R.}\ \bibnamefont
  {Hamann}},\ }\bibfield  {title} {\bibinfo {title} {{Optimized norm-conserving
  Vanderbilt pseudopotentials}},\ }\href
  {https://doi.org/0.1103/PhysRevB.88.085117} {\bibfield  {journal} {\bibinfo
  {journal} {Phys. Rev. B}\ }\textbf {\bibinfo {volume} {88}},\ \bibinfo
  {pages} {085117} (\bibinfo {year} {2013})}\BibitemShut {NoStop}%
\bibitem [{\citenamefont {Hamann}(2017)}]{Hamann2017}%
  \BibitemOpen
  \bibfield  {author} {\bibinfo {author} {\bibfnamefont {D.~R.}\ \bibnamefont
  {Hamann}},\ }\bibfield  {title} {\bibinfo {title} {{Erratum: Optimized
  norm-conserving Vanderbilt pseudopotentials [Phys. Rev. B 88, 085117
  (2013)]}},\ }\href {https://doi.org/10.1103/PhysRevB.95.239906} {\bibfield
  {journal} {\bibinfo  {journal} {Phys. Rev. B}\ }\textbf {\bibinfo {volume}
  {95}},\ \bibinfo {pages} {239906(E)} (\bibinfo {year} {2017})}\BibitemShut
  {NoStop}%
\bibitem [{\citenamefont {{van Setten}}\ \emph {et~al.}(2018)\citenamefont
  {{van Setten}}, \citenamefont {Giantomassi}, \citenamefont {Bousquet},
  \citenamefont {Verstraete}, \citenamefont {Hamann}, \citenamefont {Gonze},\
  and\ \citenamefont {Rignanese}}]{vanSetten2018}%
  \BibitemOpen
  \bibfield  {author} {\bibinfo {author} {\bibfnamefont {M.}~\bibnamefont {{van
  Setten}}}, \bibinfo {author} {\bibfnamefont {M.}~\bibnamefont {Giantomassi}},
  \bibinfo {author} {\bibfnamefont {E.}~\bibnamefont {Bousquet}}, \bibinfo
  {author} {\bibfnamefont {M.}~\bibnamefont {Verstraete}}, \bibinfo {author}
  {\bibfnamefont {D.}~\bibnamefont {Hamann}}, \bibinfo {author} {\bibfnamefont
  {X.}~\bibnamefont {Gonze}},\ and\ \bibinfo {author} {\bibfnamefont {G.-M.}\
  \bibnamefont {Rignanese}},\ }\bibfield  {title} {\bibinfo {title} {{The
  PseudoDojo}: Training and grading a 85 element optimized norm-conserving
  pseudopotential table},\ }\href
  {https://doi.org/https://doi.org/10.1016/j.cpc.2018.01.012} {\bibfield
  {journal} {\bibinfo  {journal} {Comput. Phys. Commun.}\ }\textbf {\bibinfo
  {volume} {226}},\ \bibinfo {pages} {39} (\bibinfo {year} {2018})}\BibitemShut
  {NoStop}%
\bibitem [{\citenamefont {Heyd}\ \emph {et~al.}(2003)\citenamefont {Heyd},
  \citenamefont {Scuseria},\ and\ \citenamefont {Ernzerhof}}]{HSE}%
  \BibitemOpen
  \bibfield  {author} {\bibinfo {author} {\bibfnamefont {J.}~\bibnamefont
  {Heyd}}, \bibinfo {author} {\bibfnamefont {G.~E.}\ \bibnamefont {Scuseria}},\
  and\ \bibinfo {author} {\bibfnamefont {M.}~\bibnamefont {Ernzerhof}},\
  }\bibfield  {title} {\bibinfo {title} {Hybrid functionals based on a screened
  coulomb potential},\ }\href {https://doi.org/10.1063/1.1564060} {\bibfield
  {journal} {\bibinfo  {journal} {J. Chem. Phys.}\ }\textbf {\bibinfo {volume}
  {118}},\ \bibinfo {pages} {8207} (\bibinfo {year} {2003})}\BibitemShut
  {NoStop}%
\bibitem [{\citenamefont {Lin}(2016)}]{lin_ace_2016}%
  \BibitemOpen
  \bibfield  {author} {\bibinfo {author} {\bibfnamefont {L.}~\bibnamefont
  {Lin}},\ }\bibfield  {title} {\bibinfo {title} {Adaptively compressed
  exchange operator},\ }\href {https://doi.org/10.1021/acs.jctc.6b0009}
  {\bibfield  {journal} {\bibinfo  {journal} {J. Chem. Theory Comput.}\
  }\textbf {\bibinfo {volume} {12}},\ \bibinfo {pages} {2242} (\bibinfo {year}
  {2016})}\BibitemShut {NoStop}%
\bibitem [{\citenamefont {Marrazzo}\ and\ \citenamefont
  {Colonna}(2024)}]{MC_archive}%
  \BibitemOpen
  \bibfield  {author} {\bibinfo {author} {\bibfnamefont {A.}~\bibnamefont
  {Marrazzo}}\ and\ \bibinfo {author} {\bibfnamefont {N.}~\bibnamefont
  {Colonna}},\ }\bibfield  {title} {\bibinfo {title} {Spin-dependent
  interactions in orbital-density-dependent functionals: non-collinear koopmans
  spectral functionals},\ }\bibfield  {journal} {\bibinfo  {journal} {Materials
  Cloud Archive [will be made available at time of publication]}\ }\href
  {https://doi.org/10.24435/materialscloud:kp-2v}
  {10.24435/materialscloud:kp-2v} (\bibinfo {year} {2024})\BibitemShut
  {NoStop}%
\bibitem [{\citenamefont {Madelung}(2004)}]{semicon_book}%
  \BibitemOpen
  \bibfield  {author} {\bibinfo {author} {\bibfnamefont {O.}~\bibnamefont
  {Madelung}},\ }\href {https://doi.org/10.1007/978-3-642-18865-7} {\emph
  {\bibinfo {title} {Semiconductors}}}\ (\bibinfo  {publisher} {Springer
  Berlin, Heidelberg},\ \bibinfo {year} {2004})\BibitemShut {NoStop}%
\bibitem [{\citenamefont {Shevchik}\ \emph {et~al.}(1974)\citenamefont
  {Shevchik}, \citenamefont {Tejeda},\ and\ \citenamefont
  {Cardona}}]{cardona_prb_1974}%
  \BibitemOpen
  \bibfield  {author} {\bibinfo {author} {\bibfnamefont {N.~J.}\ \bibnamefont
  {Shevchik}}, \bibinfo {author} {\bibfnamefont {J.}~\bibnamefont {Tejeda}},\
  and\ \bibinfo {author} {\bibfnamefont {M.}~\bibnamefont {Cardona}},\
  }\bibfield  {title} {\bibinfo {title} {{Densities of valence states of
  amorphous and crystalline III-V and II-VI semiconductors}},\ }\href
  {https://doi.org/10.1103/PhysRevB.9.2627} {\bibfield  {journal} {\bibinfo
  {journal} {Phys. Rev. B}\ }\textbf {\bibinfo {volume} {9}},\ \bibinfo {pages}
  {2627} (\bibinfo {year} {1974})}\BibitemShut {NoStop}%
\bibitem [{\citenamefont {Riley}\ \emph {et~al.}(2014)\citenamefont {Riley},
  \citenamefont {Mazzola}, \citenamefont {Dendzik}, \citenamefont {Michiardi},
  \citenamefont {Takayama}, \citenamefont {Bawden}, \citenamefont
  {Graner{\o}d}, \citenamefont {Leandersson}, \citenamefont {Balasubramanian},
  \citenamefont {Hoesch}, \citenamefont {Kim}, \citenamefont {Takagi},
  \citenamefont {Meevasana}, \citenamefont {Hofmann}, \citenamefont {Bahramy},
  \citenamefont {Wells},\ and\ \citenamefont {King}}]{riley_natphys_2014}%
  \BibitemOpen
  \bibfield  {author} {\bibinfo {author} {\bibfnamefont {J.~M.}\ \bibnamefont
  {Riley}}, \bibinfo {author} {\bibfnamefont {F.}~\bibnamefont {Mazzola}},
  \bibinfo {author} {\bibfnamefont {M.}~\bibnamefont {Dendzik}}, \bibinfo
  {author} {\bibfnamefont {M.}~\bibnamefont {Michiardi}}, \bibinfo {author}
  {\bibfnamefont {T.}~\bibnamefont {Takayama}}, \bibinfo {author}
  {\bibfnamefont {L.}~\bibnamefont {Bawden}}, \bibinfo {author} {\bibfnamefont
  {C.}~\bibnamefont {Graner{\o}d}}, \bibinfo {author} {\bibfnamefont
  {M.}~\bibnamefont {Leandersson}}, \bibinfo {author} {\bibfnamefont
  {T.}~\bibnamefont {Balasubramanian}}, \bibinfo {author} {\bibfnamefont
  {M.}~\bibnamefont {Hoesch}}, \bibinfo {author} {\bibfnamefont {T.~K.}\
  \bibnamefont {Kim}}, \bibinfo {author} {\bibfnamefont {H.}~\bibnamefont
  {Takagi}}, \bibinfo {author} {\bibfnamefont {W.}~\bibnamefont {Meevasana}},
  \bibinfo {author} {\bibfnamefont {P.}~\bibnamefont {Hofmann}}, \bibinfo
  {author} {\bibfnamefont {M.~S.}\ \bibnamefont {Bahramy}}, \bibinfo {author}
  {\bibfnamefont {J.~W.}\ \bibnamefont {Wells}},\ and\ \bibinfo {author}
  {\bibfnamefont {P.~D.~C.}\ \bibnamefont {King}},\ }\bibfield  {title}
  {\bibinfo {title} {{Direct} observation of spin-polarized bulk bands in an
  inversion-symmetric semiconductor},\ }\href
  {https://doi.org/10.1038/nphys3105} {\bibfield  {journal} {\bibinfo
  {journal} {Nature Physics}\ }\textbf {\bibinfo {volume} {10}},\ \bibinfo
  {pages} {835} (\bibinfo {year} {2014})}\BibitemShut {NoStop}%
\bibitem [{\citenamefont {Manzeli}\ \emph {et~al.}(2017)\citenamefont
  {Manzeli}, \citenamefont {Ovchinnikov}, \citenamefont {Pasquier},
  \citenamefont {Yazyev},\ and\ \citenamefont {Kis}}]{kis_natrevmat_2017}%
  \BibitemOpen
  \bibfield  {author} {\bibinfo {author} {\bibfnamefont {S.}~\bibnamefont
  {Manzeli}}, \bibinfo {author} {\bibfnamefont {D.}~\bibnamefont
  {Ovchinnikov}}, \bibinfo {author} {\bibfnamefont {D.}~\bibnamefont
  {Pasquier}}, \bibinfo {author} {\bibfnamefont {O.~V.}\ \bibnamefont
  {Yazyev}},\ and\ \bibinfo {author} {\bibfnamefont {A.}~\bibnamefont {Kis}},\
  }\bibfield  {title} {\bibinfo {title} {{2D} transition metal
  dichalcogenides},\ }\href {https://doi.org/10.1038/natrevmats.2017.33}
  {\bibfield  {journal} {\bibinfo  {journal} {Nature Reviews Materials}\
  }\textbf {\bibinfo {volume} {2}},\ \bibinfo {pages} {17033} (\bibinfo {year}
  {2017})}\BibitemShut {NoStop}%
\bibitem [{\citenamefont {Zhang}\ \emph {et~al.}(2014)\citenamefont {Zhang},
  \citenamefont {Liu}, \citenamefont {Luo}, \citenamefont {Freeman},\ and\
  \citenamefont {Zunger}}]{zhang_hidden_2014}%
  \BibitemOpen
  \bibfield  {author} {\bibinfo {author} {\bibfnamefont {X.}~\bibnamefont
  {Zhang}}, \bibinfo {author} {\bibfnamefont {Q.}~\bibnamefont {Liu}}, \bibinfo
  {author} {\bibfnamefont {J.-W.}\ \bibnamefont {Luo}}, \bibinfo {author}
  {\bibfnamefont {A.~J.}\ \bibnamefont {Freeman}},\ and\ \bibinfo {author}
  {\bibfnamefont {A.}~\bibnamefont {Zunger}},\ }\bibfield  {title} {\bibinfo
  {title} {{Hidden} spin polarization in inversion-symmetric bulk crystals},\
  }\href {https://doi.org/10.1038/nphys2933} {\bibfield  {journal} {\bibinfo
  {journal} {Nature Physics}\ }\textbf {\bibinfo {volume} {10}},\ \bibinfo
  {pages} {387} (\bibinfo {year} {2014})}\BibitemShut {NoStop}%
\bibitem [{\citenamefont {Schutte}\ \emph {et~al.}(1987)\citenamefont
  {Schutte}, \citenamefont {{De Boer}},\ and\ \citenamefont
  {Jellinek}}]{schutte_jssc_1987}%
  \BibitemOpen
  \bibfield  {author} {\bibinfo {author} {\bibfnamefont {W.}~\bibnamefont
  {Schutte}}, \bibinfo {author} {\bibfnamefont {J.}~\bibnamefont {{De Boer}}},\
  and\ \bibinfo {author} {\bibfnamefont {F.}~\bibnamefont {Jellinek}},\
  }\bibfield  {title} {\bibinfo {title} {{Crystal} structures of tungsten
  disulfide and diselenide},\ }\href
  {https://doi.org/10.1016/0022-4596(87)90057-0} {\bibfield  {journal}
  {\bibinfo  {journal} {Journal of Solid State Chemistry}\ }\textbf {\bibinfo
  {volume} {70}},\ \bibinfo {pages} {207} (\bibinfo {year} {1987})}\BibitemShut
  {NoStop}%
\bibitem [{\citenamefont {Bourezg}\ \emph {et~al.}(1992)\citenamefont
  {Bourezg}, \citenamefont {Couturier}, \citenamefont {Salardenne},
  \citenamefont {Doumerc},\ and\ \citenamefont {L\'evy}}]{bourzeg_prb_1992}%
  \BibitemOpen
  \bibfield  {author} {\bibinfo {author} {\bibfnamefont {R.}~\bibnamefont
  {Bourezg}}, \bibinfo {author} {\bibfnamefont {G.}~\bibnamefont {Couturier}},
  \bibinfo {author} {\bibfnamefont {J.}~\bibnamefont {Salardenne}}, \bibinfo
  {author} {\bibfnamefont {J.~P.}\ \bibnamefont {Doumerc}},\ and\ \bibinfo
  {author} {\bibfnamefont {F.}~\bibnamefont {L\'evy}},\ }\bibfield  {title}
  {\bibinfo {title} {{Interface} of n-type {WSe}$_{2}$ photoanodes in aqueous
  solution. {II}. {Photoelectrochemical} properties},\ }\href
  {https://doi.org/10.1103/PhysRevB.46.15411} {\bibfield  {journal} {\bibinfo
  {journal} {Phys. Rev. B}\ }\textbf {\bibinfo {volume} {46}},\ \bibinfo
  {pages} {15411} (\bibinfo {year} {1992})}\BibitemShut {NoStop}%
\bibitem [{\citenamefont {Kam}\ \emph {et~al.}(1984)\citenamefont {Kam},
  \citenamefont {Chang},\ and\ \citenamefont {Lynch}}]{kam_jpcssp_1984}%
  \BibitemOpen
  \bibfield  {author} {\bibinfo {author} {\bibfnamefont {K.-K.}\ \bibnamefont
  {Kam}}, \bibinfo {author} {\bibfnamefont {C.-L.}\ \bibnamefont {Chang}},\
  and\ \bibinfo {author} {\bibfnamefont {D.~W.}\ \bibnamefont {Lynch}},\
  }\bibfield  {title} {\bibinfo {title} {{Fundamental} absorption edges and
  indirect band gaps in {W}$_{1-x}${Mo}$_x${Se}$_2$ (o$\leq${x}$\geq$1)},\
  }\href {https://doi.org/10.1088/0022-3719/17/22/021} {\bibfield  {journal}
  {\bibinfo  {journal} {Journal of Physics C: Solid State Physics}\ }\textbf
  {\bibinfo {volume} {17}},\ \bibinfo {pages} {4031} (\bibinfo {year}
  {1984})}\BibitemShut {NoStop}%
\bibitem [{\citenamefont {Traving}\ \emph {et~al.}(1997)\citenamefont
  {Traving}, \citenamefont {Boehme}, \citenamefont {Kipp}, \citenamefont
  {Skibowski}, \citenamefont {Starrost}, \citenamefont {Krasovskii},
  \citenamefont {Perlov},\ and\ \citenamefont {Schattke}}]{traving_prb_1997}%
  \BibitemOpen
  \bibfield  {author} {\bibinfo {author} {\bibfnamefont {M.}~\bibnamefont
  {Traving}}, \bibinfo {author} {\bibfnamefont {M.}~\bibnamefont {Boehme}},
  \bibinfo {author} {\bibfnamefont {L.}~\bibnamefont {Kipp}}, \bibinfo {author}
  {\bibfnamefont {M.}~\bibnamefont {Skibowski}}, \bibinfo {author}
  {\bibfnamefont {F.}~\bibnamefont {Starrost}}, \bibinfo {author}
  {\bibfnamefont {E.~E.}\ \bibnamefont {Krasovskii}}, \bibinfo {author}
  {\bibfnamefont {A.}~\bibnamefont {Perlov}},\ and\ \bibinfo {author}
  {\bibfnamefont {W.}~\bibnamefont {Schattke}},\ }\bibfield  {title} {\bibinfo
  {title} {{Electronic structure} of {WSe}$_{2}$: A combined photoemission and
  inverse photoemission study},\ }\href
  {https://doi.org/10.1103/PhysRevB.55.10392} {\bibfield  {journal} {\bibinfo
  {journal} {Phys. Rev. B}\ }\textbf {\bibinfo {volume} {55}},\ \bibinfo
  {pages} {10392} (\bibinfo {year} {1997})}\BibitemShut {NoStop}%
\bibitem [{\citenamefont {Finteis}\ \emph {et~al.}(1997)\citenamefont
  {Finteis}, \citenamefont {Hengsberger}, \citenamefont {Straub}, \citenamefont
  {Fauth}, \citenamefont {Claessen}, \citenamefont {Auer}, \citenamefont
  {Steiner}, \citenamefont {H\"ufner}, \citenamefont {Blaha}, \citenamefont
  {V\"ogt}, \citenamefont {Lux-Steiner},\ and\ \citenamefont
  {Bucher}}]{finteis_prb_1997}%
  \BibitemOpen
  \bibfield  {author} {\bibinfo {author} {\bibfnamefont {T.}~\bibnamefont
  {Finteis}}, \bibinfo {author} {\bibfnamefont {M.}~\bibnamefont
  {Hengsberger}}, \bibinfo {author} {\bibfnamefont {T.}~\bibnamefont {Straub}},
  \bibinfo {author} {\bibfnamefont {K.}~\bibnamefont {Fauth}}, \bibinfo
  {author} {\bibfnamefont {R.}~\bibnamefont {Claessen}}, \bibinfo {author}
  {\bibfnamefont {P.}~\bibnamefont {Auer}}, \bibinfo {author} {\bibfnamefont
  {P.}~\bibnamefont {Steiner}}, \bibinfo {author} {\bibfnamefont
  {S.}~\bibnamefont {H\"ufner}}, \bibinfo {author} {\bibfnamefont
  {P.}~\bibnamefont {Blaha}}, \bibinfo {author} {\bibfnamefont
  {M.}~\bibnamefont {V\"ogt}}, \bibinfo {author} {\bibfnamefont
  {M.}~\bibnamefont {Lux-Steiner}},\ and\ \bibinfo {author} {\bibfnamefont
  {E.}~\bibnamefont {Bucher}},\ }\bibfield  {title} {\bibinfo {title}
  {{Occupied} and unoccupied electronic band structure of {WSe}$_2$},\ }\href
  {https://doi.org/10.1103/PhysRevB.55.10400} {\bibfield  {journal} {\bibinfo
  {journal} {Phys. Rev. B}\ }\textbf {\bibinfo {volume} {55}},\ \bibinfo
  {pages} {10400} (\bibinfo {year} {1997})}\BibitemShut {NoStop}%
\bibitem [{\citenamefont {Jiang}(2012)}]{jiang_JPCC_2012}%
  \BibitemOpen
  \bibfield  {author} {\bibinfo {author} {\bibfnamefont {H.}~\bibnamefont
  {Jiang}},\ }\bibfield  {title} {\bibinfo {title} {{Electronic} {Band}
  {Structures} of {Molybdenum} and {Tungsten} {Dichalcogenides} by the {GW}
  {Approach}},\ }\href {https://doi.org/10.1021/jp300079d} {\bibfield
  {journal} {\bibinfo  {journal} {The Journal of Physical Chemistry C}\
  }\textbf {\bibinfo {volume} {116}},\ \bibinfo {pages} {7664} (\bibinfo {year}
  {2012})}\BibitemShut {NoStop}%
\bibitem [{\citenamefont {Perdew}\ \emph {et~al.}(1996)\citenamefont {Perdew},
  \citenamefont {Burke},\ and\ \citenamefont {Ernzerhof}}]{perdew_prl_1996}%
  \BibitemOpen
  \bibfield  {author} {\bibinfo {author} {\bibfnamefont {J.~P.}\ \bibnamefont
  {Perdew}}, \bibinfo {author} {\bibfnamefont {K.}~\bibnamefont {Burke}},\ and\
  \bibinfo {author} {\bibfnamefont {M.}~\bibnamefont {Ernzerhof}},\ }\bibfield
  {title} {\bibinfo {title} {{Generalized Gradient Approximation Made
  Simple}},\ }\href {https://doi.org/10.1103/PhysRevLett.77.3865} {\bibfield
  {journal} {\bibinfo  {journal} {Phys. Rev. Lett.}\ }\textbf {\bibinfo
  {volume} {77}},\ \bibinfo {pages} {3865} (\bibinfo {year}
  {1996})}\BibitemShut {NoStop}%
\bibitem [{\citenamefont {Stoumpos}\ \emph {et~al.}(2013)\citenamefont
  {Stoumpos}, \citenamefont {Malliakas}, \citenamefont {Peters}, \citenamefont
  {Liu}, \citenamefont {Sebastian}, \citenamefont {Im}, \citenamefont
  {Chasapis}, \citenamefont {Wibowo}, \citenamefont {Chung}, \citenamefont
  {Freeman}, \citenamefont {Wessels},\ and\ \citenamefont
  {Kanatzidis}}]{stoumpos_crystal_2013}%
  \BibitemOpen
  \bibfield  {author} {\bibinfo {author} {\bibfnamefont {C.~C.}\ \bibnamefont
  {Stoumpos}}, \bibinfo {author} {\bibfnamefont {C.~D.}\ \bibnamefont
  {Malliakas}}, \bibinfo {author} {\bibfnamefont {J.~A.}\ \bibnamefont
  {Peters}}, \bibinfo {author} {\bibfnamefont {Z.}~\bibnamefont {Liu}},
  \bibinfo {author} {\bibfnamefont {M.}~\bibnamefont {Sebastian}}, \bibinfo
  {author} {\bibfnamefont {J.}~\bibnamefont {Im}}, \bibinfo {author}
  {\bibfnamefont {T.~C.}\ \bibnamefont {Chasapis}}, \bibinfo {author}
  {\bibfnamefont {A.~C.}\ \bibnamefont {Wibowo}}, \bibinfo {author}
  {\bibfnamefont {D.~Y.}\ \bibnamefont {Chung}}, \bibinfo {author}
  {\bibfnamefont {A.~J.}\ \bibnamefont {Freeman}}, \bibinfo {author}
  {\bibfnamefont {B.~W.}\ \bibnamefont {Wessels}},\ and\ \bibinfo {author}
  {\bibfnamefont {M.~G.}\ \bibnamefont {Kanatzidis}},\ }\bibfield  {title}
  {\bibinfo {title} {{Crystal} {Growth} of the {Perovskite} {Semiconductor}
  {CsPbBr$_3$}: {A} {New} {Material} for {High}-{Energy} {Radiation}
  {Detection}},\ }\href {https://doi.org/10.1021/cg400645t} {\bibfield
  {journal} {\bibinfo  {journal} {Crystal Growth \& Design}\ }\textbf {\bibinfo
  {volume} {13}},\ \bibinfo {pages} {2722} (\bibinfo {year}
  {2013})}\BibitemShut {NoStop}%
\bibitem [{\citenamefont {Hoffman}\ \emph {et~al.}(2016)\citenamefont
  {Hoffman}, \citenamefont {Schleper},\ and\ \citenamefont
  {Kamat}}]{hoffman_transformation_2016}%
  \BibitemOpen
  \bibfield  {author} {\bibinfo {author} {\bibfnamefont {J.~B.}\ \bibnamefont
  {Hoffman}}, \bibinfo {author} {\bibfnamefont {A.~L.}\ \bibnamefont
  {Schleper}},\ and\ \bibinfo {author} {\bibfnamefont {P.~V.}\ \bibnamefont
  {Kamat}},\ }\bibfield  {title} {\bibinfo {title} {{Transformation} of
  {Sintered} {CsPbBr$_3$} {Nanocrystals} to {Cubic} {CsPbI$_3$} and {Gradient}
  {CsPbBr$_x$I$_{3-x}$} through {Halide} {Exchange}},\ }\href
  {https://doi.org/10.1021/jacs.6b04661} {\bibfield  {journal} {\bibinfo
  {journal} {Journal of the American Chemical Society}\ }\textbf {\bibinfo
  {volume} {138}},\ \bibinfo {pages} {8603} (\bibinfo {year}
  {2016})}\BibitemShut {NoStop}%
\bibitem [{\citenamefont {{McGuire}}\ \emph {et~al.}()\citenamefont
  {{McGuire}}, \citenamefont {Dixit}, \citenamefont {Cooper},\ and\
  \citenamefont {Sales}}]{mcguire_coupling_2015}%
  \BibitemOpen
  \bibfield  {author} {\bibinfo {author} {\bibfnamefont {M.~A.}\ \bibnamefont
  {{McGuire}}}, \bibinfo {author} {\bibfnamefont {H.}~\bibnamefont {Dixit}},
  \bibinfo {author} {\bibfnamefont {V.~R.}\ \bibnamefont {Cooper}},\ and\
  \bibinfo {author} {\bibfnamefont {B.~C.}\ \bibnamefont {Sales}},\ }\bibfield
  {title} {\bibinfo {title} {Coupling of crystal structure and magnetism in the
  layered, ferromagnetic insulator {CrI}3},\ }\href
  {https://doi.org/10.1021/cm504242t} {\bibfield  {journal} {\bibinfo
  {journal} {Chemistry of Materials}\ }\textbf {\bibinfo {volume} {27}},\
  \bibinfo {pages} {612}},\ \bibinfo {note} {publisher: American Chemical
  Society}\BibitemShut {NoStop}%
\bibitem [{\citenamefont {Acharya}\ \emph {et~al.}(2021)\citenamefont
  {Acharya}, \citenamefont {Pashov}, \citenamefont {Cunningham}, \citenamefont
  {Rudenko}, \citenamefont {R\"osner}, \citenamefont {Gr\"uning}, \citenamefont
  {van Schilfgaarde},\ and\ \citenamefont {Katsnelson}}]{acharya_prb_2021}%
  \BibitemOpen
  \bibfield  {author} {\bibinfo {author} {\bibfnamefont {S.}~\bibnamefont
  {Acharya}}, \bibinfo {author} {\bibfnamefont {D.}~\bibnamefont {Pashov}},
  \bibinfo {author} {\bibfnamefont {B.}~\bibnamefont {Cunningham}}, \bibinfo
  {author} {\bibfnamefont {A.~N.}\ \bibnamefont {Rudenko}}, \bibinfo {author}
  {\bibfnamefont {M.}~\bibnamefont {R\"osner}}, \bibinfo {author}
  {\bibfnamefont {M.}~\bibnamefont {Gr\"uning}}, \bibinfo {author}
  {\bibfnamefont {M.}~\bibnamefont {van Schilfgaarde}},\ and\ \bibinfo {author}
  {\bibfnamefont {M.~I.}\ \bibnamefont {Katsnelson}},\ }\bibfield  {title}
  {\bibinfo {title} {Electronic structure of chromium trihalides beyond density
  functional theory},\ }\href {https://doi.org/10.1103/PhysRevB.104.155109}
  {\bibfield  {journal} {\bibinfo  {journal} {Phys. Rev. B}\ }\textbf {\bibinfo
  {volume} {104}},\ \bibinfo {pages} {155109} (\bibinfo {year}
  {2021})}\BibitemShut {NoStop}%
\bibitem [{\citenamefont {Acharya}\ \emph {et~al.}(2022)\citenamefont
  {Acharya}, \citenamefont {Pashov}, \citenamefont {Rudenko}, \citenamefont
  {R\"{o}sner}, \citenamefont {Schilfgaarde},\ and\ \citenamefont
  {Katsnelson}}]{acharya_npj2D_2022}%
  \BibitemOpen
  \bibfield  {author} {\bibinfo {author} {\bibfnamefont {S.}~\bibnamefont
  {Acharya}}, \bibinfo {author} {\bibfnamefont {D.}~\bibnamefont {Pashov}},
  \bibinfo {author} {\bibfnamefont {A.~N.}\ \bibnamefont {Rudenko}}, \bibinfo
  {author} {\bibfnamefont {M.}~\bibnamefont {R\"{o}sner}}, \bibinfo {author}
  {\bibfnamefont {M.~v.}\ \bibnamefont {Schilfgaarde}},\ and\ \bibinfo {author}
  {\bibfnamefont {M.~I.}\ \bibnamefont {Katsnelson}},\ }\bibfield  {title}
  {\bibinfo {title} {Real- and momentum-space description of the excitons in
  bulk and monolayer chromium tri-halides},\ }\bibfield  {journal} {\bibinfo
  {journal} {npj 2D Materials and Applications}\ }\textbf {\bibinfo {volume}
  {6}},\ \href {https://doi.org/10.1038/s41699-022-00307-7}
  {10.1038/s41699-022-00307-7} (\bibinfo {year} {2022})\BibitemShut {NoStop}%
\bibitem [{\citenamefont {Dillon}\ and\ \citenamefont
  {Olson}(1965)}]{dillon_jap_1965}%
  \BibitemOpen
  \bibfield  {author} {\bibinfo {author} {\bibfnamefont {J.}~\bibnamefont
  {Dillon}, \bibfnamefont {J.~F.}}\ and\ \bibinfo {author} {\bibfnamefont
  {C.~E.}\ \bibnamefont {Olson}},\ }\bibfield  {title} {\bibinfo {title}
  {{Magnetization, Resonance, and Optical Properties of the Ferromagnet
  CrI$_3$}},\ }\href {https://doi.org/10.1063/1.1714194} {\bibfield  {journal}
  {\bibinfo  {journal} {Journal of Applied Physics}\ }\textbf {\bibinfo
  {volume} {36}},\ \bibinfo {pages} {1259} (\bibinfo {year}
  {1965})}\BibitemShut {NoStop}%
\bibitem [{\citenamefont {Kutepov}(2021)}]{kutepov_prm_2021}%
  \BibitemOpen
  \bibfield  {author} {\bibinfo {author} {\bibfnamefont {A.~L.}\ \bibnamefont
  {Kutepov}},\ }\bibfield  {title} {\bibinfo {title} {Electronic structure of
  van der {Waals} ferromagnet {CrI}$_{3}$ from self-consistent vertex corrected
  ${GW}$ approaches},\ }\href
  {https://doi.org/10.1103/PhysRevMaterials.5.083805} {\bibfield  {journal}
  {\bibinfo  {journal} {Phys. Rev. Mater.}\ }\textbf {\bibinfo {volume} {5}},\
  \bibinfo {pages} {083805} (\bibinfo {year} {2021})}\BibitemShut {NoStop}%
\bibitem [{\citenamefont {Lee}\ \emph {et~al.}(2020)\citenamefont {Lee},
  \citenamefont {Kotani},\ and\ \citenamefont {Ke}}]{lee_prb_2020}%
  \BibitemOpen
  \bibfield  {author} {\bibinfo {author} {\bibfnamefont {Y.}~\bibnamefont
  {Lee}}, \bibinfo {author} {\bibfnamefont {T.}~\bibnamefont {Kotani}},\ and\
  \bibinfo {author} {\bibfnamefont {L.}~\bibnamefont {Ke}},\ }\bibfield
  {title} {\bibinfo {title} {{Role of nonlocality in exchange correlation for
  magnetic two-dimensional van der Waals materials}},\ }\href
  {https://doi.org/10.1103/PhysRevB.101.241409} {\bibfield  {journal} {\bibinfo
   {journal} {Phys. Rev. B}\ }\textbf {\bibinfo {volume} {101}},\ \bibinfo
  {pages} {241409(R)} (\bibinfo {year} {2020})}\BibitemShut {NoStop}%
\bibitem [{\citenamefont {Deguchi}\ \emph {et~al.}(2016)\citenamefont
  {Deguchi}, \citenamefont {Sato}, \citenamefont {Kino},\ and\ \citenamefont
  {Kotani}}]{deguchi_JJAP_2016}%
  \BibitemOpen
  \bibfield  {author} {\bibinfo {author} {\bibfnamefont {D.}~\bibnamefont
  {Deguchi}}, \bibinfo {author} {\bibfnamefont {K.}~\bibnamefont {Sato}},
  \bibinfo {author} {\bibfnamefont {H.}~\bibnamefont {Kino}},\ and\ \bibinfo
  {author} {\bibfnamefont {T.}~\bibnamefont {Kotani}},\ }\bibfield  {title}
  {\bibinfo {title} {Accurate energy bands calculated by the hybrid
  quasiparticle self-consistent {GW} method implemented in the {ECALJ}
  package},\ }\href {https://doi.org/10.7567/jjap.55.051201} {\bibfield
  {journal} {\bibinfo  {journal} {Japanese Journal of Applied Physics}\
  }\textbf {\bibinfo {volume} {55}},\ \bibinfo {pages} {051201} (\bibinfo
  {year} {2016})}\BibitemShut {NoStop}%
\bibitem [{\citenamefont {Chantis}\ \emph {et~al.}(2006)\citenamefont
  {Chantis}, \citenamefont {van Schilfgaarde},\ and\ \citenamefont
  {Kotani}}]{chantis_prl_2006}%
  \BibitemOpen
  \bibfield  {author} {\bibinfo {author} {\bibfnamefont {A.~N.}\ \bibnamefont
  {Chantis}}, \bibinfo {author} {\bibfnamefont {M.}~\bibnamefont {van
  Schilfgaarde}},\ and\ \bibinfo {author} {\bibfnamefont {T.}~\bibnamefont
  {Kotani}},\ }\bibfield  {title} {\bibinfo {title} {{Ab Initio Prediction of
  Conduction Band Spin Splitting in Zinc Blende Semiconductors}},\ }\href
  {https://doi.org/10.1103/PhysRevLett.96.086405} {\bibfield  {journal}
  {\bibinfo  {journal} {Phys. Rev. Lett.}\ }\textbf {\bibinfo {volume} {96}},\
  \bibinfo {pages} {086405} (\bibinfo {year} {2006})}\BibitemShut {NoStop}%
\bibitem [{\citenamefont {Martin}\ \emph {et~al.}(2016)\citenamefont {Martin},
  \citenamefont {Reining},\ and\ \citenamefont {Ceperley}}]{Martin2016}%
  \BibitemOpen
  \bibfield  {author} {\bibinfo {author} {\bibfnamefont {R.~M.}\ \bibnamefont
  {Martin}}, \bibinfo {author} {\bibfnamefont {L.}~\bibnamefont {Reining}},\
  and\ \bibinfo {author} {\bibfnamefont {D.~M.}\ \bibnamefont {Ceperley}},\
  }\href {https://doi.org/10.1017/cbo9781139050807} {\emph {\bibinfo {title}
  {Interacting Electrons: Theory and Computational Approaches}}}\ (\bibinfo
  {publisher} {Cambridge University Press},\ \bibinfo {year}
  {2016})\BibitemShut {NoStop}%
\bibitem [{\citenamefont {Giuliani}\ and\ \citenamefont
  {Vignale}(2005)}]{Vignale_2005}%
  \BibitemOpen
  \bibfield  {author} {\bibinfo {author} {\bibfnamefont {G.}~\bibnamefont
  {Giuliani}}\ and\ \bibinfo {author} {\bibfnamefont {G.}~\bibnamefont
  {Vignale}},\ }\href@noop {} {\emph {\bibinfo {title} {Quantum Theory of the
  Electron Liquid}}}\ (\bibinfo  {publisher} {Cambridge University Press},\
  \bibinfo {year} {2005})\BibitemShut {NoStop}%
\bibitem [{\citenamefont {Del~Sole}\ \emph {et~al.}(1994)\citenamefont
  {Del~Sole}, \citenamefont {Reining},\ and\ \citenamefont
  {Godby}}]{del_sole_gwg_1994}%
  \BibitemOpen
  \bibfield  {author} {\bibinfo {author} {\bibfnamefont {R.}~\bibnamefont
  {Del~Sole}}, \bibinfo {author} {\bibfnamefont {L.}~\bibnamefont {Reining}},\
  and\ \bibinfo {author} {\bibfnamefont {R.~W.}\ \bibnamefont {Godby}},\
  }\bibfield  {title} {\bibinfo {title} {{GW$\Gamma$} approximation for
  electron self-energies in semiconductors and insulators},\ }\href
  {https://doi.org/10.1103/PhysRevB.49.8024} {\bibfield  {journal} {\bibinfo
  {journal} {Physical Review B}\ }\textbf {\bibinfo {volume} {49}},\ \bibinfo
  {pages} {8024} (\bibinfo {year} {1994})}\BibitemShut {NoStop}%
\bibitem [{\citenamefont {Hybertsen}\ and\ \citenamefont
  {Louie}(1987)}]{hybertsen_ab_1987}%
  \BibitemOpen
  \bibfield  {author} {\bibinfo {author} {\bibfnamefont {M.~S.}\ \bibnamefont
  {Hybertsen}}\ and\ \bibinfo {author} {\bibfnamefont {S.~G.}\ \bibnamefont
  {Louie}},\ }\bibfield  {title} {\bibinfo {title} {{Ab initio static
  dielectric matrices from the density-functional approach. {I}. {Formulation}
  and application to semiconductors and insulators}},\ }\href
  {https://doi.org/10.1103/PhysRevB.35.5585} {\bibfield  {journal} {\bibinfo
  {journal} {Physical Review B}\ }\textbf {\bibinfo {volume} {35}},\ \bibinfo
  {pages} {5585} (\bibinfo {year} {1987})}\BibitemShut {NoStop}%
\bibitem [{\citenamefont {Bruneval}\ \emph {et~al.}(2005)\citenamefont
  {Bruneval}, \citenamefont {Sottile}, \citenamefont {Olevano}, \citenamefont
  {Del~Sole},\ and\ \citenamefont {Reining}}]{bruneval_many-body_2005}%
  \BibitemOpen
  \bibfield  {author} {\bibinfo {author} {\bibfnamefont {F.}~\bibnamefont
  {Bruneval}}, \bibinfo {author} {\bibfnamefont {F.}~\bibnamefont {Sottile}},
  \bibinfo {author} {\bibfnamefont {V.}~\bibnamefont {Olevano}}, \bibinfo
  {author} {\bibfnamefont {R.}~\bibnamefont {Del~Sole}},\ and\ \bibinfo
  {author} {\bibfnamefont {L.}~\bibnamefont {Reining}},\ }\bibfield  {title}
  {\bibinfo {title} {{Many}-{Body} {Perturbation} {Theory} {Using} the
  {Density}-{Functional} {Concept}: {Beyond} the ${GW}$ {Approximation}},\
  }\href {https://doi.org/10.1103/PhysRevLett.94.186402} {\bibfield  {journal}
  {\bibinfo  {journal} {Physical Review Letters}\ }\textbf {\bibinfo {volume}
  {94}},\ \bibinfo {pages} {186402} (\bibinfo {year} {2005})}\BibitemShut
  {NoStop}%
\bibitem [{\citenamefont {Aryasetiawan}\ \emph {et~al.}(2012)\citenamefont
  {Aryasetiawan}, \citenamefont {Sakuma},\ and\ \citenamefont
  {Karlsson}}]{self-screen_prb_2012}%
  \BibitemOpen
  \bibfield  {author} {\bibinfo {author} {\bibfnamefont {F.}~\bibnamefont
  {Aryasetiawan}}, \bibinfo {author} {\bibfnamefont {R.}~\bibnamefont
  {Sakuma}},\ and\ \bibinfo {author} {\bibfnamefont {K.}~\bibnamefont
  {Karlsson}},\ }\bibfield  {title} {\bibinfo {title} {${GW}$ approximation
  with self-screening correction},\ }\href
  {https://doi.org/10.1103/PhysRevB.85.035106} {\bibfield  {journal} {\bibinfo
  {journal} {Phys. Rev. B}\ }\textbf {\bibinfo {volume} {85}},\ \bibinfo
  {pages} {035106} (\bibinfo {year} {2012})}\BibitemShut {NoStop}%
\bibitem [{\citenamefont {Christiansson}\ and\ \citenamefont
  {Aryasetiawan}(2023)}]{self-screen_prb_2023}%
  \BibitemOpen
  \bibfield  {author} {\bibinfo {author} {\bibfnamefont {V.}~\bibnamefont
  {Christiansson}}\ and\ \bibinfo {author} {\bibfnamefont {F.}~\bibnamefont
  {Aryasetiawan}},\ }\bibfield  {title} {\bibinfo {title} {Self-screening
  corrections beyond the random-phase approximation: Applications to band gaps
  of semiconductors},\ }\href {https://doi.org/10.1103/PhysRevB.107.125105}
  {\bibfield  {journal} {\bibinfo  {journal} {Phys. Rev. B}\ }\textbf {\bibinfo
  {volume} {107}},\ \bibinfo {pages} {125105} (\bibinfo {year}
  {2023})}\BibitemShut {NoStop}%
\bibitem [{\citenamefont {Sj\"ostedt}\ and\ \citenamefont
  {Nordstr\"om}(2002)}]{nordstrom_prb_2002}%
  \BibitemOpen
  \bibfield  {author} {\bibinfo {author} {\bibfnamefont {E.}~\bibnamefont
  {Sj\"ostedt}}\ and\ \bibinfo {author} {\bibfnamefont {L.}~\bibnamefont
  {Nordstr\"om}},\ }\bibfield  {title} {\bibinfo {title} {{Noncollinear}
  full-potential studies of $\ensuremath{\gamma}\ensuremath{-}\mathrm{Fe}$},\
  }\href {https://doi.org/10.1103/PhysRevB.66.014447} {\bibfield  {journal}
  {\bibinfo  {journal} {Phys. Rev. B}\ }\textbf {\bibinfo {volume} {66}},\
  \bibinfo {pages} {014447} (\bibinfo {year} {2002})}\BibitemShut {NoStop}%
\bibitem [{\citenamefont {Ullrich}(2018)}]{ullrich_prb_2018}%
  \BibitemOpen
  \bibfield  {author} {\bibinfo {author} {\bibfnamefont {C.~A.}\ \bibnamefont
  {Ullrich}},\ }\bibfield  {title} {\bibinfo {title} {{Density}-functional
  theory for systems with noncollinear spin: {Orbital}-dependent
  exchange-correlation functionals and their application to the hubbard
  dimer},\ }\href {https://doi.org/10.1103/PhysRevB.98.035140} {\bibfield
  {journal} {\bibinfo  {journal} {Phys. Rev. B}\ }\textbf {\bibinfo {volume}
  {98}},\ \bibinfo {pages} {035140} (\bibinfo {year} {2018})}\BibitemShut
  {NoStop}%
\bibitem [{\citenamefont {Sharma}\ \emph {et~al.}(2018)\citenamefont {Sharma},
  \citenamefont {Gross}, \citenamefont {Sanna},\ and\ \citenamefont
  {Dewhurst}}]{sharma_jctc_2018}%
  \BibitemOpen
  \bibfield  {author} {\bibinfo {author} {\bibfnamefont {S.}~\bibnamefont
  {Sharma}}, \bibinfo {author} {\bibfnamefont {E.~K.~U.}\ \bibnamefont
  {Gross}}, \bibinfo {author} {\bibfnamefont {A.}~\bibnamefont {Sanna}},\ and\
  \bibinfo {author} {\bibfnamefont {J.~K.}\ \bibnamefont {Dewhurst}},\
  }\bibfield  {title} {\bibinfo {title} {Source-free exchange-correlation
  magnetic fields in density functional theory},\ }\href
  {https://doi.org/10.1021/acs.jctc.7b01049} {\bibfield  {journal} {\bibinfo
  {journal} {Journal of Chemical Theory and Computation}\ }\textbf {\bibinfo
  {volume} {14}},\ \bibinfo {pages} {1247} (\bibinfo {year}
  {2018})}\BibitemShut {NoStop}%
\bibitem [{\citenamefont {Dewhurst}\ \emph {et~al.}(2018)\citenamefont
  {Dewhurst}, \citenamefont {Sanna},\ and\ \citenamefont
  {Sharma}}]{dewhurst_eplb_2018}%
  \BibitemOpen
  \bibfield  {author} {\bibinfo {author} {\bibfnamefont {J.~K.}\ \bibnamefont
  {Dewhurst}}, \bibinfo {author} {\bibfnamefont {A.}~\bibnamefont {Sanna}},\
  and\ \bibinfo {author} {\bibfnamefont {S.}~\bibnamefont {Sharma}},\
  }\bibfield  {title} {\bibinfo {title} {Effect of exchange-correlation
  spin--torque on spin dynamics},\ }\href
  {https://doi.org/10.1140/epjb/e2018-90146-1} {\bibfield  {journal} {\bibinfo
  {journal} {The European Physical Journal B}\ }\textbf {\bibinfo {volume}
  {91}},\ \bibinfo {pages} {218} (\bibinfo {year} {2018})}\BibitemShut
  {NoStop}%
\bibitem [{\citenamefont {Perdew}\ and\ \citenamefont
  {Zunger}(1981)}]{perdew_self-interaction_1981}%
  \BibitemOpen
  \bibfield  {author} {\bibinfo {author} {\bibfnamefont {J.~P.}\ \bibnamefont
  {Perdew}}\ and\ \bibinfo {author} {\bibfnamefont {A.}~\bibnamefont
  {Zunger}},\ }\bibfield  {title} {\bibinfo {title} {{Self-interaction
  correction to density-functional approximations for many-electron systems}},\
  }\href {https://doi.org/10.1103/PhysRevB.23.5048} {\bibfield  {journal}
  {\bibinfo  {journal} {Physical Review B}\ }\textbf {\bibinfo {volume} {23}},\
  \bibinfo {pages} {5048} (\bibinfo {year} {1981})}\BibitemShut {NoStop}%
\bibitem [{\citenamefont {Tancogne-Dejean}\ \emph {et~al.}(2023)\citenamefont
  {Tancogne-Dejean}, \citenamefont {Lüders},\ and\ \citenamefont
  {Ullrich}}]{tancogne-dejean_self-interaction_2023}%
  \BibitemOpen
  \bibfield  {author} {\bibinfo {author} {\bibfnamefont {N.}~\bibnamefont
  {Tancogne-Dejean}}, \bibinfo {author} {\bibfnamefont {M.}~\bibnamefont
  {Lüders}},\ and\ \bibinfo {author} {\bibfnamefont {C.~A.}\ \bibnamefont
  {Ullrich}},\ }\bibfield  {title} {\bibinfo {title} {{Self-interaction
  correction schemes for non-collinear spin-density-functional theory}},\
  }\href {https://doi.org/10.1063/5.0179087} {\bibfield  {journal} {\bibinfo
  {journal} {The Journal of Chemical Physics}\ }\textbf {\bibinfo {volume}
  {159}},\ \bibinfo {pages} {224110} (\bibinfo {year} {2023})}\BibitemShut
  {NoStop}%
\bibitem [{\citenamefont {Vitale}\ \emph {et~al.}(2020)\citenamefont {Vitale},
  \citenamefont {Pizzi}, \citenamefont {Marrazzo}, \citenamefont {Yates},
  \citenamefont {Marzari},\ and\ \citenamefont {Mostofi}}]{Vitale2020}%
  \BibitemOpen
  \bibfield  {author} {\bibinfo {author} {\bibfnamefont {V.}~\bibnamefont
  {Vitale}}, \bibinfo {author} {\bibfnamefont {G.}~\bibnamefont {Pizzi}},
  \bibinfo {author} {\bibfnamefont {A.}~\bibnamefont {Marrazzo}}, \bibinfo
  {author} {\bibfnamefont {J.~R.}\ \bibnamefont {Yates}}, \bibinfo {author}
  {\bibfnamefont {N.}~\bibnamefont {Marzari}},\ and\ \bibinfo {author}
  {\bibfnamefont {A.~A.}\ \bibnamefont {Mostofi}},\ }\bibfield  {title}
  {\bibinfo {title} {{Automated} high-throughput {Wannierisation}},\ }\href
  {https://doi.org/10.1038/s41524-020-0312-y} {\bibfield  {journal} {\bibinfo
  {journal} {npj Comput. Mater.}\ }\textbf {\bibinfo {volume} {6}},\ \bibinfo
  {pages} {1} (\bibinfo {year} {2020})}\BibitemShut {NoStop}%
\bibitem [{\citenamefont {Qiao}\ \emph
  {et~al.}(2023{\natexlab{a}})\citenamefont {Qiao}, \citenamefont {Pizzi},\
  and\ \citenamefont {Marzari}}]{Qiao2023}%
  \BibitemOpen
  \bibfield  {author} {\bibinfo {author} {\bibfnamefont {J.}~\bibnamefont
  {Qiao}}, \bibinfo {author} {\bibfnamefont {G.}~\bibnamefont {Pizzi}},\ and\
  \bibinfo {author} {\bibfnamefont {N.}~\bibnamefont {Marzari}},\ }\bibfield
  {title} {\bibinfo {title} {{Projectability disentanglement for accurate and
  automated electronic-structure Hamiltonians}},\ }\href
  {https://doi.org/10.1038/s41524-023-01146-w} {\bibfield  {journal} {\bibinfo
  {journal} {npj Comput. Mater.}\ }\textbf {\bibinfo {volume} {9}},\ \bibinfo
  {pages} {208} (\bibinfo {year} {2023}{\natexlab{a}})}\BibitemShut {NoStop}%
\bibitem [{\citenamefont {Qiao}\ \emph
  {et~al.}(2023{\natexlab{b}})\citenamefont {Qiao}, \citenamefont {Pizzi},\
  and\ \citenamefont {Marzari}}]{Qiao2023a}%
  \BibitemOpen
  \bibfield  {author} {\bibinfo {author} {\bibfnamefont {J.}~\bibnamefont
  {Qiao}}, \bibinfo {author} {\bibfnamefont {G.}~\bibnamefont {Pizzi}},\ and\
  \bibinfo {author} {\bibfnamefont {N.}~\bibnamefont {Marzari}},\ }\bibfield
  {title} {\bibinfo {title} {{Automated mixing of maximally localized Wannier
  functions into target manifolds}},\ }\href
  {https://doi.org/10.1038/s41524-023-01147-9} {\bibfield  {journal} {\bibinfo
  {journal} {npj Comput. Mater.}\ }\textbf {\bibinfo {volume} {9}},\ \bibinfo
  {pages} {206} (\bibinfo {year} {2023}{\natexlab{b}})}\BibitemShut {NoStop}%
\bibitem [{\citenamefont {Monserrat}(2016)}]{Bart_PRB_2016}%
  \BibitemOpen
  \bibfield  {author} {\bibinfo {author} {\bibfnamefont {B.}~\bibnamefont
  {Monserrat}},\ }\bibfield  {title} {\bibinfo {title} {Vibrational averages
  along thermal lines},\ }\href {https://doi.org/10.1103/PhysRevB.93.014302}
  {\bibfield  {journal} {\bibinfo  {journal} {Phys. Rev. B}\ }\textbf {\bibinfo
  {volume} {93}},\ \bibinfo {pages} {014302} (\bibinfo {year}
  {2016})}\BibitemShut {NoStop}%
\bibitem [{\citenamefont {Zacharias}\ and\ \citenamefont
  {Giustino}(2020)}]{ZG_PRR_2022}%
  \BibitemOpen
  \bibfield  {author} {\bibinfo {author} {\bibfnamefont {M.}~\bibnamefont
  {Zacharias}}\ and\ \bibinfo {author} {\bibfnamefont {F.}~\bibnamefont
  {Giustino}},\ }\bibfield  {title} {\bibinfo {title} {Theory of the special
  displacement method for electronic structure calculations at finite
  temperature},\ }\href {https://doi.org/10.1103/PhysRevResearch.2.013357}
  {\bibfield  {journal} {\bibinfo  {journal} {Phys. Rev. Res.}\ }\textbf
  {\bibinfo {volume} {2}},\ \bibinfo {pages} {013357} (\bibinfo {year}
  {2020})}\BibitemShut {NoStop}%
\bibitem [{\citenamefont {Szilva}\ \emph
  {et~al.}(2023{\natexlab{b}})\citenamefont {Szilva}, \citenamefont {Kvashnin},
  \citenamefont {Stepanov}, \citenamefont {Nordstr\"om}, \citenamefont
  {Eriksson}, \citenamefont {Lichtenstein},\ and\ \citenamefont
  {Katsnelson}}]{rmp_attila_2023}%
  \BibitemOpen
  \bibfield  {author} {\bibinfo {author} {\bibfnamefont {A.}~\bibnamefont
  {Szilva}}, \bibinfo {author} {\bibfnamefont {Y.}~\bibnamefont {Kvashnin}},
  \bibinfo {author} {\bibfnamefont {E.~A.}\ \bibnamefont {Stepanov}}, \bibinfo
  {author} {\bibfnamefont {L.}~\bibnamefont {Nordstr\"om}}, \bibinfo {author}
  {\bibfnamefont {O.}~\bibnamefont {Eriksson}}, \bibinfo {author}
  {\bibfnamefont {A.~I.}\ \bibnamefont {Lichtenstein}},\ and\ \bibinfo {author}
  {\bibfnamefont {M.~I.}\ \bibnamefont {Katsnelson}},\ }\bibfield  {title}
  {\bibinfo {title} {{Quantitative} theory of magnetic interactions in
  solids},\ }\href {https://doi.org/10.1103/RevModPhys.95.035004} {\bibfield
  {journal} {\bibinfo  {journal} {Rev. Mod. Phys.}\ }\textbf {\bibinfo {volume}
  {95}},\ \bibinfo {pages} {035004} (\bibinfo {year}
  {2023}{\natexlab{b}})}\BibitemShut {NoStop}%
\end{thebibliography}%

\newpage
\onecolumngrid
\section*{Supplementary Material}

\renewcommand{\figurename}{Supplementary Fig.}
\renewcommand{\tablename}{Supplementary Tab.}
\renewcommand\thefigure{\arabic{figure}}
\renewcommand\thetable{\arabic{table}}

\renewcommand\thesection{S. \Alph{section}}
\renewcommand\thesubsection{S. \arabic{subsection}}

\renewcommand\theequation{S. \arabic{equation}}

\setcounter{equation}{0}
\setcounter{figure}{0}
\setcounter{table}{0}
\setcounter{subsection}{0}
\setcounter{section}{0}

\section{The 2$^{nd}$ order expansion of non-collinear Koopmans-compliant functionals}
In the following we provide more details about the derivation of the 2$^{nd}$ order expansion for the Koopmans-complaint functional in the non-collinear framework that is discussed in the main text.
\subsection{Derivatives with respect to occupations}
 We start by evaluating Eq.~(3) of the main text, first we consider the scalar term:
\begin{eqnarray}
  \braket{\psi_i| \frac{dV_{\rm Hxc}}{df_i}|\psi_i} &=& \sum_{\sigma,\sigma'}\int d\mbf{r}d\mbf{r}' \, \psi_i^*(\mbf{r},\sigma)\frac{dV_{\rm Hxc}(\mbf{r})}{df_i}\delta_{\sigma,\sigma'}\delta(\mbf{r}-\mbf{r}')\psi_i(\mbf{r}',\sigma')\\
  &=& \sum_{\sigma}\int d\mbf{r} \, n_{i, \rho}^{\sigma}(\mbf{r})\frac{dV_{\rm Hxc}(\mbf{r})}{df_i}\\
  &=& \int d\mbf{r} \, n_{i, \rho}(\mbf{r})\int d\mbf{r}' \, \left(\frac{\delta V_{\rm Hxc}(\mbf{r})}{\delta \rho(\mbf{r}')}\frac{d \rho(\mbf{r}')}{d f_i} + \sum_{\alpha}\frac{\delta V_{\rm Hxc}(\mbf{r})}{\delta m_\alpha(\mbf{r}')}\frac{d m_{\alpha}(\mbf{r}')}{d f_i}\right)\\
  &=& \int d\mbf{r}d\mbf{r}' \, n_{i, \rho}(\mbf{r})\left(F_{\rm Hxc}^{\rho,\rho}(\mbf{r},\mbf{r}')\frac{d \rho(\mbf{r}')}{d f_i} + \sum_{\alpha}F_{\rm xc}^{\rho,m_{\alpha}}(\mbf{r},\mbf{r}')\frac{d m_{\alpha}(\mbf{r'})}{d f_i}\right),\\
\end{eqnarray}
where we used completeness and the chain rule for functional derivatives. The quantity $F_{\rm Hxc}$ represents the Hartree and exchange-correlation kernel. Now we consider the non-collinear exchange-correlation terms:
\begin{eqnarray}
  \braket{\psi_i|\frac{d{W}_{xc,\alpha}}{df_i}\sigma_{\alpha}|\psi_i} &=& \sum_{\sigma,\sigma'}\int d\mbf{r}d\mbf{r}' \, \psi_i^*(\mbf{r},\sigma)\frac{dW_{xc,\alpha}(\mbf{r})}{df_i}\sigma_{\alpha}\delta(\mbf{r}-\mbf{r}')\psi_i(\mbf{r}',\sigma')\\
    &=& \int d\mbf{r} \, n_{i,m_\alpha}(\mbf{r})\frac{dW_{xc,\alpha}(\mbf{r})}{df_i}\\
    &=& \int d\mbf{r} \, n_{i,m_\alpha}(\mbf{r})\int d\mbf{r}'\left(\frac{\delta W_{xc,\alpha}(\mbf{r})}{\delta \rho(\mbf{r}')}\frac{d \rho(\mbf{r}')}{d f_i} + \sum_{\beta}\frac{\delta W_{xc,\alpha}(\mbf{r})}{\delta m_\beta(\mbf{r}')}\frac{d m_{\beta}(\mbf{r'})}{d f_i}\right)\\
    &=& \int d\mbf{r}d\mbf{r}' \, n_{i,m_\alpha}(\mbf{r})\left(F_{\rm xc}^{m_{\alpha},\rho}(\mbf{r},\mbf{r}')\frac{d \rho(\mbf{r})}{d f_i} + \sum_{\beta}F_{\rm xc}^{m_{\alpha},m_{\beta}}(\mbf{r},\mbf{r}')\frac{d m_{\beta}(\mbf{r'})}{d f_i}\right).
\end{eqnarray}
Now, we sum the two terms and obtain:
\begin{eqnarray}
\braket{\psi_i| \frac{dV_{\rm Hxc}}{df_i}+\sum_{\alpha}\frac{d{W}_{xc,\alpha}}{df_i}\sigma_{\alpha}|\psi_i} &=& \int d\mbf{r}d\mbf{r}' \left[n_{i, \rho}(\mbf{r})\left(F_{\rm xc}^{\rho,\rho}(\mbf{r},\mbf{r}')\frac{d \rho(\mbf{r})}{d f_i} + \sum_{\beta}F_{\rm xc}^{\rho,m_{\beta}}(\mbf{r},\mbf{r}')\frac{d m_{\beta}(\mbf{r'})}{d f_i}\right) \right.+ \\
&+& \left. n_{i,m_\alpha}(\mbf{r})\left(F_{\rm xc}^{m_{\alpha},\rho}(\mbf{r},\mbf{r}')\frac{d \rho(\mbf{r})}{d f_i} + \sum_{\beta}F_{\rm xc}^{m_{\alpha},m_{\beta}}(\mbf{r},\mbf{r}')\frac{d m_{\beta}(\mbf{r'})}{d f_i}\right)\right].
\end{eqnarray}
We leverage the symmetry between charge and magnetization in the equations above, and introduce a more compact notation:
\begin{eqnarray}
\bm{\rho}(\mbf{r}) &=& \begin{pmatrix}\rho(\mbf{r}),m_x(\mbf{r}),m_y(\mbf{r}),m_z(\mbf{r})\end{pmatrix}\\
\bm{n}_i(\mbf{r}) &=& \begin{pmatrix}n_{i, \rho}(\mbf{r}),n_{i, m_x}(\mbf{r}),n_{i,m_y}(\mbf{r}),n_{i,m_z}(\mbf{r})\end{pmatrix}\\
\tilde{\bm{\sigma}} &=& \begin{pmatrix}\sigma_0,\sigma_x,\sigma_y,\sigma_z\end{pmatrix} \\
\mathbf{V}_{\rm KS}(\mbf{r}) &=& \begin{pmatrix}V_{\rm KS}(\mbf{r}),V_{{\rm KS},x}(\mbf{r}),V_{{\rm KS},y}(\mbf{r}),V_{{\rm KS},z}(\mbf{r})\end{pmatrix}\\
\mathbf{V}_{\rm Hxc}(\mbf{r}) &=& \begin{pmatrix}V_{\rm Hxc}(\mbf{r}),W_{xc,x}(\mbf{r}),W_{xc,y}(\mbf{r}),W_{xc,z}(\mbf{r})\end{pmatrix}\\
\boldsymbol{F}_{\rm Hxc}(\mbf{r},\mbf{r}') &=& \begin{pmatrix}F_{\rm Hxc}^{\rho,\rho}& F_{\rm xc}^{\rho,m_x}& F_{\rm xc}^{\rho,m_y} &F_{\rm xc}^{\rho,m_z}\\
  F_{\rm xc}^{m_x,\rho}& F_{\rm xc}^{m_x,m_x}& F_{\rm xc}^{m_x,m_y}& F_{\rm xc}^{m_x,m_z}\\
  F_{\rm xc}^{m_y,\rho}& F_{\rm xc}^{m_y,m_x}& F_{\rm xc}^{m_y,m_y}& F_{\rm xc}^{m_y,m_z}\\
  F_{\rm xc}^{m_z,\rho}& F_{\rm xc}^{m_z,m_x}& F_{\rm xc}^{m_z,m_y}& F_{\rm xc}^{m_z,m_z}
\end{pmatrix}.
\end{eqnarray}
With this compact notation we can rewrite everything as matrix-vector product and obtain Eq.(6) of the main text:
\begin{equation}
  \label{compact_2nd_order_deriv_NC}
  \braket{\psi_i| \frac{d\mbf{V}_{\rm Hxc}}{df_i}\cdot\tilde{\bm{\sigma}} |\psi_i}=\int d\mbf{r}d\mbf{r}' \bm{n}_i(\mbf{r}) \mbf{F}_{\rm Hxc}(\mbf{r},\mbf{r}') \frac{d \bm{\rho}(\mbf{r}')}{d f_i}
\end{equation}
which make transparent that the NC case can be recast in the same form of a collinear problem for four-vector densities and a four-by-four matrix for exchange-correlation kernel. 
\subsection{Derivatives with respect to occupations}
Now we derive a non-collinear Dyson equation for the four-density, the first step is to explicitly calculate the total derivative
\begin{eqnarray}
\frac{d \bm{\rho}(\mbf{r})}{d f_i} &=& \bm{n}_i(\mbf{r}) + \int d\mbf{r}' \sum_j f_j\sum_{\alpha}\frac{\delta \bm{n}_j(\mbf{r})}{\delta V_{{\rm KS},\alpha}(\mbf{r}')}\frac{d V_{{\rm KS},\alpha}(\mbf{r}')}{df_i}\\
&=&\bm{n}_i(\mbf{r}) + \int d\mbf{r}' \sum_{\alpha}\frac{\delta \bm{\rho}(\mbf{r})}{\delta V_{{\rm {\rm KS}},\alpha}(\mbf{r}')}\frac{d V_{ {\rm KS},\alpha}(\mbf{r}')}{df_i}\\
&=&\bm{n}_i(\mbf{r}) + \int d\mbf{r}' \bm{\chi}_0(\mbf{r},\mbf{r}')\frac{d \mbf{V}_{\rm Hxc}(\mbf{r}')}{df_i},
\end{eqnarray}
where we used that only the Hartree and exchange correlation part of the total Kohn-Sham potential depend on occupancies, and introduced the response function as a four-by-four matrix
\begin{equation}
  \bm{\chi}_0(\mbf{r},\mbf{r}')= \begin{pmatrix}\chi_0^{\rho,\rho}& \chi_0^{\rho,m_x}& \chi_0^{\rho,m_y} &\chi_0^{\rho,m_z}\\
    \chi_0^{m_x,\rho}& \chi_0^{m_x,m_x}& \chi_0^{m_x,m_y}& \chi_0^{m_x,m_z}\\
    \chi_0^{m_y,\rho}& \chi_0^{m_y,m_x}& \chi_0^{m_y,m_y}& \chi_0^{m_y,m_z}\\
    \chi_0^{m_z,\rho}& \chi_0^{m_z,m_x}& \chi_0^{m_z,m_y}& \chi_0^{m_z,m_z}\end{pmatrix}.
\end{equation}
We use the chain rule applied in Eq.~(\ref{compact_2nd_order_deriv_NC}) and insert it into the equation above to obtain a Dyson equation:

\begin{equation}
  \label{eq:Dyson_init}
  \frac{d \bm{\rho}(\mbf{r})}{d f_i} = \bm{n}_i(\mbf{r}) + \int d\mbf{r}' \bm{\chi}_0(\mbf{r},\mbf{r}')\int d\mbf{r''}\mbf{F}_{\rm Hxc}(\mbf{r}',\mbf{r}'') \frac{d \bm{\rho}(\mbf{r}'')}{d f_i},
\end{equation}
Its iterative solution can be recast in a compact form by introducing the interacting response function 

\begin{equation}
  \bm{\chi}(\mbf{r},\mbf{r}') = \bm{\chi}_0(\mbf{r},\mbf{r}') + \int d\mbf{r''} \bm{\chi}_0(\mbf{r},\mbf{r}'')  \int d\mbf{r'''} \mbf{F}_{\rm Hxc}(\mbf{r}'',\mbf{r}''') \ \bm{\chi}(\mbf{r}''',\mbf{r}').
\end{equation}
which allows to write Eq.~(\ref{eq:Dyson_init}) as in Eq.(7) of the main text:

\begin{equation}
 \frac{d \bm{\rho}(\mbf{r})}{d f_i} = \bm{n}_i(\mbf{r}) + \int d\mbf{r}' \bm{\chi}(\mbf{r},\mbf{r}')\int d\mbf{r''}\mbf{F}_{\rm Hxc}(\mbf{r}',\mbf{r}'') \bm{n}_i(\mbf{r}'').
\end{equation}
Finally, we obtain the second-order non-collinear Koopmans-compliant functional:

\begin{eqnarray}
  \Pi_i^{(2)rKI}  &=& \frac{1}{2}f_i(1-f_i) \int d\mbf{r}d\mbf{r}' \bm{n}_i(\mbf{r}) \mbf{F}_{\rm Hxc}(\mbf{r},\mbf{r}') \left(\bm{n}_i(\mbf{r}') + \int d\mbf{r}'' \bm{\chi}(\mbf{r}',\mbf{r}'')\int d\mbf{r}'''\mbf{F}_{\rm Hxc}(\mbf{r}'',\mbf{r}''') \bm{n}_i(\mbf{r}''')\right)\\
  &=& \frac{1}{2}f_i(1-f_i) \int d\mbf{r}d\mbf{r}' \bm{n}_i(\mbf{r}) \mathbb{F}_{\rm Hxc}(\mbf{r},\mbf{r}') \bm{n}_i(\mbf{r}')
\end{eqnarray}
where we defined the screened Hartree and exchange-correlation kernel as in Eq.(10) of the main text:
\begin{equation}
 \mathbb{F}_{\rm Hxc}(\mbf{r},\mbf{r}')  =  \left(\mbf{F}_{\rm Hxc}(\mbf{r},\mbf{r}')+ \int d\mbf{r}'' \mbf{F}_{\rm Hxc}(\mbf{r},\mbf{r}'')\int d\mbf{r}''' \bm{\chi}(\mbf{r}'',\mbf{r}''')\mbf{F}_{\rm Hxc}(\mbf{r}''',\mbf{r}')\right).
\end{equation}

\section{Non-collinear Koopmans-Wannier Hamiltonian}
As discussed in the main text, in the following we use Wannier functions as a proxy for localized variational orbitals and calculate the corresponding matrix elements for the Koopmans-Wannier Hamiltonian:

\begin{eqnarray}
  \Delta H_{ij}^{KI(2)}(\mbf{R}) &=&  \sum_{\alpha}\int d\mbf{r}\, \braket{w_i^*(\mbf{r}+\mbf{R})|\sigma_\alpha|w_j(\mbf{r})} \mathcal{V}^{KI(2)}_{\mbf{0}j,\alpha}[\boldsymbol{\rho}_{\mbf{0}j}](\mbf{r})\\
  &=& \sum_{\alpha}\int d\mbf{r} \, n_{ij}^{\alpha}(\mbf{R}) \mathcal{V}^{KI(2)}_{\mbf{0}j,\alpha}[\boldsymbol{\rho}_{\mbf{0}j}](\mbf{r})\\
  &=& \int d\mbf{r} \,\mbf{n}_{ij}(\mbf{R}) \boldsymbol{\mathcal{V}}^{KI(2)}_{\mbf{0}j}[\boldsymbol{\rho}_{\mbf{0}j}](\mbf{r}),
 \end{eqnarray}
where the Koopmans potential is defined in Eq.~(11) of the main text and expanded on Pauli matrices, so we define
\begin{equation}
  \mbf{n}_{ij}(\mbf{R}) = \frac{1}{N_{\mbf{k}}} \sum_{\mbf{k}}e^{i\mbf{k}\cdot\mbf{R}}\frac{1}{N_{\mbf{q}}}\sum_{\mbf{q}}e^{i\mbf{q}\cdot\mbf{r}}\bm{n}_{\mbf{k},\mbf{k}+\mbf{q}}^{ij},
\end{equation}
and
\begin{equation}
  \bm{n}^{ij}_{\mbf{k},\mbf{k}+\mbf{q}}(\mbf{r}) = (n^{ij,\rho}_{\mbf{k},\mbf{k}+\mbf{q}}(\mbf{r}), n^{ij,m_x}_{\mbf{k},\mbf{k}+\mbf{q}}(\mbf{r}),n^{ij,m_y}_{\mbf{k},\mbf{k}+\mbf{q}}(\mbf{r}),n^{ij,m_z}_{\mbf{k},\mbf{k}+\mbf{q}}(\mbf{r})),
\end{equation}
with
\begin{equation}
  n^{ij,\alpha}_{\mbf{k},\mbf{k}+\mbf{q}}(\mbf{r}) = \braket{u^{\rm W}_{i,\mbf{k}}(\mbf{r})|\sigma_{\alpha}|u^{\rm W}_{j,\mbf{k}+\mbf{q}}(\mbf{r})},
\end{equation}
where $u^{W}$ are the periodic part of Bloch states in the Wanier gauge. Now let us define the perturbing potential $\mbf{V}_{pert}$, which we expand in monochromatic perturbations
\begin{equation}
  \mbf{V}_{pert,\mbf{q}}^{\mbf{0}j} = \int d\mbf{r}' \left[\boldsymbol{F}^{\mbf{q}}_{\rm Hxc}(\mbf{r},\mbf{r}') \bm{n}_{i,\mbf{q}}^{\mbf{0}j}(\mbf{r}')\right].
  \label{eq:kcw_pert}
\end{equation}
That allows us to rewrite the KI potential as a purely scalar term plus a $\mbf{r}-$dependent contribution, which can be written as a sum of monochromatic perturbations
\begin{equation}
  \mathcal{V}^{KI(2)}_{i}(\mbf{r}) = -\frac{1}{2}\Delta_{\mbf{0}\mbf{j}}^{KI(2)} + (1-f_i)\sum_{\mbf{q}}e^{i\mbf{q}\cdot\mbf{r}} \mbf{V}_{pert,\mbf{q}}^{\mbf{0}j}\cdot\bm{\tilde{\sigma}}.
\end{equation}
The first term reads:
\begin{eqnarray}
  \Delta_{\mbf{0}\mbf{j}}^{KI(2)} &=& \int d\mbf{r}d\mbf{r}' \bm{n}_{\mbf{0}j}(\mbf{r}) \boldsymbol{F}_{\rm Hxc}(\mbf{r},\mbf{r}') \bm{n}_{\mbf{0}j}(\mbf{r}')\\
  &=& \frac{1}{N_{\mbf{q}}}\sum_{\mbf{q}}\int d\mbf{r} \bm{n}_{\mbf{0}j,\mbf{q}}(\mbf{r})  \mbf{V}_{pert,\mbf{q}}^{\mbf{0}j}.
\end{eqnarray}
So the matrix elements of the Koopmans-Wannier Hamiltonian are:
\begin{equation}
  \Delta H_{ij}^{KI(2)}(\mbf{R}) = -\frac{1}{2}\Delta_{\mbf{0}\mbf{j}}^{KI(2)} \delta_{\mbf{R},\mbf{0}}\delta_{i,j}+ \Delta H_{ij,\mbf{r}}^{KI(2)}(\mbf{R}),
\end{equation}
where the second term comes from the $\mbf{r}-$dependent part of the potential and reads:
\begin{eqnarray}
  \Delta H_{ij,\mbf{r}}^{KI(2)}(\mbf{R}) &=& (1-f_i)\int d\mbf{r}\sum_{\mbf{q}}e^{i\mbf{q}\cdot\mbf{r}} \mbf{V}_{pert,\mbf{q}}^{\mbf{0}j} \frac{1}{N_{\mbf{k}}} \sum_{\mbf{k}}e^{i\mbf{k}\cdot\mbf{R}}\frac{1}{N_{\mbf{q}'}}\sum_{\mbf{q}'}e^{i\mbf{q}'\cdot\mbf{r}}\bm{n}_{\mbf{k},\mbf{k}+\mbf{q}'}^{ij}\\
  &=&(1-f_i)\frac{1}{N_{\mbf{k}}}\sum_{\mbf{k}} e^{i\mbf{k}\cdot\mbf{R}}\frac{1}{N_{\mbf{q}}^2}\sum_{\mbf{q}\mbf{q}'}\int d\mbf{r} e^{i(\mbf{q}+\mbf{q}')\cdot\mbf{r}} \mbf{V}_{pert,\mbf{q}}^{\mbf{0}j}\cdot\bm{n}_{\mbf{k},\mbf{k}+\mbf{q}'}^{ij}\\
  &=&(1-f_i)\frac{1}{N_{\mbf{k}}}\sum_{\mbf{k}} e^{i\mbf{k}\cdot\mbf{R}}\frac{1}{N_{\mbf{q}}^2}\sum_{\mbf{q}\mbf{q}'}\sum_{\mbf{G}\mbf{G}'}\int d\mbf{r} e^{i(\mbf{q}+\mbf{q}')\cdot\mbf{r}} e^{i(\mbf{G}+\mbf{G}')\cdot\mbf{r}}\mbf{V}_{pert,\mbf{q}}^{\mbf{0}j}(\mbf{G})\cdot\bm{n}_{\mbf{k},\mbf{k}+\mbf{q}'}^{ij}(\mbf{G}')\\
&=&(1-f_i)\frac{1}{N_{\mbf{k}}}\sum_{\mbf{k}} e^{i\mbf{k}\cdot\mbf{R}}\frac{1}{N_{\mbf{q}}}\sum_{\mbf{q}}\sum_{\mbf{G}}\mbf{V}_{pert,\mbf{q}}^{\mbf{0}j}(\mbf{G})\bm{n}_{\mbf{k},\mbf{k}-\mbf{q}}^{ij}(-\mbf{G})\\
  &=&(1-f_i)\frac{1}{N_{\mbf{k}}}\sum_{\mbf{k}} e^{i\mbf{k}\cdot\mbf{R}}\frac{1}{N_{\mbf{q}}}\sum_{\mbf{q}}\sum_{\mbf{G}}\mbf{V}_{pert,\mbf{q}}^{\mbf{0}j}(\mbf{G})[\bm{n}_{\mbf{k}-\mbf{q},\mbf{k}}^{ji}(\mbf{G})]^*.
\end{eqnarray}
In the last line we use the following:
\begin{align}
\bm{n}_{\mbf{k},\mbf{k-q}}^{ij}(-\mbf{G})
& = \int d\mathbf{r} \, \bm{n}_{\mbf{k},\mbf{k-q}}^{ij}(\mbf{r}) e^{-i\mbf{G}\cdot \mbf{r}}  \\
& = \left\{ \int d\mathbf{r} \, [\bm{n}_{\mbf{k},\mbf{k-q}}^{ij}(\mbf{r})]^* e^{i\mbf{G}\cdot \mbf{r}} \right\} ^* \\
& = \left\{ \int d\mathbf{r} \, \langle {u}^{\rm W}_{\mbf{k-q},j}(\mbf{r}) | \boldsymbol{\tilde{\sigma}} | {u}^{\rm W}_{\mbf{k},i}(\mbf{r}) \rangle  e^{i\mbf{G}\cdot \mbf{r}} \right\} ^* \\
& = [\bm{n}_{\mbf{k}-\mbf{q},\mbf{k}}^{ji}(\mbf{G})]^*
\end{align}
with $\boldsymbol{\tilde{\sigma}}$ being the four Pauli matrices.

\section{Corrections beyond the second-order approximation: detailed derivation}
We give here additional details about the correction beyond the second order expansion introduced in Sec. III D of the main text. 
Without loss of generality and to keep the notation simple we neglect the spin degrees of freedom here. The final and general result reported in the main text can be obtained by upgrading the single-component objects (orbitals, orbital densities, charge densities) of the following derivation to the corresponding four-components and the scalar Hxc kernel $F_{\rm Hxc}(\mathbf{r},\mathbf{r}')$ to a four-by-four matrix (as discussed in the main text). 
As discussed in the main text, the renormalization of the screening coefficient is given in terms of the correction $\Delta^{\rm u}_i =  \Pi_i^{\rm uKI} -  \Pi_i^{\rm (2)uKI}$:
\begin{equation}
    \Delta \alpha_i = \frac{\Delta_i^{\rm u}}{\Pi_i^{\rm (2)uKI}}
\end{equation}
This term depends on the occupation of the orbital at hand and it is ill defined at integer occupations as both the numerator and the denominator are exactly zero (this is because both $\Pi_i^{\rm uKI}$ and $\Pi_i^{\rm (2)uKI}$ are exactly zero in this limit). For the case of occupied states, we avoid this ambiguity by evaluating this term in the limit of the occupation tending to one from the left (for empty states the derivation requires to take the limit for the occupation tending to zero from the right): 
\begin{equation}
    \Delta \alpha_i = \lim_{\delta \rightarrow 0} \frac{[\Pi_i^{\rm uKI}(\delta) -  \Pi_i^{\rm (2)uKI}(\delta)]}{\Pi_i^{\rm (2)uKI}(\delta)}
\end{equation}
where we defined $\delta = 1-f_i$. In this limit the full $\Pi^{\rm uKI}$ correction reads
\begin{align}
    \Pi_i^{\rm uKI}(\delta) & = -\left[ E_i(N-1+f_i) -E(N-1) \right] + f_i \left[ E(N) - E_i(N-1) \right] \nonumber \\ 
    & = E_i(N-1) - E_i(N-\delta) + (1-\delta)[E(N) -E_i(N-1)] \nonumber \\
    & = E(N) - E_i(N-\delta) - \delta[E(N) -E_i(N-1)] \nonumber \\
    & = \delta \lambda_i -\frac{1}{2}\delta^2 k_i +o(\delta^3) -  \delta(\lambda_i -\frac{1}{2}k_i + \tilde{\Delta}^{\rm u}_i) = \nonumber \\
    & = \frac{1}{2}\delta(1-\delta)k_i - \delta \tilde{\Delta}^{\rm u}_i +o(\delta^3)= \nonumber \\ 
    & = \Pi_i^{\rm (2)uKI} - \delta \tilde{\Delta}^{\rm u}_i + o(\delta^3)
\end{align}
where $\lambda_i = \left. \frac{\partial E^{\rm DFT}}{\partial\delta}\right|_0 = \langle \phi_i | \hat{H}^{\rm DFT} | \phi_i \rangle$ , $k_i = \left. \frac{\partial^2 E^{\rm DFT}}{\partial \delta^2}\right|_0 = \langle n_i | F_{\rm Hxc} | n_i \rangle$, and $\tilde{\Delta}^{\rm u}_i = E(N)-E_i(N-1)-\lambda_i + \frac{1}{2}k_i$. In the expression above its understood that all the DFT total energies at different number of particle are computed in the frozen orbital approximation, i.e., assuming that the orbitals are the one from the reference N-electron calculation. 
The renormalization of the screening coefficient becomes:
\begin{equation}
    \Delta \alpha_i = \lim_{\delta \rightarrow 0} \frac{[\Pi_i^{\rm uKI}(\delta) -  \Pi_i^{\rm (2)uKI}(\delta)]}{\Pi_i^{\rm (2)uKI}(\delta)} = 
     \lim_{\delta \rightarrow 0} \frac{-\delta \tilde{\Delta}^{\rm u}_i}{\frac{1}{2}\delta(1-\delta)k_i} = \frac{-2\tilde{\Delta}^{\rm u}_i}{\langle n_i | F_{\rm Hxc} | n_i \rangle}
\end{equation}
which is the final result reported in the main text.

Equivalently, from the generalized piece-wise linearity condition of KI:
\begin{align}
   & \langle \phi_i | \hat{H^{\rm (2)KI}} | \phi_i \rangle = \frac{dE^{\rm KI}}{df_i} = E(N) - E_i(N-1) = \lambda_i -\frac{1}{2}k_i + \tilde{\Delta}_i^{r} \nonumber \\
   & \lambda_i + \tilde{\alpha_i} \langle \phi | v_i^{\rm (2)KI} | \phi_i \rangle = \lambda_i -\frac{1}{2}k^{\rm r}_i + \tilde{\Delta}_i^{\rm r} \nonumber \\
   & -\tilde{\alpha}_i \frac{1}{2} k_i^{\rm u} = - \frac{1}{2} k^{\rm r}_i + \tilde{\Delta}_i^{\rm r} \simeq - \frac{1}{2} k^{\rm r}_i + \tilde{\Delta}_i^{\rm u} \nonumber \\
   & \tilde{\alpha}_i = \frac{k^{\rm r}_i - \tilde{\Delta}_i^{\rm u}}{k^{\rm u}_i}  = \alpha_i - \frac{2\tilde{\Delta}^{\rm u}_i}{\langle n_i | F_{\rm Hxc} | n_i \rangle}\nonumber \\
   & \Delta\alpha^{\rm u}_i = \frac{-2\tilde{\Delta}_i}{\langle n_i | F_{\rm Hxc} | n_i \rangle}
\end{align}
This show that neglecting $\tilde{\Delta}_i$ reduces to the linear response screening coefficients~\cite{colonna_jctc_2018, colonna_jctc_2022}. The additional contribution would account for the deviation of the second order Taylor expansion from the $\Delta$SCF DFT energy $E(N)-E_i(N-1)$.

\end{document}